\documentclass[11pt,a4paper]{article}
\usepackage[british]{babel}
\usepackage[utf8]{inputenc}

\usepackage{dsfont}
\usepackage{bm}
\usepackage{multirow}
\usepackage{mathtools}
\usepackage{amsfonts}
\usepackage{setspace}

\usepackage{jheppub}

\newcommand{\Z}{\mathbb{Z}}			%set of integers/cyclic group
\newcommand{\perm}{\mathcal{S}}		%permutation group

\renewcommand{\O}{\mathcal{O}}		%big-O notation
\newcommand{\p}{\partial}               %partial derivative
\newcommand{\imp}{\quad\Rightarrow\quad}%implication
\newcommand{\e}{\varepsilon}		%shorthand varepsilon
\newcommand{\inv}{{-1}}			%inverse
\newcommand{\1}{\mathds{1}}			%unit matrix
\newcommand{\id}{\text{id}}			%identity permutation

\newcommand{\lagr}{\mathcal{L}}		%Lagrangian
\newcommand{\flav}{\mathcal{F}}		%flavour structure
\newcommand{\ampl}{\mathcal{M}}		%(stripped) amplitude
\newcommand{\fsplit}{\mathcal{R}}		%set of flavour splits
\newcommand{\vrtf}{\mathcal{V}}		%(stripped) vertex factor
\newcommand{\basis}{\mathcal{B}}		%mandelstam basis
\newcommand{\transf}[1]{\overset{#1}{\longrightarrow}}	%transformation arrow
\newcommand{\lagt}[2]{L_{#1,#2}}		%LEC for general Lagrangian term
\newcommand{\lagm}[2]{\O_{#1,#2}}		%monomial for general Lagrangian term

\newcommand{\kd}[1]{\delta^{#1}}					%kronecker delta
\newcommand{\tr}[1]{{\left\langle#1\right\rangle}}	%short-form trace
\newcommand{\comm}[2]{{\left[#1,#2\right]}}			%commutator
\newcommand{\acomm}[2]{{\left\{#1,#2\right\}}}		%anticommutator

\newcommand{\card}[1]{{\left|#1\right|}}			%cardinality
\renewcommand{\mod}[1]{\:(\mathrm{mod}\:#1)}			%modulo

\newcommand{\chpt}{{$\chi$PT}}					%ChPT or xPT?

\newcommand{\ol}[1]{{\overline{#1}}}				%abbreviated index lists
\newcommand{\A}{\mathrm{A}}						%hexadecimal digits
\newcommand{\B}{\mathrm{B}}
\newcommand{\C}{\mathrm{C}}

\newcommand{\form}{\textsc{form}}
\newcommand{\fodge}{\textsc{fodge}}
\newcommand{\TikZ}{{Ti\emph{k}Z}}
\newcommand{\cpp}{C%
	\nolinebreak\hspace{-.05em}\raisebox{.4ex}{\tiny\bf +}%
	\nolinebreak\hspace{-.05em}\raisebox{.4ex}{\tiny\bf +}%
	}

\newlength{\eqindent}
\newlength{\peqindent}
\newlength{\ppeqindent}
\newlength{\pppeqindent}

\newcommand{\eqbreak}[1]{%
	\notag\\&\hspace{\eqindent}#1\,}
\newcommand{\peqbreak}[1]{%
	\right.\notag\\&\hspace{\peqindent}\left.#1\,}
\newcommand{\ppeqbreak}[1]{%
	\right.\right.\notag\\&\hspace{\ppeqindent}\left.\left.#1\,}
\newcommand{\pppeqbreak}[1]{%
	\right.\right.\right.\notag\\&\hspace{\pppeqindent}\left.\left.\left.#1\,}
	
\newcommand{\mvphantom}{\vphantom{\frac{(s_{12})}{s_{12}}}}

\usepackage{tikzexternal}

\newcommand{\externalprefix}{figures/fig-}
\tikzsetexternalprefix{\externalprefix}

\newcommand{\tikzineq}[3][]{%
	\raisebox{-0.5\height}{\includegraphics{\externalprefix#3.pdf}}}

\newcommand{\ordidx}[1]{\textbf{\tiny #1}}
\newcommand{\fodgespace}{\quad}

\title{Higher-order tree-level amplitudes\\in the nonlinear sigma model}
\author[a]{Johan Bijnens,}
\author[b]{Karol Kampf,}
\author[a]{Mattias Sjö}

\affiliation[a]{Department of Astronomy and Theoretical Physics, Lund University,\\
	Sölvegatan 14A, Lund, Sweden}
\affiliation[b]{Institute of Particle and Nuclear Physics, Charles University,\\
	V Holešovičkách 2, Prague, Czech Republic}

\emailAdd{bijnens@thep.lu.se}
\emailAdd{karol.kampf@mff.cuni.cz}
\emailAdd{mattias.sjo@thep.lu.se}

\abstract{We present a generalisation of the flavour-ordering method applied
to the chiral nonlinear sigma model with any number of flavours. We use an
extended Lagrangian with terms containing any number of derivatives,
organised in a power-counting hierarchy. The method allows diagrammatic
computations at tree-level with any number of legs at any order in the
power-counting. Using an automated implementation of the method,
we calculate amplitudes ranging from 12 legs at leading order, $\O(p^2)$,
to 6 legs at next-to-next-to-next-to-leading order, $\O(p^8)$. In addition
to this, we generalise several properties of amplitudes in the nonlinear
sigma model to higher orders. These include the double soft limit and the
uniqueness of stripped amplitudes.}

\preprint{\begin{minipage}{3cm}LU TP 19-46\\September 2019
  \end{minipage}}

\begin{document}

\maketitle

\section{Introduction}
\label{sec:introduction}

In 1960, Gell-Mann and Lévy~\cite{gell-mann-levy} proposed a number of models
for mesons and nucleons. Two of these, the linear and nonlinear sigma models,
were extended to highly general quantum field theories with many different
applications. One of the most important application is interaction of mesons
described by the nonlinear sigma model (NLSM) extended by
Weinberg~\cite{weinberg-chpt} and Gasser and
Leutwyler~\cite{gasser-leutwyler-1, gasser-leutwyler-2} into chiral
perturbation theory (\chpt). 
A recent introductory review is~\cite{Pich:2018ltt} and more
introductory literature can be found at~\cite{chpthomepage}.
This effective field theory (EFT) of low-energy QCD is not only widely used
today in many phenomenological applications, but also motivated further
theoretical avenues for the beyond-standard-model physics such as technicolour
and little Higgs models. Examples of recent work in \chpt~is the calculation
of meson-meson scattering for a general number of
flavours at two loops \cite{meson-meson}, and masses and decays up to
next-to-next-to-leading order \cite{pi-mass-decay}. In this paper, we will 
push the study of this type of models in a different direction.

Even at tree-level, diagrammatic many-particle calculations in EFTs become very
complicated due to the rapidly increasing number of terms in the effective
Lagrangian, but can be facilitated with tools similar to those used for gluon
scattering in perturbative QCD. In recent years, the renewal of interest
in the $S$-matrix program for the gauge theory and gravity has in fact led to
progress in both simplification of complicated technical calculations as well as
discoveries of new properties~\cite{Elvang:2013cua}. The possibility to apply
similar amplitude methods to EFTs started recently and is mainly connected
with studies of the NLSM. First, it was demonstrated that it is indeed possible
to employ recursive methods in~\cite{Kampf:2012fn}, further studied and
developed in~\cite{Cheung:2015ota}. The crucial ingredient in developing the
recursive formula is the existence of the so-called Adler zero~\cite{adler},
the vanishing of scattering amplitudes for soft momenta of Goldstone bosons
(pions for NLSM), as a consequence of a spontaneous symmetry breaking in EFT.
The argument can be also inverted and used for classification of the allowed
space of EFT theories based on their soft properties. It turned out that the
leading order of NLSM is one important representative of exceptional EFTs.
% (besides DBI and special Galileon [???]). 
The exceptional status of those theories is connected with the fact that all
their interaction vertices are uniquely fixed by a single coupling constant,
most conveniently the lowest four-point vertex. This can be labelled as a
soft-bootstrap program, studied and developed in recent years by several
groups \cite{Cheung:2016drk,Elvang:2018dco,Low:2019ynd}.
It represents a rebirth of similar attempts at the end of the 1960s
\cite{Osborn:1969ku,Susskind:1970gf,Ellis:1970nt}.

The exceptional theories have also appeared in completely different context,
the so-called CHY scattering equation \cite{Cachazo:2014xea}, studied more
recently also in \cite{Gomez:2019cik}. This indeed suggests their uniqueness,
and though of completely different nature, it hints to deeper connections with
gauge theory and gravity. It is probably one of the main motivation behind
the recent increase of activities  in studying theoretical properties of NLSM:
\cite{Chen:2013fya,Chen:2014dfa,Du:2015esa,Low:2015ogb,Du:2016tbc,Carrasco:2016ldy,Du:2016njc,Cheung:2017yef,Low:2017mlh,Low:2018acv,Rodina:2018pcb,Mizera:2018jbh,Bjerrum-Bohr:2018jqe}.
This effort demonstrates the importance of NLSM; however, these studies mainly
concentrated only on the leading, two-derivative ($\O(p^2)$) order. As pointed
out in \cite{Low:2019ynd}, it is important to expand the on-shell soft bootstrap
program to higher orders. Our work aims in this direction. An early attempt is
\cite{Cornwell:1971sp} and one that appeared during the writing up of this
paper is \cite{Carrillo-Gonzalez:2019aao}. 

We will mainly focus on the problem of calculating scattering
amplitudes at  tree-level with increasing number of legs and orders,
with possible flavour
splitting, i.e. beyond single-trace amplitudes. 
Using recursion relations, tree-level amplitudes based on the
leading-order term in the Lagrangian have been computed with up to 10 external
particles \cite{Kampf:2013vha}. Using more general recursion relations based
on soft limits \cite{Cheung:2015ota}, 6-particle tree-level interactions have
been computed using the next-to-leading-order Lagrangian \cite{Low:2019ynd}.
These methods suffer limitations when higher-order Lagrangian terms are
used, and can not handle loops. 

In this paper, we generalise an enhanced diagrammatic method called
flavour-ordering, which was introduced in \cite{Kampf:2013vha}. We apply it to a
generalised version of the $SU(N)$ or $U(N)$ chiral NLSM, which includes terms
with arbitrarily high power-counting order in the effective Lagrangian. This
generalisation corresponds to removing all external fields from the general
\chpt\ Lagrangian. The method allows computation of tree-level amplitudes with
any number of external particles using Lagrangian terms of any order, and is
valid also beyond tree-level. It is significantly more efficient than a
brute-force Feynman diagram approach, and the caveats that appear beyond the
leading order can be handled with simple rules. Preliminary results can be found
in the Lund university master thesis \cite{masterthesis}.

In section~\ref{sec:NLSM}, we describe the NLSM and introduce our notation. Our
main new results on the method side are described in section~\ref{sec:flavour}
and \ref{sec:ampl}. Section \ref{sec:flavour} discusses our generalization of
flavour-ordering, while section~\ref{sec:ampl} discusses how this can be used to
calculate more complex amplitudes as well as the kinematic methods needed.
Section~\ref{sec:examples} discusses the amplitudes we have calculated using our
methods; the longer expressions are relegated to appendix \ref{app:results} and
the supplementary material \cite{supplementary}. Our main conclusions are
reviewed in section~\ref{sec:conclusions}. The Lagrangians are given in
appendix~\ref{app:p6lagr}, together with some results regarding renormalisation
of the amplitudes. Appendix \ref{app:ortho} contains the proof of the
orthogonality of flavour structures. The double soft limit with multiple traces
is derived in appendix \ref{app:dslim}, and appendix \ref{app:closed-basis}
derives the minimal bases of kinematic variables used in the amplitude
calculations.
%%%%%%%%%%%%%%%%%%%%%%%%%%%%%%%%%%%%%%%%%%%%%%%%%%%%%%%%%%%%%%%%%%%%%%%%%%%%%

\section{The nonlinear sigma model}
\label{sec:NLSM}

The nonlinear sigma model describes the Nambu-Goldstone bosons that arise when
 a global symmetry group $G$ is broken to a subgroup $H$. Each configuration
 of the Nambu-Goldstone fields can be uniquely mapped to an element of the
 coset space $G/H$, and from each such coset, a representative $\xi(\phi)$ may
 be chosen to represent the field configuration $\phi$. 

In the context of low-energy QCD, the group $G$ is the chiral
 group $SU(N_f)_L\times SU(N_f)_R$, which is a global symmetry of the massless 
QCD Lagrangian with $N_f$ quark flavours. It is broken to the diagonal
 subgroup $H=SU(N_f)_V$, so the coset space $G/H$ is isomorphic to $SU(N_f)$.
With a chiral decomposition of the coset representatives, $\xi=(\xi_L,\xi_R)$,
we may represent the Nambu-Goldstone fields with the unitary
matrix $u(\phi) = \xi_R(\phi) = \xi_L^\dag(\phi)$ parametrised as
\begin{equation}
	u(\phi) = \exp\left(\frac{i\Phi(\phi)}{F\sqrt2}\right),\qquad
          \Phi(\phi) = t^a\phi^a
	\label{eq:udef}
\end{equation}
with the flavour index with $a$ running from 1 to $N_f^2-1$. Here, $t^a$ are
the generators of $SU(N_f)$, and $F$ is a constant.\footnote{The above
expression for $u(\phi)$ is only one of many possible parametrisations,
but is the most common.} We use Einstein's summation notation without
distinction between upper and lower flavour indices, and use the following
normalisation for the generators:
\begin{equation}
	\tr{t^at^b} = \kd{ab},\qquad \comm{t^a}{t^b} = i f^{abc}t^c,
\end{equation}
where $\tr{\cdots}$ denotes a trace over internal indices. Here, $f^{abc}$
are the totally antisymmetric structure constants of $SU(N_f)$. With this
convention, the $SU(2)$ generators can be chosen such that they relate to the
Pauli matrices as $t^a = \sigma^a/\sqrt2, f^{abc}=\epsilon^{abc}\sqrt2$.
Likewise, the $SU(3)$ generators can be chosen in terms of the Gell-Mann
matrices like $t^a = \lambda^a/\sqrt2$.

Under a chiral transformation $g=(g_L,g_R)$, $u(\phi)$ transforms as
\begin{equation}
	u \transf{g} g_R\, u\, h^\dag(u,g) = h(u,g)\, u\, g_L,
\end{equation}
where the compensating transformation $h(u,g)\in H$ is defined by the above
relation.

When constructing the most general symmetry-consistent Lagrangian, it is more
con\-venient to replace $u$ by
\begin{equation}
	u_\mu = i\left(u^\dag\p_\mu u - u \p_\mu u^\dag\right),\qquad
       u_\mu \transf{g} h(u,g) u_\mu h(u,g)^\dag,
	\label{eq:def-u}
\end{equation}
which was introduced in this context by \cite{better-basis}. Higher derivatives
are applied through the covariant derivative
\begin{equation}
	\nabla_\mu X = \p_\mu X + \comm{\Gamma_\mu}{X},\qquad
       \Gamma_\mu = \frac12\left(u^\dag \p_\mu u + u\p_\mu u^\dag\right),
	\label{eq:cov-deriv}
\end{equation}
which has the convenient properties
\begin{equation}
 X \transf{g} h X h^\dag \imp \nabla_\mu X \transf{g} h \nabla_\mu X h^\dag,\qquad
	\nabla_\mu u_\nu - \nabla_\nu u_\mu = 0.
	\label{eq:cov-comm}
\end{equation}
Note that we do not include the external fields that are used in chiral
perturbation theory.

The Lagrangian is often written using derivatives of $U(\phi) = u(\phi)^2$
and its conjugate. It is possible to convert directly
between $\p_\mu U^{(\dag)}$ and $u_\mu$ by using unitarity:
\begin{equation}
\p_\mu U^\dag \p_\nu U = -(U^\dag \p_\mu U) (U^\dag \p_\nu U),\qquad
U^\dag \p_\mu U = u^\dag u^\dag\p_\mu u u - \p_\mu u^\dag u = -i u^\dag u_\mu u.
\end{equation}
This makes $\p_\mu U^\dag\p_\nu U$ wholly interchangeable with $u_\mu u_\nu$
inside a trace.

With the above definitions, the simplest valid term in the NLSM Lagrangian is
\begin{equation}
	\lagr_2 = \frac{F^2}{4}\tr{u_\mu u^\mu},
	\label{eq:lagr2}
\end{equation}
where the constant in front is fixed by the canonical normalisation of the
kinetic term.

Beyond this, there is an infinite sequence of increasingly complex terms
permitted by the chiral symmetries\footnote{Many authors refer
to $\lagr_2$ as the
full Lagrangian of the NLSM. We instead use ``the NLSM'' to refer to the more
general version, which includes also terms with more derivatives.}.
We also impose parity ($P$), charge-conjugation ($C$) and Lorentz invariance.
We restrict to the sector that involves an even number of
Levi-Civita tensors ($\epsilon_{\mu\nu\alpha\beta}$), which can always be
rewritten in terms of the Minkowski metric only.
The terms can be organised into a hierarchy based on power counting in the
momentum scale $p$. Since each derivative in the Lagrangian brings down one
factor of $p$ into an amplitude, both $u_\mu$ and $\nabla_\mu$ are $\O(p)$
and the power-counting at the Lagrangian level is simply counting derivatives.
Thus, we may split the Lagrangian as
\begin{equation}
	\lagr = \sum_{n=1}^\infty \lagr_{2n},
\end{equation}
where $\lagr_{2n}$ is $\O(p^{2n})$ and contains $2n$ derivatives carrying $n$
pairs of Lorentz indices. Assuming a low momentum scale, we may then ignore
all terms above a certain $n$.

The four-derivative $\O(p^4)$ Lagrangian is, for
general $N_f$ \cite{gasser-leutwyler-1, gasser-leutwyler-2, p6lagr},
\begin{equation}
	\lagr_4 = 
		  L_0\tr{u_\mu u_\nu u^\mu u^\nu} 
		+ L_1\tr{u_\mu u^\mu}\tr{u_\nu u^\nu} 
		+ L_2\tr{u_\mu u_\nu}\tr{u^\mu u^\nu} 
		+ L_3\tr{u_\mu u^\mu u_\nu u^\nu}.
		\label{eq:lagr4}
\end{equation}
The $L_i$ are independent coupling constants, so-called low-energy
constants (LECs). It is in principle possible to derive the LECs from
any underlying theory (e.g.\ QCD), but in practice, they are unknown
parameters that must be measured by experiments or lattice simulations.

The Lagrangian is known also at $\O(p^6)$ and $\O(p^8)$. The latter is the
first 135 terms in the \chpt\ Lagrangian of \cite{p8lagr}; the former has
only been published with different notation and formulated in a way that
gives redundant terms when naïvely reduced to the NLSM \cite{p6lagr}. A more
compatible version, constructed in conjunction with \cite{p8lagr}, is given in
appendix~\ref{app:p6lagr}. The Lagrangian at $\O(p^{10})$ and above has not
been studied. 

\subsection{Restrictions due to fixed $N_f$ and dimensionality}
\label{sec:small-Nf}

The Lagrangians discussed above are the most general ones. They are valid in
any dimension and for a generic number of flavours.

When $N_f$ is small, the Cayley-Hamilton theorem gives additional linear
relations that reduce the number of independent terms. The theorem states that
for any $N_f\times N_f$ matrix $M$, the characteristic polynomial
\begin{equation}
	p(\lambda) = \det(\lambda\1 - M),
\end{equation}
which is zero whenever $\lambda$ is an eigenvalue of $M$, 
is also satisfied by $M$, i.e.\ $p(M) = 0$ when viewed as a matrix polynomial.
When $N_f=2$, this implies the identity
\begin{equation}
  \acomm{A}{B} = \tr{AB},
\end{equation}
for traceless $2\times2$ matrices $A,B$. When $N_f=3$, the identity is
\begin{equation}
\sum_{\substack{\text{permutations of}\\\{ABC\}}}\tr{ABCD}\quad
= \sum_{\substack{\text{cyclic permutations}\\\text{of }\{ABC\}}}\tr{AB}\tr{CD},
\end{equation}
for traceless matrices. The relations when $\tr{A}\ne0$ used
in \cite{p6lagr,p8lagr} contain many more terms.

As an example, we may choose $A=C=u_\mu$ and $B=D=u_\nu$; these are traceless
as a consequence of the identity $\p_\mu\det(A) = \det(A)\tr{A^\inv\p_\mu A}$,
which holds for any invertible $A$, and which reduces
to $\tr{A^\dag\p_\mu A}=0$ when $A\in SU(N_f)$.

The $N_f=2$ identity allows for the elimination of all Lagrangian terms
containing a product of two or more traces from any $\lagr_{2n}$; for
instance, $L_1$ and $L_2$ may be eliminated from $\lagr_4$.
The $N_f=3$ identity allows for the removal of a single term from $\lagr_4$,
7 terms from $\lagr_6$, and so on. The standard choice is to remove the
$L_0$-term of \eqref{eq:lagr4} for $N_f=3$ \cite{gasser-leutwyler-2}, and the $L_0$- and $L_3$-terms
for $N_f=2$ \cite{gasser-leutwyler-1}.

When the spacetime dimension $D$ is finite, the Schouten identity
implies
\begin{equation}
	\left(f^{\mu_1\mu_2\cdots\mu_k}u_{\mu_1}u_{\mu_2}\cdots u_{\mu_k}\right)^2 = 0\quad \text{if $k > D$},
\end{equation}
where $f^{\mu_1\mu_2\cdots\mu_k}$ is antisymmetric in all its indices.
This results in additional linear relations among the terms in $\lagr_{2k}$
for $k > D$. For $D=4$, this does not affect any of the currently known orders.
In the sector involving a single Levi-Civita tensor it already removes
a large number of terms at $\O(p^6)$.

%%%%%%%%%%%%%%%%%%%%%%%%%%%%%%%%%%%%%%%%%%%%%%%%%%%%%%%%%%%%%%%%%%%%%%%%%%%%

\section{Flavour-ordering}
\label{sec:flavour}

With the structure of the NLSM established, we are ready to use it for
perturbative calculations of scattering amplitudes. However, the infinite
number of interaction terms requires the use of some scheme for restricting
it to a manageable subset. Even then, the resulting vertex factors are very
intricate, both in their dependence on the particle momenta, and in their
group-algebraic structure. This leaves only the simplest Feynman diagrams
tractable by hand, and even computer algebra becomes highly time-consuming
when tackling more complicated cases directly.

In this section, we will direct much effort towards the development of
simpler ways to perform these calculations. As we will see, the
group-algebraic structure of the flavour indices carried by the particles
can be used to condense an amplitude into a much more easily manageable
expression, for which simpler calculation rules exist. We will mostly follow
the derivation of $\O(p^2)$ flavour-ordering as presented
in~\cite{Kampf:2013vha}, but insert the notation to support
our own generalisations to higher-order vertices.

\subsection{Some notation}
\label{sec:notation}

In this section, we will need a compact notation for writing the flavour
structures of scattering amplitudes. A flavour structure is a product of one
or more traces containing group generators carrying the flavour indices of
the external particles in some order. We will represent this as
\begin{equation}
	\flav_\sigma(r_1,r_2,\ldots) = 
		\tr{t^{a_{\sigma(1)}}\cdots t^{a_{\sigma(r_1)}}}
		\tr{t^{a_{\sigma(r_1+1)}}\cdots t^{a_{\sigma(r_1+r_2)}}}\cdots.
\end{equation}
%where we write $a_i$ rather than $t^{a_i}$ for readability. 
The $i$th trace contains $r_i$ generators, ordered by a permutation
$\sigma\in\perm_n$. For example,
$\tr{{a_1}{a_3}}\tr{{a_2}{a_4}} = \flav_{1324}(2,2)$, and
$\tr{{a_1}{a_2}\cdots {a_n}}=\flav_\id(n)$, where $\id(i)=i$ is the identity
permutation.

We encapsulate the $r_i$ in
 \mbox{$R = \{r_1,r_2,\ldots,r_\card{R}\}$}.
We call $R$ a \emph{flavour splitting}.
$\card{R}$ is the number of traces in the flavour structure, and we 
write $\flav_\sigma(R)$ rather than $\flav_\sigma(r_1,\ldots)$. For a structure
with $n$ indices, we impose the restrictions
\begin{equation}
	\sum_{i=1}^\card{R} r_i = n,\qquad r_1\leq r_2\leq\ldots\leq r_\card{R}.
	\label{eq:fsplit-sort}
\end{equation}
The latter limits the number of equivalent ways to write a flavour structure.

Since traces are cyclic, $\flav_\sigma(R)$ will be invariant under cyclic
permutations of the indices inside each trace. If $r_i=r_j$, it will also be
invariant under swapping the contents of the $i$th and $j$th trace. As a
generalisation of the cyclic group $\Z_n$, we define $\Z_R$ to be the group
of all permutations under which $\flav_\sigma(R)$ is invariant. For instance,
\begin{equation}
	\begin{split}
		\Z_{\{2,2\}} &= \{12\,34, 21\,34, 12\,43, 21\,43, 34\,12, 43\,12, 34\,21, 43\,21\},\\
		\Z_{\{2,4\}} &= \{12\,3456, 21\,3456, 12\,4563, 21\,4563, 12\,5634, 21\,5634, 12\,6345, 21\,6345\},
	\end{split}
\end{equation}
where we label a permutation by how $12\ldots n$ ends up. We have inserted 
spaces between blocks of indices corresponding to different traces to make 
it more legible.

In this notation, we generalise the notion of two permutations being
equivalent modulo a cyclic permutation: we write
$\sigma \equiv \rho \mod{\Z_R}$ if $\flav_\sigma(R) = \flav_\rho(R)$. For
instance, $1234\equiv 2341\mod{\Z_{\{4\}}}$ and
$1234\equiv2134\mod{\Z_{\{2,2\}}}$.

$\Z_{\{2,2\}}$ is isomorphic to the dihedral group $D_4$. Other $\Z_R$ are not
isomorphic to such well-known groups, but $\Z_{\{2,4\}}\simeq \Z_2\times\Z_4$,
and in general, $\Z_R\simeq \Z_{r_1}\times \Z_{r_2}\times\cdots$ whenever
all $r_i$ are different. When some $r_i$ are equal (say, $m$ in a row), the
group will be non-abelian and isomorphic to a semidirect product,
e.g. $\Z_{\{2,2,2\}}\simeq (\Z_2\times\Z_2\times\Z_2)\rtimes \perm_3$.
In general, $\Z_R\simeq (\Z_{r_1}\times \Z_{r_2}\times\cdots)\rtimes
  (\perm_{m_1}\times \perm_{m_2}\times\cdots)$, where each $m_j$ is the length
of a stretch of equal $r_i$.\footnote{The proof follows from the following
definition of the semidirect product: if a group $G$ has a subgroup $H$ and a
normal subgroup $N$, then $G=N\rtimes H$ if $G=\{nh\:|\: n\in N,h\in H\}$ and
$N\cap H = e$, the identity element. The groups $N\simeq(\Z_{r_1}\times\cdots)$
of cyclings within traces and $H\simeq(\perm_{m_1}\times\cdots)$ of swaps of
equal-size traces are clearly subgroups of $G=\Z_R$, and $N$ is normal since
$gng^\inv \in N$ for any $n\in N$, $g\in G$ --- any trace swaps in $g$ are
cancelled by $g^\inv$, leaving only cyclings. Any permutation in $\Z_R$ is the
composition of a cycling and a trace swap, and the only element shared
by $N$ and $H$ is $\id$, which completes the proof.}

\subsection{Stripped vertex factors}

Each term in the Lagrangian will produce an infinite tower of interaction
vertices with increasingly many legs. Due to parity and the absence of
Levi-Civita tensors, only terms with an even number of legs are produced.
If the Lagrangian term contains a product of several traces,
the flavour indices of the corresponding vertices will be distributed between
the same number of traces in multiple ways. If a trace contains an even number
of $u_\mu$'s in the Lagrangian, the corresponding trace in the vertices will
only contain an even number of indices, again from parity.

We will organise the vertices by their power-counting order and flavour
splitting. For instance, in the expansion of $\lagr_2$ \eqref{eq:lagr2},
\begin{equation}
	\lagr_2 = \frac12\tr{t^at^b}\p_\mu\phi^a\p^\mu\phi^b + \frac{1}{F^2}\tr{t^at^bt^ct^d}\left(\frac{1}{6}\phi^a\p_\mu\phi^b\phi^c\p^\mu\phi^d - \frac{1}{12}\phi^a\phi^b\p_\mu\phi^c\p^\mu\phi^d\right) + \ldots,
	\label{eq:p2lagr-exp}
\end{equation}
both terms attached to the 4-index trace will be part of the $\O(p^2)$ vertex
with splitting \mbox{$R=\{4\}$}, which we label $\vrtf^{abcd}_{2,\{4\}}$,
a vertex (factor).
At order $p^m$, a specific flavour splitting $R$ for a vertex with $n$ legs, and thus $n$ flavour indices $a_i$, will have a vertex
factor $\vrtf^{a_1\ldots a_n}_{m,R}$.
It will in general contain contributions from many different Lagrangian terms,
but we treat it as a single factor for the purposes of Feynman diagrams.

We can further organise the contents of
an $n$-point $\O(p^m)$ vertex by flavour structure, i.e. all possible
distributions of the $n$-flavour indices over the flavour splitting $R$:
\begin{equation}
	\vrtf_{m,R}^{a_1a_2\cdots a_n}(p_1,p_2\ldots, p_n) = \sum_{\sigma\in\perm_n/\Z_R} \flav_\sigma(R)\,\vrtf_{m,R,\sigma}(p_1,p_2,\ldots,p_n),
	\label{eq:vertex}
\end{equation}
where $\vrtf_{m,R,\sigma}$ contains whatever kinematic factors come attached to $\flav_\sigma(R)$. 
Due to the derivatives, the kinematic factors $\vrtf_{m,R,\sigma}$ are functions
of the momenta $p_i$ of the interacting particles. Here and in all other
places, we treat all momenta as ingoing. 
Since $\flav_\sigma(R)$ is invariant under $\Z_R$, the kinematic factors must
also have this symmetry, i.e.
\begin{equation}
	\vrtf_{m,R,\sigma}(p_1,p_2,\ldots,p_n) = \vrtf_{m,R,\sigma}(p_{\rho(1)},p_{\rho(2)},\ldots,p_{\rho(n)})
	\label{eq:vert-cycle}
\end{equation}
for any $\rho\in \Z_R$. Also, Bose symmetry implies that the act of rearranging
the legs of the vertex by any permutation $\rho\in\perm_n$ must have the effect
\begin{equation}
	\vrtf_{m,R,\sigma\circ\rho}(p_1,p_2,\ldots,p_n) = \vrtf_{m,R,\sigma}(p_{\rho(1)},p_{\rho(2)},\ldots,p_{\rho(n)}),
	\label{eq:vert-bose}
\end{equation}
where $\circ$ denotes composition of permutations. Specifically,
\begin{equation}
	\vrtf_{m,R,\sigma}(p_1,p_2,\ldots,p_n) = \vrtf_{m,R}(p_{\sigma(1)},p_{\sigma(2)},\ldots,p_{\sigma(n)}),
	\label{eq:vert-strip}
\end{equation}
where $\vrtf_{m,R} = \vrtf_{m,R,\id}$ is called a \emph{stripped} vertex
factor.\footnote{The word ``stripped'' is typical in the context of EFTs.
For the analogous concept in perturbative QCD (where ``flavour'' is replaced
by ``colour''), the word ``primitive'' is used instead;
see e.g. \cite{reuschle,colour-order}. In older literature,
the word ``dual'' is common.}
It contains all the necessary information of the full vertex factor, but is
only a kinematic factor with no flavour structure. It can be ``dressed'' into
a full vertex factor by the simple act of multiplying by $\flav_\id(R)$ and
then summing over all $\sigma\in\perm_n/\Z_R$. 

A stripped vertex factor has the property of being \emph{flavour-ordered},
since it is the kinematic factor attached to $\flav_\id(R)$, where all flavour
indices are sorted in ascending order. Thanks to this, its explicit form can
be derived by expanding the relevant Lagrangian terms and discarding all terms
where any flavour index appears out of order. This saves a significant amount
of work for the more complicated vertices.

Stripped vertices serve as the first ingredient in our method. In the
following sections, we treat diagrams and amplitudes along the same lines.

\subsection{Stripped amplitudes}

Like the vertices, we may organise the diagrams by their power-counting order
and flavour structure. The order can be determined by using Weinberg's
power-counting formula,
\begin{equation}
D = 2 + 2L + \sum_d(d-2)N_d,
\label{eq:powercount}
\end{equation}
which states that a diagram containing $L$ loops and $N_d$ $\O(p^d)$ vertices
is $\O(p^D)$ overall. Due to the form $(d-2)$, a diagram may contain any
number of $\O(p^2)$ vertices without changing its order.
%An $\O(p^m)$ tree-level diagram may contain a single $\O(p^m)$ vertex,
% or an $\O(p^{m-2})$ vertex and an $\O(p^4)$ vertex, and so on. Each loop
% increases the power as much as the addition of an $\O(p^4)$ vertex.

As for the vertex factors, we may decompose the $\O(p^m)$ $n$-point amplitude
as
\begin{equation}
	\ampl_{m,n}^{a_1a_2\cdots a_n}(p_1,p_2,\ldots,p_n) = 
		\sum_{R\in\fsplit(m,n)}\sum_{\sigma\in\perm_n/\Z_R}
		\flav_{\sigma}(R)
		\ampl_{m,R,\sigma}(p_1,p_2,\ldots,p_n),
\end{equation}
where $\ampl_{m,R,\sigma}$ carries all kinematic factors, and $\fsplit(m,n)$
contains all flavour splittings that contribute to the amplitude. Its contents
will become apparent when drawing diagrams.

The direct analogues of (\ref{eq:vert-cycle}--\ref{eq:vert-strip}) hold also for $\ampl_{m,R,\sigma}$, and we may similarly define the \emph{stripped amplitude} $\ampl_{m,R}$ with the property
\begin{equation}
	\ampl_{N,R,\sigma}(p_1,p_2,\ldots,p_n) = \ampl_{m,R}(p_{\sigma(1)},p_{\sigma(2)},\ldots,p_{\sigma(n)}).
\end{equation}
It is sufficient to compute the stripped amplitude, since summing
over flavour splittings and permutations,
\begin{equation}
	\ampl_{m,n}^{a_1a_2\cdots a_n}(p_1,p_2,\ldots,p_n) = 
		\sum_{R\in\fsplit(N,n)}\sum_{\sigma\in\perm_n/\Z_R}
		\flav_{\sigma}(R)
		\ampl_{m,R}(p_{\sigma(1)},p_{\sigma(2)},\ldots,p_{\sigma(n)})\,,
	\label{eq:ampl}
\end{equation}
gives the full amplitude.

\subsection{Flavour-ordered diagrams}

Due to its relative simplicity, the stripped amplitude serves as the target
of our methods. Like the stripped vertex factors, it is flavour-ordered, so
when calculating it, we may discard all terms where two flavour indices
appear out of order. We can derive further simplifications by studying how
the flavour structures behave when two sub-diagrams are joined by propagators.
The NLSM Feynman rule for a propagator with momentum $q$ is
\begin{equation}
	\tikzineq{
		\draw [thick] (0,0) -- (2,0) node [midway, above] {$q$};
		\draw [fill = black] (0,0) circle[radius = 0.03] node [below] {$a$};
		\draw [fill = black] (2,0) circle[radius = 0.03] node [below] {$b$};
	}{propagator}
        = \frac{i\kd{ab}}{q^2 + i\epsilon},
\end{equation}
so the flavour structures are simply contracted by the delta. For $SU(N_f)$,
the contraction can be performed through the Fierz identity,
\begin{equation}
t^a_{ij}t^a_{k\ell} = \delta_{i\ell}\delta_{jk}
   - \frac{1}{N_f}\delta_{ij}\delta_{k\ell},
\end{equation}
where $ijk\ell$ are the internal indices of the generators. Inside traces,
the identity implies
\begin{align}
\tr{Xt^a}\tr{t^a Y} &= \tr{XY} - \frac{1}{N_f}\tr{X}\tr{Y},
	\label{eq:contr}\\
\tr{Xt^aYt^a} &= \tr{X}\tr{Y} - \frac{1}{N_f}\tr{XY}
	\label{eq:contr-alt}
\end{align}
for arbitrary $X$ and $Y$. For future reference, we will name the first term
on the right-hand side the \emph{multiplet term} and the second term
(containing $N_f^\inv$) the \emph{singlet term}. In $U(N_f)$, the
corresponding identities contain only the multiplet term.

For tree-level diagrams, \eqref{eq:contr} is the relevant identity. Its
multiplet term preserves the ordering of $X$ and $Y$; the singlet term does
not, but we will ignore it for now and deal with it in
section~\ref{sec:singlet}. We then see that the stripped amplitude only gets
contributions from stripped vertex factors (if $X$ or $Y$ is not
flavour-ordered, neither is $XY$) that are combined in ways that maintain
their flavour-ordering. In a diagrammatic view, this is rather intuitive to
achieve; for instance, the following constitutes all the distinct ways to
assemble two 4-point vertices into an $\O(p^2)$ 6-point diagram:
\begin{equation}
	\tikzineq{
		\draw [thick] (-.5,.5)  node [anchor=east] {\tiny 1}  -- (0,0) -- (-.5,0)             node [anchor=east] {\tiny 2} ;
		\draw [thick] (-.5,-.5) node [anchor=east] {\tiny 3}  -- (0,0) -- (1,0) -- (1.5,-.5)  node [anchor=west] {\tiny 4} ;
		\draw [thick] (1.5,0)   node [anchor=west] {\tiny 5}  -- (1,0) -- (1.5,.5)            node [anchor=west] {\tiny 6} ;
	}{label0}\qquad
	\tikzineq{
		\draw [thick] (-.5,.5)  node [anchor=east] {\tiny 2}  -- (0,0) -- (-.5,0)             node [anchor=east] {\tiny 3} ;
		\draw [thick] (-.5,-.5) node [anchor=east] {\tiny 4}  -- (0,0) -- (1,0) -- (1.5,-.5)  node [anchor=west] {\tiny 5} ;
		\draw [thick] (1.5,0)   node [anchor=west] {\tiny 6}  -- (1,0) -- (1.5,.5)            node [anchor=west] {\tiny 1} ;
	}{label1}\qquad
	\tikzineq{
		\draw [thick] (-.5,.5)  node [anchor=east] {\tiny 3}  -- (0,0) -- (-.5,0)             node [anchor=east] {\tiny 4} ;
		\draw [thick] (-.5,-.5) node [anchor=east] {\tiny 5}  -- (0,0) -- (1,0) -- (1.5,-.5)  node [anchor=west] {\tiny 6} ;
		\draw [thick] (1.5,0)   node [anchor=west] {\tiny 1}  -- (1,0) -- (1.5,.5)            node [anchor=west] {\tiny 2} ;
	}{label2}.\label{eq:6-labellings}
\end{equation} 
The labels on the legs refer to external momenta and flavour indices.
Flavour-ordering corresponds to having all indices in cyclic order around the
diagram labelled counter\-clockwise; we will keep this convention in the
remainder. These three labellings give distinct kinematic factors, e.g.
they have distinct propagator momenta \mbox{$(p_1+p_2+p_3)^2$},
\mbox{$(p_2+p_3+p_4)^2$}, and \mbox{$(p_3+p_4+p_5)^2$}, respectively. Due to
the symmetry of the diagram, the remaining three cyclic permutations of the
labels are not distinct from these three. All other labellings are not
flavour-ordered, and can be ignored.

For compactness, we will draw flavour-ordered diagrams with unlabelled legs.
These are defined as the sum over all distinct flavour-ordered ways to label
them. Equivalently, they can be defined as any flavour-ordered labelling,
summed over $\Z_R$, and divided by the factor needed to account for symmetry.
For 4, 6 and 8 particles at $\O(p^2)$, the flavour-ordered diagrams are
\begin{equation}
	\begin{gathered}
		\tikzineq{
			\draw [thick] (-.5,.5) -- (0,0) -- (.5,-.5);
			\draw [thick] (-.5,-.5) -- (0,0) -- (.5,.5);
		}{4pt}\qquad
		\qquad
		\tikzineq{
			\draw [thick] (-.5,.5)   -- (0,0) -- (-.5,0);
			\draw [thick] (-.5,-.5)  -- (0,0) -- (.5,-.5);
			\draw [thick] (.5,0)     -- (0,0) -- (.5,.5);
		}{6pt1}\qquad
		\tikzineq{
			\draw [thick] (-.5,.5)   -- (0,0) -- (-.5,0);
			\draw [thick] (-.5,-.5)  -- (0,0) -- (1,0) -- (1.5,-.5);
			\draw [thick] (1.5,0)    -- (1,0) -- (1.5,.5);
		}{6pt2}
	\\
		\tikzineq{
			\draw [thick] (-.5,.5)   -- (0,0) -- (-.5,0);
			\draw [thick] (-.5,-.5)  -- (0,0) -- (.5,-.5);
			\draw [thick] (.5,0)     -- (0,0) -- (.5,.5);
			\draw [thick] (0,.5)     -- (0,0) -- (0,-.5);
		}{8pt1}\qquad
		\tikzineq{
			\draw [thick] (-.5,.5)   -- (0,0) -- (-.5,0);
			\draw [thick] (0,.5)     -- (0,0) -- (0,-.5);
			\draw [thick] (-.5,-.5)  -- (0,0) -- (1,0) -- (1.5,-.5);
			\draw [thick] (1.5,0)    -- (1,0) -- (1.5,.5);
		}{8pt2}\qquad
		\tikzineq{
			\draw [thick] (-.5,.5)   -- (0,0) -- (-.5,0);
			\draw [thick] (-.5,-.5)  -- (0,0) -- (.5,0) -- (1,0) -- (1.5,-.5);
			\draw [thick] (1.5,0)    -- (1,0) -- (1.5,.5);
			\draw [thick] (.5,.5) -- (.5,0) -- (.5,-.5);
		}{8pt3}\qquad
		\tikzineq{
			\draw [thick] (-.5,.5)   -- (0,0) -- (-.5,0);
			\draw [thick] (-.5,-.5)  -- (0,0) -- (.5,0) -- (1,0) -- (1.5,-.5);
			\draw [thick] (1.5,0)    -- (1,0) -- (1.5,.5);
			\draw [thick] (.25,.5) -- (.5,0) -- (.75,.5);
		}{8pt4}\quad,
	\end{gathered}
	\label{eq:p2diagrs}
\end{equation}
respectively. The second 6-point diagram is the sum of the three
in \eqref{eq:6-labellings}. Stripped vertex factors are completely symmetric
under their respective $\Z_R$ by virtue of \eqref{eq:vert-cycle}, so
single-vertex diagrams always have only one distinct labelling. Therefore,
the 4-point diagram and the first 6-point diagram in \eqref{eq:p2diagrs}
should not be summed over other labellings. The 8-point diagrams have
1, 8, 4, and 8 distinct labellings, respectively, as can be seen from their
symmetry. Note that since the order of the legs of a stripped vertex factor
matters, the last two diagrams are distinct.%, and have different properties.

Above $\O(p^2)$, we begin to encounter flavour-split vertices, but they can
be integrated into the flavour-ordering routine. We still label the legs
according to the identity permu\-tation, but instead of summing over cyclic
permutations, we sum over $\Z_R$, and once again only consider distinct
labellings.

At higher orders, we also need to distinguish vertices of different order,
which is done by attaching a number to all vertices above $\O(p^2)$. In order
to distinguish vertices with split flavour structures, we leave a gap in the
vertex, so that each contiguous piece of a diagram resides in a single trace.
For instance, the 4-point $\O(p^4)$ diagrams are
\begin{equation}
	\tikzineq{
		\draw [thick] (-.5,.5) -- (0,0) -- (.5,-.5);
		\draw [thick] (-.5,-.5) -- (0,0) -- (.5,.5);
		\draw (0,0) node [anchor=south] {\ordidx{4}};
	}{p4unsplit}\qquad
	\tikzineq{
		\draw [thick, rounded corners=3pt] (-.5, .5) -- (0,0) -- (-.5,-.5);
		\draw [thick, rounded corners=3pt] ( .5,-.5) -- (0,0) -- ( .5, .5);
		\draw (0,0) node [anchor=south] {\ordidx{4}};
	}{p4split}
\end{equation}
for $R=\{4\}$ and $R=\{2,2\}$, respectively. Neither diagram has more than one
distinct labelling, since they contain only a single vertex each. The four
lines in the right diagram are still kinematically connected, but are
separated flavour-wise. Since there is a direct correspondence between traces
in a flavour structure and contiguous pieces of a diagram, we will
simply refer to the pieces as \emph{traces}. 

Some adjustment is needed when handling split diagrams. 
Since $\tr{X}\tr{Y} = \tr{Y}\tr{X}$, the traces may ``float''
to different positions around the same vertex. For instance, 
\begin{equation}
	\tikzineq{
		\draw [thick, rounded corners=4pt] (-.5,.5) -- (0,0) -- (-.5,0);
		\draw [thick, rounded corners=4pt] (-.5,-.5)  -- (0,0) -- (1,0) -- (1.5,0);
		\draw [thick] (1.5,-.5)    -- (1,0) -- (1.5,.5);
		\draw (0,0) node [anchor=south] {\ordidx{4}};
	}{float0}\qquad
	\tikzineq{
		\draw [thick, rounded corners=4pt] (-.5,-.5) -- (0,0) -- (-.5,0);
		\draw [thick, rounded corners=4pt] (-.5,.5)  -- (0,0) -- (1,0) -- (1.5,0);
		\draw [thick] (1.5,-.5)    -- (1,0) -- (1.5,.5);
		\draw (0,0) node [anchor=south] {\ordidx{4}};
	}{float1}
\end{equation}
are the same. By our conventions, the distinct labellings of this diagram
are
\begin{equation}
	\tikzineq{
		\draw [thick, rounded corners=4pt] (-.5,.5)  node [anchor=east] {\tiny 1} -- (0,0) -- (-.5,0)          node [anchor=east] {\tiny 2};
		\draw [thick, rounded corners=4pt] (-.5,-.5) node [anchor=east] {\tiny 3} -- (0,0) -- (1,0) -- (1.5,0) node [anchor=west] {\tiny 5};
		\draw [thick]                      (1.5,-.5) node [anchor=west] {\tiny 4} -- (1,0) -- (1.5,.5)         node [anchor=west] {\tiny 6};
		\draw (0,0) node [anchor=south] {\ordidx{4}};
	}{splitlabel0}\qquad
	\tikzineq{
		\draw [thick, rounded corners=4pt] (-.5,.5)  node [anchor=east] {\tiny 1} -- (0,0) -- (-.5,0)          node [anchor=east] {\tiny 2};
		\draw [thick, rounded corners=4pt] (-.5,-.5) node [anchor=east] {\tiny 4} -- (0,0) -- (1,0) -- (1.5,0) node [anchor=west] {\tiny 6};
		\draw [thick]                      (1.5,-.5) node [anchor=west] {\tiny 5} -- (1,0) -- (1.5,.5)         node [anchor=west] {\tiny 3};
		\draw (0,0) node [anchor=south] {\ordidx{4}};
	}{splitlabel1}\qquad
	\tikzineq{
		\draw [thick, rounded corners=4pt] (-.5,.5)  node [anchor=east] {\tiny 1} -- (0,0) -- (-.5,0)          node [anchor=east] {\tiny 2};
		\draw [thick, rounded corners=4pt] (-.5,-.5) node [anchor=east] {\tiny 5} -- (0,0) -- (1,0) -- (1.5,0) node [anchor=west] {\tiny 3};
		\draw [thick]                      (1.5,-.5) node [anchor=west] {\tiny 6} -- (1,0) -- (1.5,.5)         node [anchor=west] {\tiny 4};
		\draw (0,0) node [anchor=south] {\ordidx{4}};
	}{splitlabel2}\qquad
	\tikzineq{
		\draw [thick, rounded corners=4pt] (-.5,.5)  node [anchor=east] {\tiny 1} -- (0,0) -- (-.5,0)          node [anchor=east] {\tiny 2};
		\draw [thick, rounded corners=4pt] (-.5,-.5) node [anchor=east] {\tiny 6} -- (0,0) -- (1,0) -- (1.5,0) node [anchor=west] {\tiny 4};
		\draw [thick]                      (1.5,-.5) node [anchor=west] {\tiny 3} -- (1,0) -- (1.5,.5)         node [anchor=west] {\tiny 5};
		\draw (0,0) node [anchor=south] {\ordidx{4}};
	}{splitlabel3}
	\label{eq:24-labels}
\end{equation}
Labels 1 and 2 are applied to the smaller trace
(as per \eqref{eq:fsplit-sort}), and no cycling is needed due to the symmetry
of the vertex. Labels 3456 must be summed over all four cyclings, since each
cycling gives a different propagator. No other labelling is
flavour-ordered; in particular,
\begin{equation}
	\tikzineq{
		\draw [thick, rounded corners=4pt] (-.5,.5)  node [anchor=east] {\tiny 5} -- (0,0) -- (-.5,0)          node [anchor=east] {\tiny 6};
		\draw [thick, rounded corners=4pt] (-.5,-.5) node [anchor=east] {\tiny 1} -- (0,0) -- (1,0) -- (1.5,0) node [anchor=west] {\tiny 3};
		\draw [thick]                      (1.5,-.5) node [anchor=west] {\tiny 2} -- (1,0) -- (1.5,.5)         node [anchor=west] {\tiny 4};
		\draw (0,0) node [anchor=south] {\ordidx{4}};
	}{splitlabel-wrong},
\end{equation}
which would be valid on a single-trace diagram, should not be counted, since
it has flavour structure $\flav_\id(4,2)$ in disagreement
with \eqref{eq:fsplit-sort}. Including it would be double-counting when
summing over all $\sigma$ in \eqref{eq:ampl}, since it is obtained
from \eqref{eq:24-labels} via a permutation in $\perm_6/\Z_{\{2,4\}}$.

Extra caveats sometimes show up. For instance, the two $\O(p^4)$ diagrams
\begin{equation}
	\tikzineq{
		\draw [thick] (-.5,.5)   -- (0,0) -- (-.5,0);
		\draw [thick] (-.5,-.5)  -- (0,0) -- (.5,0) -- (1,0) -- (1.5,-.5);
		\draw [thick] (1.5,0)    -- (1,0) -- (1.5,.5);
		\draw [thick, rounded corners=4pt] (.25,.5) -- (.5,0) -- (.75,.5);
		\draw (.5,0) node [anchor=north] {\ordidx{4}};
	}{8ptsplit0}\qquad
	\tikzineq{
		\draw [thick] (-.5,.5) -- (0,0) -- (-.5,-.5);
		\draw [thick, rounded corners=3pt] (-.5,0)   -- (0,0) -- (.5,0) -- (.5,-.5);
		
		\draw [thick] (1.5,.5) -- (1,0) -- (1.5,-.5);
		\draw [thick, rounded corners=3pt] (1.5,0)   -- (1,0) -- (.5,0) -- (.5, .5);
		\draw (.5,0) node [anchor=south east] {\ordidx{4}};
	}{8ptsplit1}
\end{equation}
emerge from different orientations of the same three vertices, but have
completely different flavour structure and properties. In the first diagram,
the smaller trace should not be cycled at all, and the larger trace only
halfway, since it is symmetric (compare to the $\O(p^2)$ 6-point diagram).
In the second diagram, all $4\cdot 4$ combined cyclings of the two traces are
distict, but due to the symmetry of the diagram, swapping them, e.g.
\begin{equation}
	\tikzineq{
		\draw [thick] (-.5,.5) node [anchor=east] {\tiny 1} -- (0,0) -- (-.5,-.5) node [anchor=east] {\tiny 3};
		\draw [thick, rounded corners=3pt] (-.5,0) node [anchor=east] {\tiny 2} -- (0,0) -- (.5,0) -- (.5,-.5) node [anchor=north] {\tiny 4};
		
		\draw [thick] (1.5,.5) node [anchor=west] {\tiny 7} -- (1,0) -- (1.5,-.5) node [anchor=west] {\tiny 5};
		\draw [thick, rounded corners=3pt] (1.5,0) node [anchor=west] {\tiny 6} -- (1,0) -- (.5,0) -- (.5, .5) node [anchor=south] {\tiny 8};
		\draw (.5,0) node [anchor=south east] {\ordidx{4}};
	}{equivlabel0}\qquad \longleftrightarrow \qquad
	\tikzineq{
		\draw [thick] (-.5,.5) node [anchor=east] {\tiny 5} -- (0,0) -- (-.5,-.5) node [anchor=east] {\tiny 7};
		\draw [thick, rounded corners=3pt] (-.5,0) node [anchor=east] {\tiny 6} -- (0,0) -- (.5,0) -- (.5,-.5) node [anchor=north] {\tiny 8};
		
		\draw [thick] (1.5,.5) node [anchor=west] {\tiny 3} -- (1,0) -- (1.5,-.5) node [anchor=west] {\tiny 1};
		\draw [thick, rounded corners=3pt] (1.5,0) node [anchor=west] {\tiny 2} -- (1,0) -- (.5,0) -- (.5, .5) node [anchor=south] {\tiny 4};
		\draw (.5,0) node [anchor=south east] {\ordidx{4}};
	}{equivlabel1}
\end{equation}
does not produce a distinct kinematic structure and should not be counted.

In the $\O(p^6)$ diagrams
\begin{equation}
	\tikzineq{
		\draw [thick, rounded corners=4pt, name path=ad] (0,.5)   -- (0,0) -- (-.5,-.5);
		
		\path [name path=b] (-.5,.5) -- (0,0);
		\path [name path=c] (-.5,0) -- (0,0);
		
		\draw [thick, name intersections={of=ad and b}] (-.5,.5) -- (intersection-1);
		\draw [thick, name intersections={of=ad and c}] (-.5, 0) -- (intersection-1);
		
		\draw [thick, rounded corners=3pt] (0,-.5)  -- (0,0) -- (1,0) -- (1.5,-.5);
		\draw [thick, rounded corners=4pt] (1.5,0)    -- (1,0) -- (1.5,.5);
		\draw (0,0) node [anchor=south west] {\ordidx{4}};
		\draw (1,0) node [anchor=south] {\ordidx{4}};
	}{traceswap0}\qquad
	\tikzineq{
		\draw [thick, rounded corners=4pt] (-.5, .5) -- (0,0) -- (0, .5);
		\draw [thick, rounded corners=4pt] (-.5,-.5) -- (0,0) -- (0,-.5);
		\draw [thick] (-.5,0)  -- (0,0) -- (1,0) -- (1.5,-.5);
		\draw [thick] (1.5,.5) -- (1,0) -- (1.5,0);
		\draw (0,0) node [anchor=south west] {\ordidx{6}};
	}{traceswap1}\quad,
\end{equation}
the component of $\Z_R$ that swaps equal-size traces does play a role. In the
first diagram, we may place either 12 or 34 in the trace straddling the
propagator, and we must sum over both placements. In addition to that, we must
sum over cyclings of the trace that straddles the propagator. In the second
diagram, the two smaller traces are equivalent under the $\Z_{\{2,2,2\}}$
symmetry of the vertex, and we should not sum over both ways of placing the
labels 12 and 34.

\subsection{The singlet problem and its solution}
\label{sec:singlet}

The construction of flavour-ordered diagrams hinges heavily on the use
of \eqref{eq:contr}, or specifically the muliplet term, $\tr{XY}$. The
singlet term, $\tr{X}\tr{Y}/N_f$, threatens the notion that the stripped
amplitude is given by exactly the flavour-ordered diagrams. Consider the
diagrams
\begin{equation}
	\tikzineq{
		\draw [thick] (-.5,.5)  node [anchor=east] {\tiny 1}  -- (0,0) -- (-.5,0)             node [anchor=east] {\tiny 2} ;
		\draw [thick] (-.5,-.5) node [anchor=east] {\tiny 3}  -- (0,0) -- (1,0) -- (1.5,-.5)  node [anchor=west] {\tiny 4} ;
		\draw [thick] (1.5,0)   node [anchor=west] {\tiny 5}  -- (1,0) -- (1.5,.5)            node [anchor=west] {\tiny 6} ;
	}{singlet0}\qquad
	\tikzineq{
		\draw [thick] (-.5,.5)  node [anchor=east] {\tiny 2}  -- (0,0) -- (-.5,0)             node [anchor=east] {\tiny 3} ;
		\draw [thick] (-.5,-.5) node [anchor=east] {\tiny 4}  -- (0,0) -- (1,0) -- (1.5,-.5)  node [anchor=west] {\tiny 5} ;
		\draw [thick] (1.5,0)   node [anchor=west] {\tiny 6}  -- (1,0) -- (1.5,.5)            node [anchor=west] {\tiny 1} ;
	}{singlet1}\qquad
	\tikzineq{
		\draw [thick] (-.5,.5)  node [anchor=east] {\tiny 3}  -- (0,0) -- (-.5,0)             node [anchor=east] {\tiny 1} ;
		\draw [thick] (-.5,-.5) node [anchor=east] {\tiny 2}  -- (0,0) -- (1,0) -- (1.5,-.5)  node [anchor=west] {\tiny 6} ;
		\draw [thick] (1.5,0)   node [anchor=west] {\tiny 4}  -- (1,0) -- (1.5,.5)            node [anchor=west] {\tiny 5} ;
	}{singlet2}.
	\label{eq:singlet-example}
\end{equation}
The first diagram is flavour-ordered according to both the multiplet and
singlet terms. The second diagram is also flavour-ordered according to our
definitions, but gives the non-flavour-ordered structure $\tr{234}\tr{561}$
under the singlet. The third diagram is not flavour-ordered, but the singlet
gives the flavour-ordered structure $\tr{123}\tr{456}$. Since the only
permutation contained in both $\Z_{\{6\}}$ and $\Z_{\{3,3\}}$ is $\id$, the
behavour of the singlet and multiplet terms is clearly very different and
must be treated carefully.

There is, however, an elegant solution. As stated previously, the singlet term
in \eqref{eq:contr} is not present in $U(N_f)$. Therefore, in the $U(N_f)$
NLSM we may always do flavour-ordering without singlet issues.
We can extend this to $SU(N_f)$ by using its similarity to $U(N_f)$.

The $U(N_f)$ algebra differs from the $SU(N_f)$ algebra by a non-traceless
generator $t^0$ that commutes with all other generators. Due to the latter
property, its associated field $\phi^0$ forms a $U(1)$ singlet separate from
the $SU(N_f)$ mutliplet $\phi^a$. With this in mind, a more elucidating form
of \eqref{eq:contr} is
\begin{equation}
	\sum_{a=1}^{N_f^2-1}\tr{Xt^a}\tr{t^a Y} = \sum_{a=0}^{N_f^2-1}\tr{Xt^a}\tr{t^a Y} - \tr{Xt^0}\tr{t^0Y},
	\label{eq:singlet-sub}
\end{equation}
where we temporarily suppress Einstein summation. This expression suggests
that a $SU(N_f)$ propagator (left) represents a $U(N_f)$ propagator (right)
minus the singlet propa\-gator, and explains our naming of the terms
in \eqref{eq:contr}. The $N_f^\inv$ is absorbed into $t^0$ since
$t^0=\1/\sqrt{N_f}$.

Now, if we extend our Lagrangian-building field like
\begin{equation}
	\hat\Phi(\phi^0,\phi) = t^0\phi^0 + \Phi(\phi),\qquad
	\hat u(\phi^0,\phi) 
		= \exp\left(\frac{i\hat\Phi}{F\sqrt{2}}\right) 
		= \exp\left(\frac{i\phi^0t^0}{F\sqrt{2}}\right)u(\phi),
\end{equation}
where $u(\phi)\in SU(N_f)$ and $\hat u(\phi^0,\phi)\in U(N_f)$, we see that
\begin{equation}
	\hat U = \hat u^2 \imp 
	\hat U^\dag\p_\mu\hat U = \left(\frac{i\sqrt2}{F\sqrt{N_f}}\right)\p_\mu \phi^0 + U^\dag\p_\mu U
\end{equation}
(remembering that $U^\dag \p_\mu U$ is equivalent to $u_\mu$), and therefore
\begin{equation}
	\hat\lagr_2 = \frac12\p_\mu\phi^0\p^\mu\phi^0 + \frac{F^2}{4}\tr{u_\mu u^\mu}.
\end{equation}
At this order, the singlet decouples from the other fields and forms a free
theory. Therefore, no $\O(p^2)$ vertex involves the singlet, so there is no
distinction between $U(N_f)$ and $SU(N_f)$ amplitudes at this order, and we
may ignore the singlet term in \eqref{eq:contr}.

This observation was sufficient in \cite{Kampf:2013vha}, but we must handle
the singlet problem beyond $\O(p^2)$. $\lagr_4$ and all higher-order
Lagrangians introduce vertices that couple the singlet to the other particles.
However, a singlet propagator can only exist if both vertices at its ends
couple to it. Since this requires at least two vertices of at least $\O(p^4)$,
the diagram as a whole must be at least $\O(p^6)$ to include such
complications.\footnote{If the singlet forms a loop, only one $\O(p^4)$ vertex
is necessary, but the loop itself increases the power counting, so $\O(p^6)$
is needed in this case as well.}$^,$\footnote{An interesting parallel can be
seen in \cite{reuschle}, where $U(1)$ gluons similar to our singlets must be
introduced in perturbative QCD. While our singlets only emerge with at least
two higher-order vertices, their $U(1)$ gluons cancel unless the diagram
contains at least two quark lines. In general, there are several intriguing
analogies between the inclusion of quark lines in gluon scattering (where
there are no higher-order vertices) and the inclusion of higher-order vertices
in the NLSM (where there are no quark lines).} Therefore, flavour-ordering at
$\O(p^4)$ works with no other complications than the introduction of split
vertices.

At $\O(p^6)$ and above, the singlet term in \eqref{eq:contr} can not be
avoided in $SU(N_f)$, but the interpretation of \eqref{eq:singlet-sub} still
holds. In order to build a $SU(N_f)$ amplitude, we first work in $U(N_f)$ to
build flavour-ordered diagrams using only the multiplet term. Then, we
construct all diagrams with singlet propagators in a similar fashion,
maintaining flavour-ordering independently. For instance, the full suite
of $\O(p^6)$ 6-point diagrams is
\begin{equation}
	\begin{gathered}
		\tikzineq{
			\draw [thick] (-.5,.5)   -- (0,0) -- (-.5,0);
			\draw [thick] (-.5,-.5)  -- (0,0) -- (.5,-.5);
			\draw [thick] (.5,0)     -- (0,0) -- (.5,.5);
			\draw (0,0) node [anchor=south] {\ordidx 6};
		}{Op6-6pt0}\qquad
		\tikzineq{
			\draw [thick, rounded corners=3pt, name path=ad] (-.5, .5) -- (0,0) -- (-.5,-.5);
			\draw [thick, rounded corners=3pt, name path=eh] ( .5,-.5) -- (0,0) -- ( .5, .5);
			
			\path [name path=bf] (-.5, .2) -- (.5,-.2);
			\path [name path=cg] (-.5,-.2) -- (.5, .2);
			
			\draw [thick, name intersections={of=cg and eh}] ( .5, .2) -- (intersection-1);
			\draw [thick, name intersections={of=bf and eh}] ( .5,-.2) -- (intersection-1);
			\draw (0,0) node [anchor=south] {\ordidx 6};
		}{Op6-6pt1}\qquad
		\tikzineq{
			\draw [thick, rounded corners=3pt, name path=ac] (-.5, .5) -- (0,0) -- (-.5,-.5);
			\draw [thick, rounded corners=3pt, name path=df] ( .5,-.5) -- (0,0) -- ( .5, .5);
			
			\path [name path=be] (-.5, 0) -- (.5,0);
			
			\draw [thick, name intersections={of=be and ac}] (-.5,0) -- (intersection-1);
			\draw [thick, name intersections={of=be and df}] ( .5,0) -- (intersection-1);
			\draw (0,0) node [anchor=south] {\ordidx 6};
		}{Op6-6pt2}\qquad
		\tikzineq{
			\draw [thick, rounded corners=4pt] (-.5,.5) -- (0,0) -- (-.5,0);
			\draw [thick, rounded corners=4pt] ( .5,.5) -- (0,0) -- ( .5,0);
			\draw [thick, rounded corners=4pt] (-.5,-.5) -- (0,0) -- (.5,-.5);
			\draw (0,0) node [anchor=south] {\ordidx 6};
		}{Op6-6pt3}
	\\
		\tikzineq{
			\draw [thick] (-.5,.5)   -- (0,0) -- (-.5,0);
			\draw [thick] (-.5,-.5)  -- (0,0) -- (1,0) -- (1.5,-.5);
			\draw [thick] (1.5,0)    -- (1,0) -- (1.5,.5);
			\draw (0,0) node [anchor=south] {\ordidx 6};
		}{Op6-6pt4}\qquad
		\tikzineq{
			\draw [thick] (-.5,.5)   -- (0,0) -- (-.5,0);
			\draw [thick] (-.5,-.5)  -- (0,0) -- (1,0) -- (1.5,-.5);
			\draw [thick] (1.5,0)    -- (1,0) -- (1.5,.5);
			\draw (0,0) node [anchor=south] {\ordidx 4};
			\draw (1,0) node [anchor=south] {\ordidx 4};
		}{Op6-6pt5}\qquad
		\tikzineq{
			\draw [thick, rounded corners=6pt] (-.5,.5) -- (0,0) -- (-.5,0);
			\draw [thick, rounded corners=4pt] (-.5,-.5)  -- (0,0) -- (1,0) -- (1.5,0);
			\draw [thick] (1.5,-.5)    -- (1,0) -- (1.5,.5);
			\draw (0,0) node [anchor=south] {\ordidx 6};
		}{Op6-6pt6}\qquad
		\tikzineq{
			\draw [thick, rounded corners=6pt] (-.5,.5) -- (0,0) -- (-.5,0);
			\draw [thick, rounded corners=4pt] (-.5,-.5)  -- (0,0) -- (1,0) -- (1.5,0);
			\draw [thick] (1.5,-.5) -- (1,0) -- (1.5,.5);
			\draw (0,0) node [anchor=south] {\ordidx 4};
			\draw (1,0) node [anchor=south] {\ordidx 4};
		}{Op6-6pt7}\qquad
		\tikzineq{
			\draw [thick, rounded corners=4pt] (-.5,.5) -- (0,0) -- (-.5,0);
			\draw [thick, rounded corners=4pt] (-.5,-.5)  -- (0,0) -- (1,0) -- (1.5,-.5);
			\draw [thick, rounded corners=4pt] (1.5,.5)    -- (1,0) -- (1.5,0);
			\draw (0,0) node [anchor=south] {\ordidx 4};
			\draw (1,0) node [anchor=south] {\ordidx 4};
		}{Op6-6pt8}
	\\
		\tikzineq{
			\draw [thick] (-.5,.5)   -- (0,0) -- (-.5,0);
			\draw [thick] (-.5,-.5)  -- (0,0) (1,0) -- (1.5,-.5);
			\draw [thick] (1.5,0)    -- (1,0) -- (1.5,.5);
			\draw [thick, dashed] (0,0) -- (1,0);
			\draw (0,0) node [anchor=south] {\ordidx 4};
			\draw (1,0) node [anchor=south] {\ordidx 4};
		}{Op6-6pt9}\quad,
	\end{gathered}
	\label{eq:p6-diagrs}
\end{equation}
including one singlet propagator, indicated by a dashed line. It implicitly
includes a factor of $-N_f^\inv$, and its flavour structure is split $\{3,3\}$
over the propagator. All cyclings of the two traces should be counted as
distinct, since the vertices are invariant under $\Z_4$, not $\Z_3$. By adding
the singlet diagrams to the others, we get the stripped $SU(N_f)$ amplitude.

The singlet diagram contains all contractions that are flavour-ordered under
the singlet term, like the first and last diagram in \eqref{eq:singlet-example}.
The decoupling of the singlet at $\O(p^2)$ means that these contributions must
cancel in the amplitude at this order, which is not at all obvious from the
individual diagrams. Still, recasting the singlet terms as flavour-ordered
singlet diagrams is valid, as follows from the uniqueness of the stripped
amplitude. 

\subsection{Uniqueness of stripped amplitudes}
\label{sec:unique}

Above, we have blindly trusted the definition of the stripped amplitude as
everything that comes attached to the flavour-ordered
structure $\flav_\id(R)$. If this definition is not unique, flavour-ordering
would not necessarily be valid, and we could not rely on our use of singlet
diagrams. However, we can show that the stripped amplitude is indeed unique,
using a generalisation of a method presented by \cite{mangano-parke} and
adapted to flavour-ordering by \cite{Kampf:2013vha}.

The uniqueness hinges on the orthogonality relation
\begin{equation}
	\flav_\sigma(Q)\cdot\big[\flav_\rho(R)\big]^* = N_f^n
	\begin{cases}
		1 + \O\left(N_f^{-2}\right)	&	\text{if $Q = R$ and $\sigma\equiv\rho\mod{\Z_R}$,}				\\
		\O\left(N_f^{-\gamma}\right)	&	\text{otherwise ($\gamma\geq 1$; see below)}	
	\end{cases}
	\label{eq:ortho}
\end{equation}
using the notation defined in section~\ref{sec:notation}. The dot in the
left-hand side indicates contraction over all flavour indices. If
$Q\neq R$, $\gamma\geq 1$, and if $\sigma\not\equiv\rho\mod{\Z_R}$,
 $\gamma\geq 2$; therefore, the single-trace version
 (i.e. that given in \cite{Kampf:2013vha}) has $\O(N_f^{-2})$ as its second
 case. The more different the flavour structures are, the larger $\gamma$ is.
The relation \eqref{eq:ortho} is proven in appendix~\ref{app:ortho} and
states that any given flavour structure $\flav_\sigma(Q)$ is orthogonal at
leading order in $N_f$ to all other flavour structures whose permutations are
not equivalent to $\sigma$, or whose flavour splittings are not equal to $Q$.

In the context of stripped amplitude uniqueness, \eqref{eq:ortho} can be
applied as follows. In analogy with \eqref{eq:vertex} and \eqref{eq:ampl}, we
write some arbitrary quantity $\mathcal X$ in the form
\begin{equation}
	\mathcal X^{a_1\cdots a_n} = \sum_{R\in\fsplit} \sum_{\sigma\in\perm_n/\Z_R} \flav_\sigma(R)\mathcal X_{\sigma,R},
\end{equation}
where $\fsplit$ is some appropriate selection of flavour splittings. Then,
we use the orthogonality relation \eqref{eq:ortho} to perform the projection
\begin{equation}
	\mathcal X^{a_1\cdots a_n} \left[\flav_\id(R)\right]^* = 
		N_f^n\left[\mathcal X_{\id,R} + \O\left(\frac{1}{N_f}\right)\right].
	\label{eq:project}
\end{equation}
This means that we can always project out the stripped $\mathcal X$, and that
any overlap with other terms must come suppressed by at least $N_f^\inv$. In
a stripped amplitude of $\O(p^4)$ or lower, the stripped amplitude can not
contain any powers of $N_f^\inv$ due to the decoupling of the singlet, so
there can be no overlap for arbitrary $N_f$. This proves that stripped
amplitudes are unique at $\O(p^4)$ or below.

At higher orders, things are not as simple, since there are possibly many
factors of $N_f^\inv$. This would allow mixing between different
stripped $\mathcal X$'s, threatening to break uniqueness. However, it can be
resolved by expressing $\mathcal X^{a_1\cdots a_n}$ as a polynomial in $N_f^\inv$,
\begin{equation}
	\mathcal X^{a_1\cdots a_n} = 
		\mathcal X_0^{a_1\cdots a_n} + 
		\frac{1}{N_f}\mathcal X_1^{a_1\cdots a_n} + 
		\frac{1}{N_f^2}\mathcal X_2^{a_1\cdots a_n} + \ldots
\end{equation}
such that each $\mathcal X_i^{a_1\cdots a_n}$, and therefore also its stripped
counterpart, is independent of $N_f^\inv$. We then apply the projection to
each $\mathcal X_i^{a_1\cdots a_n}$ independently, and ignore the $\O(N_f^\inv)$
completely. Thus, stripped amplitudes, vertex factors, and other analogous
quantities are unique to all orders.\footnote{This uniqueness is of course
only up to a permutation in $\Z_R$, but since we sum over those in the
definition of the stripped quantity, they are unique for our purposes.} 

The proof holds for general $N_f$, but for any specific $N_f$, there may be
additional relations between the generators that break the uniqueness. The
Cayley-Hamilton relations provide such relations for small $N_f$. However,
we always assume that the relations have been ``exhausted'' by removing terms
from the Lagrangian, so that they do not affect the uniqueness of stripped
amplitudes.

This proof in this section has significant consequences. Most importantly, it
guarantees the correctenss of our method of flavour ordering with split traces
and singlets: gathering all flavour-ordered pieces of the full amplitude is
guaranteed to equal the unique stripped amplitude. Also, uniqueness allows
many properties of the full amplitude to carry over to the stripped amplitude,
as is discussed below.

A second consequence is worthy of note. The full amplitude of
some $\O(p^N)$ $n$-particle process is constructed from
$\card{\fsplit(N,n)}$ different stripped amplitudes. When summed over
permutations according to \eqref{eq:ampl}, the total number of flavour
structures grows to
\begin{equation}
	\mathcal N_{N,n} \sim \sum_{R\in\fsplit(N,n)} \frac{\card{\perm_n}}{\card{\Z_R}},
\end{equation}
which is a very rapidly growing number --- even
at $\O(p^2)$, $\mathcal N(2,n) \sim (n-1)!$. Since the flavour structures are
not truly orthogonal, the expression for the cross section of the process,
proportional to
\mbox{$\ampl^{a_1\cdots a_n}_{N,n}[\ampl^{a_1\cdots a_n}_{N,n}]^\dag$}, grows in
length as $(\mathcal N_{N,n})^2$. However, the expression for the cross
section contracts the flavour structures as in \eqref{eq:ortho}, which
suppresses products of non-equivalent flavour structures by a factor
of $N_f^{-1}$ for each difference (or $N_f^{-2}$ in the single-trace case).
Therefore, in the limit $N_f\to\infty$, flavour structures \emph{are}
orthogonal, and the cross section only grows as $\mathcal N_{N,n}$. Even with
finite $N_f$, most cross-terms will be heavily suppressed, and can most likely
e ignored. 

An alternative approach would be to construct other bases for flavour space
that are more orthogonal than the trace bases used here, as is done in
perturbative QCD by \cite{multiplet-1}. Such methods have so far not been
applied in the present context.

%%%%%%%%%%%%%%%%%%%%%%%%%%%%%%%%%%%%%%%%%%%%%%%%%%%%%%%%%%%%%%%%%%%%%%%%%%%%%

\section{NLSM amplitudes}
\label{sec:ampl}

In this section, we introduce and generalise several concepts related to
NLSM amplitudes and flavour-ordering.

\subsection{Adler zeroes and soft limits}
\label{sec:adler}

In any effective field theory emerging from the spontanous beaking of a global
symmetry, the amplitude possesses the so-called Adler zero,
\begin{equation}
	\lim_{\e\to0}\ampl^{a_1\cdots a_n}(p_1,\ldots, \e p_i,\ldots, p_n) = 0,
	\label{eq:adler}
\end{equation}
for any $i$ \cite{adler, adler-proof}. The approach to zero will generally go
as $\e^\sigma$, where the \emph{soft degree} $\sigma\geq 1$ can be used to
classify and construct EFTs \cite{Cheung:2016drk,eft-soft}. The NLSM has
$\sigma=1$. Due to the orthogonality of flavour structures and the uniqueness
of stripped amplitudes, Adler zeroes may only exist in the full amplitude if
they also exist, with the same soft degree, in the stripped amplitudes.
Therefore, \eqref{eq:adler} and any statement relying on it can equally well
be applied to the stripped amplitudes.

The Adler zeroes may be used as a starting point to construct amplitudes
through recursion relations \cite{Cheung:2015ota,Low:2019ynd}. For our
purposes, however, their main use is in validating the correctness of
complicated stripped amplitudes. Since far from every term in the amplitude
is proportional to $p_i$, the Adler zero must manifest itself through
intricate cancellations. Therefore, any error in the amplitude is extremely
likely to give a finite right-hand side in \eqref{eq:adler}.

Beside the Adler zeroes, there also exists the \emph{double soft limit},
where two momenta are sent to zero at the same rate. It turns out that the
double soft limit of any $(n+2)$-particle amplitude can be expressed in terms
of $n$-particle amplitudes with the soft particles removed; for the NLSM,
the specific form is
\begin{multline}
	\lim_{\e\to0}\ampl_{m,n+2}^{aba_1\cdots a_n}(\e p,\e q, p_1,\ldots,p_n) = \\
		-\frac{1}{4F^2}\sum_{i=1}^n f^{abc}f^{a_i dc}\frac{p_i\cdot(p-q)}{p_i\cdot(p+q)}
		\ampl_{m,n}^{a_1\cdots a_{(i-1)}da_{(i+1)}\cdots a_n}(p_1,\cdots,p_n).
	\label{eq:dslim-full}
\end{multline}
This was conjectured in \cite{simplest-qft} and proven in \cite{Kampf:2013vha}.
Like the Adler zero, it can be projected to a relation for stripped amplitudes,
although the projection is not entirely trivial. The result for single-trace
flavour structures is given in \cite{Kampf:2013vha}. We derive the counterpart
for general flavour structures in appendix~\ref{app:dslim}, with the result
being as follows. At any order $m$ in the power counting and for any flavour split $R\in\fsplit(m,n+2)$, the double soft limit
\begin{equation}
	\lim_{\e\to 0} \ampl_{m,R}(p_1,\ldots,p_{i-1},\e p_i,\ldots,\e p_j,p_{j+1},\ldots,p_{n+2})
\end{equation}
is nonzero if the indices $i-1$, $i$, $j$ and $j+1$ are consecutive and lie
within the same trace; we will call this condition $\mathcal C$. 
It is also nonzero if the indices can be made to satisfy $\mathcal C$ by
applying a permutation in $\Z_R$ and possibly swapping $i$ and $j$.
In all other cases, the double soft limit is zero.

Since $\ampl_{m,R}$ is invariant under $\Z_R$, we can without loss of
generality assume that $\mathcal C$ holds whenever the double soft limit is
nonzero. Assuming this, the double soft limit is
\begin{multline}
	\lim_{\e\to 0} \ampl_{m,R}(p_1,\ldots,\e p_i,\e p_j,\ldots,p_{(n+2)}) \\= 
		\frac{1}{4 F^2}\left(
			\frac{p_{(j+1)}\cdot(p_i - p_j)}{p_{(j+1)}\cdot(p_i+p_j)} -
			\frac{p_{(i-1)}\cdot(p_i - p_j)}{p_{(i-1)}\cdot(p_i+p_j)}
		\right)
		\ampl_{m,R'}(p_1,\ldots,p_{(i-1)},p_{(j+1)},\ldots,p_{(n+2)}),
	\label{eq:dslim}
\end{multline}
where $R' \in \fsplit(m,n)$ is $R$ with the location of the soft particles
removed and $j=i+1$. The result, which generalises
that given in \cite{Kampf:2013vha},
is quite remarkable: for properly chosen $i,j$, the double soft limit amounts
to removing the soft particles from the amplitude and multiplying by a simple
kinetic factor. The factor is similar to those that arise in IR divergences,
which is understandable --- both arise from propagators going on-shell in the
soft (IR) limit.

\subsection{Generalised Mandelstam invariants}
\label{sec:mandel}

In order to express stripped amplitudes in a way that naturally includes
on-shellness and conservation of momentum, we will employ bases of
generalised Mandelstam invariants in the form
\begin{equation}
	s_{ijk\cdots} = (p_i + p_j + p_k + \ldots)^2,
	\label{eq:gen-mandel}
\end{equation}
In this notation, the standard 4-particle Mandelstam invariants are
\begin{equation}
	s = s_{12} = s_{34},\qquad t = s_{13} = s_{24},\qquad u = s_{23} = s_{41}.
\end{equation}
Since $s+t+u=0$, this basis is overcomplete, and one element can be removed.
We will generally use bases where $ijk\ldots$ are consecutive, so we choose
to keep $\{s,u\}$ as the 4-particle basis. 

For $n$ particles, the products of momenta are related to the invariants
with consecutive indices through
\begin{equation}
	\begin{split}
		2p_i\cdot p_{i+1} &= s_{i(i+1)},\\
		2p_i\cdot p_{i+2} &= s_{i(i+1)(i+2)} - s_{i(i+1)} - s_{(i+1)(i+2)},\\
		j > i+2:\quad 2p_i\cdot p_j 	&= s_{i \cdots j} - s_{i \cdots (j-1)} - s_{(i+1) \cdots j} + s_{(i+1) \cdots (j-1)}.
	\end{split}
\end{equation}
Based on this, a complete basis of invariants for $n=6$ is
\begin{equation}
\basis_{\{6\}} = \big\{s_{12},s_{23},s_{34},s_{45},s_{56},s_{61},~ s_{123},s_{234},s_{345}\big\},
\end{equation}
where $s_{456}$ etc.\ are not needed due to conservation of momentum
in the form
\begin{equation}
	s_{i\cdots(i+k-1)} = s_{(i+k)\cdots (i-1)},
	\label{eq:com}
\end{equation}
with indices cycling around from $n$ to 1. The form of $\basis_{\{6\}}$ can be
carried on to any even $n$, giving 
\begin{multline}
	\basis_{\{n\}} = \big\{s_{12},s_{23},\ldots,s_{n(n-1)},s_{n1},\quad s_{123},s_{234},\ldots,s_{n12},\quad \ldots, \quad \\
	s_{12\cdots (n/2)},\ldots,s_{(n/2-1)\cdots(n-1)}\big\}.
	\label{eq:mandel-basis}
\end{multline}
This contains $n(n-3)/2$ invariants, which is also the number of independent
products that can be formed from $\{p_1,\ldots,p_n\}$ with all $p_i^2=0$.%
\footnote{There are $n(n+1)/2$ ways to form products of pairs of $p_i$,
  $i\in\{1,\ldots,n\}$. Of these, $n$ vanish due to $p_i^2=0$. Conservation of
  momentum implies that $p_n=\sum_{i=1}^{n-1}p_i$, which gives $n$ linear
  combinations among the remaining products, reducing the number of
  independent ones to $n(n+1)/2 - 2n = n(n-3)/2$.}  Note that all invariants
have consecutive indices.

These bases are only linearly independent in sufficiently high spacetime
dimensions $D$. If $D<n+1$, the Gram determinant gives relations among the
basis elements. In practice, these relations are so algebraically messy 
that we have found it simpler to always work in arbitrary $D$.

Mandelstam invariants have further benefits beyond taking care of on-shellness
and conservation of momentum. In an $n$-point $\O(p^2)$ single trace
flavour-ordered tree
diagram, all propagators carry a momentum $q$ such that $q^2\in\basis_{\{n\}}$.
Therefore, $\O(p^2)$ stripped amplitudes will never contain a denominator with
a sum of several invariants, making their algebraic handling simpler. This is
not true for diagrams with multi-trace flavour structures. It is also not
possible to find a different basis that contains all squared propagator momenta
in the general case; for instance, the set of all possible $q^2$ under
$R=\{2,6\}$ is not linearly independent.

Another use of Mandelstam invariants is the shortening of stripped amplitudes.
As a consequence of invariance under $\Z_R$, any stripped amplitude can be
written in the form
\begin{equation}
\label{eq:simplified}
	\ampl_{m,R}(p_1,\ldots,p_n) = (\text{simpler expression}) + [\Z_R],
\end{equation}
where we use a shorthand for the sum over $\Z_R$, generalising
the familiar idiom ``$+\text{cycl.}$''. The simpler expression is rather
obvious for simple amplitudes, but for more complicated cases, it is an
enormous aid to readability.

Any stripped amplitude can be simplified as above by separating it into
simple terms, separating the terms into cosets under $\Z_R$, and
picking a single representative from each coset. The ``simpler expression'' 
in \eqref{eq:simplified} will then be the sum of the representatives. 
For $R=\{n\}$, this works
because for any $s_{ij\cdots}\in\basis_{\{n\}}$ and $\sigma\in\Z_{\{n\}}$,
applying $\sigma$ to the indices of $s_{ij\cdots}$ yields another element
in $\basis_{\{n\}}$. Thus, $\basis_{\{n\}}$ can be said to be \emph{closed}
under $\Z_{\{n\}}$. However, the basis given in \eqref{eq:mandel-basis} is not
closed under any $\Z_R$ with $R\neq\{n\}$
(with the sole exception of $R=\{2,2\}$), so the separation into cosets
fails. Simplifying general amplitudes therefore requires either
painstaking manual work, or a Mandelstam basis that is closed under $\Z_R$.
We have no general method of finding such bases.
%, although it seems likely that they exist for all, or at least many, $R$. 
In appendix~\ref{app:closed-basis}, we present closed bases for $\Z_{\{2,4\}}$,
$\Z_{\{3,3\}}$ and $\Z_{\{2,2,2\}}$. These cover all flavour structures that
appear for $n\leq 6$.

\subsection{Diagram generation}
\label{sec:fodge}
For most amplitudes presented here, the number of diagrams is small enough
that they are easily found by hand, but above a dozen or so diagrams, this
becomes
a slow and error-prone process. We therefore automated the process by
designing a program called \fodge\ 
(\textsc{f}lavour-\textsc{o}rdered \textsc{d}iagram \textsc{ge}nerator)
written in \cpp.%
\footnote{The source code of \fodge\ can be found at 
\url{https://github.com/mssjo/fodge}.}
It produces \TikZ\ code for drawing the diagrams, and generates the input
to a set of \form\ procedures that compute the amplitudes.%
\footnote{The \form\ procedures can be found at 
\url{https://github.com/mssjo/flavour-order}. For 
\form\ itself, see \cite{Vermaseren:2000nd,Kuipers:2012rf}.}
The same procedures were used with manual input for computing 
simpler amplitudes. Inspiration was taken from the diagram generator used
in~\cite{Bijnens:2010xg,Bijnens:2013yca}.

The diagram generation works recursively. A list of all $\O(p^M)$ $N$-point
diagrams can be generated by generating all $\O(p^m)$ $n$-point diagrams
for $m\leq M$ and $n\leq N-2$, and then listing all ways to attach a
$\O(p^{2+M-m})$ $(2+N-n)$-point vertex to their external legs. Adding a list
of $\O(p^M)$ $N$-point single-vertex diagram and removing duplicates completes
the list. The number of duplicates can be reduced by restricting $m$ and $n$.

The number of independent labellings on each diagram must then be determined.
Representing diagrams in a way that shows their symmetries turns out to be very
difficult when complicated flavour structures are involved. This was not 
entirely successfully tried in the original \fodge\ used in \cite{masterthesis}.
Here, we take a different approach: each diagram is associated with
all flavour-ordered labellings of its external legs that give unique kinematic 
structures. This removes the need to explicitly consider its symmetries;
internally, the diagrams can be represented in whatever way is convenient.

As is pointed out below \eqref{eq:6-labellings}, a kinematic structure is
uniquely determined by the propagator momenta it contains. It is easy to see
that this holds for any $\O(p^2)$ diagram. At higher orders, it is sufficient
to add the order of the vertices at the ends of each propagator. The flavour
splits of the vertices are not needed if the overall flavour split of the
diagram is provided. For singlet diagrams, we must also specify how the vertex
is cycled relative to the singlet propagator, by writing down the momentum
carried by a vertex leg adjacent to the propagator. In general, any kinematic
factor is uniquely determined by listing all vertices and the momenta carried
by their legs, but this can be shown to reduce to these simpler rules when
diagrams are flavour-ordered.

Thus, \fodge\ generates all diagrams of a given order and size, equips each
diagram with an arbitrary flavour-ordered labelling, and determines the
kinematic factor as described above. $\Z_R$ is then applied to generate all
other labellings, but only a subset that gives distinct kinematic factors is
kept. If the choice of subset is consistent, equivalent diagrams will always
give an identical list of kinematic factors, so duplicates are
easily removed.

Knowing the labellings also makes the diagram generation more efficient. There
is no need to attach a vertex to several legs that are equivalent to each
other under the symmetries. By dividing the set of labels into cosets under
$\Z_R$, it is sufficient to attach vertices to legs that, in one of the
distinct labellings, carries a coset representative as its label. This reduces
the number of generated duplicates.

%%%%%%%%%%%%%%%%%%%%%%%%%%%%%%%%%%%%%%%%%%%%%%%%%%%%%%%%%%%%%%%%%%%%%%%%%%%%%

\section{Explicit amplitudes}
\label{sec:examples}

Using the methods developed in the previous sections, we have computed several
stripped NLSM amplitudes, several of which have not previously been determined.
These we discuss in this section.

\subsection{4-point amplitudes}
\label{sec:M4}

These amplitudes are by far the simplest, since their tree-level diagrams
contain no propa\-gators and only carry two flavour structures ($\{4\}$ and
$\{2,2\}$), or only one in the $\O(p^2)$ case. At $\O(p^6)$ and above, they
only receive contributions from the Lagrangian terms with no more than
four $u_\mu$'s, which is a tiny subset of the total Lagrangian.

The $\O(p^2)$ 4-point amplitude is given by a single diagram and a simple
stripped amplitude,
\begin{equation}
	\tikzineq{
		\draw [thick] (-.5,.5) -- (0,0) -- (.5,-.5);
		\draw [thick] (-.5,-.5) -- (0,0) -- (.5,.5);
	}{M4p2}
	\qquad -iF^2\ampl_{2,\{4\}} =  \frac{t}{2},
	\label{eq:M4p2}
\end{equation}
where $t$ is the Mandelstam invariant $(p_1+p_3)^2$. We have pulled factors of
$i$ and $F$ over to the left-hand side for clarity. The only independent
$\O(p^2)$ kinematic structure that is invariant under $\Z_{\{4\}}$ is $t$,
so the form of the right-hand side could have been guessed based on symmetry.

If we plug \eqref{eq:M4p2} into \eqref{eq:ampl} and apply some $SU(2)$ group
algebra, we recover the familiar $N_f=2$ amplitude
\begin{equation}
\ampl_{2,4}^{abcd}(s,t,u) = \frac{-4i}{F^2}\left[s\kd{ab}\kd{cd} + t\kd{ac}\kd{bd} + u\kd{ad}\kd{bc}\right]
	\label{eq:M4p2-full}
\end{equation}
with the Mandelstam invariants defined as in section~\ref{sec:mandel}.

The $\O(p^4)$ 4-point amplitude consists of the two diagrams
\begin{align}
	\tikzineq{
		\draw [thick] (-.5,.5) -- (0,0) -- (.5,-.5);
		\draw [thick] (-.5,-.5) -- (0,0) -- (.5,.5);
		\draw (0,0) node [anchor=south] {\ordidx{4}};
	}{M4p4-unsplit}
	&\qquad -iF^4\ampl_{4,\{4\}} = 
		2L_3(u^2+s^2) 
		+ 4L_0 t^2,
	\\
	\tikzineq{
		\draw [thick, rounded corners=3pt] (-.5, .5) -- (0,0) -- (-.5,-.5);
		\draw [thick, rounded corners=3pt] ( .5,-.5) -- (0,0) -- ( .5, .5);
		\draw (0,0) node [anchor=south] {\ordidx{4}};
	}{M4p4-split}
	&\qquad -iF^4\ampl_{4,\{2,2\}} = 
		8L_1 s^2 
		+ 4L_2(t^2+u^2),
	\label{eq:M4p4}
\end{align}
which includes the simplest example of a flavour split. There are now two
independent $\Z_4$-invariant kinematic structures, $t^2$ and $s^2+u^2$, and
likewise two independent $\Z_{\{2,2\}}$-invariant ones, $s^2$ and $t^2+u^2$.
All four appear equipped with one LEC each. The full amplitude is analogous
to \eqref{eq:M4p2-full}, but with various linear combinations of the LECs
and Mandelstam variables in place of $s$, $t$ and $u$.
The full amplitude agrees with the known results, see \cite{meson-meson}
and references therein.

The $\O(p^6)$ 4-point amplitude, like its $\O(p^4)$ analogue, has two diagrams,
\begin{align}
	\tikzineq{
		\draw [thick] (-.5,.5) -- (0,0) -- (.5,-.5);
		\draw [thick] (-.5,-.5) -- (0,0) -- (.5,.5);
		\draw (0,0) node [anchor=south] {\ordidx{6}};
	}{M4p6-unsplit}
	&\qquad -iF^4\ampl_{6,\{4\}} = 
		- \lagt{6}{3} t(s^2+u^2) 
		- 2\lagt{6}{4} t^3,
	\\
	\tikzineq{
		\draw [thick, rounded corners=4pt] (-.5, .5) -- (0,0) -- (-.5,-.5);
		\draw [thick, rounded corners=4pt] ( .5,-.5) -- (0,0) -- ( .5, .5);
		\draw (0,0) node [anchor=south] {\ordidx{6}};
	}{M4p6-split}
	&\qquad -iF^4\ampl_{6,\{2,2\}} = 
		-2\lagt{6}{1} (t^3+u^3) 
		+ \frac23 \lagt{6}{2}(s^3+t^3+u^3).
	\label{eq:M4p6}
\end{align}
As for $\O(p^4)$, there are two independent $\Z_4$-invariant kinematic
structures, $t^3$ and $t(s^2+u^2)$, and two independent $\Z_{\{2,2\}}$-invariant
ones, $s^3$ and $s(t^2+u^2)$. These four correspond to the four
LECs --- $s^3+t^3+u^3$ is a linear combination of $s^3$ and $s(t^2+u^2)$.
The full amplitude agrees with the result in  \cite{meson-meson}.

The $\O(p^8)$ amplitude, like its lower-order analogues, has two diagrams,
\begin{align}
	\tikzineq{
		\draw [thick] (-.5,.5) -- (0,0) -- (.5,-.5);
		\draw [thick] (-.5,-.5) -- (0,0) -- (.5,.5);
		\draw (0,0) node [anchor=south] {\ordidx{8}};
	}{M4p8-unsplit}
	&\qquad -iF^4\ampl_{8,\{4\}} = 
		\lagt{8}{4} s^2u^2
		+ \frac12 \lagt{8}{5} t^2(s^2 + u^2)
		+ \lagt{8}{6} t^4,		
	\\
	\tikzineq{
		\draw [thick, rounded corners=4pt] (-.5, .5) -- (0,0) -- (-.5,-.5);
		\draw [thick, rounded corners=4pt] ( .5,-.5) -- (0,0) -- ( .5, .5);
		\draw (0,0) node [anchor=south] {\ordidx{8}};
	}{M4p8-split}
	&\qquad -iF^4\ampl_{8,\{2,2\}} = 
		\lagt{8}{1} s^2(t^2+u^2)
		+ \lagt{8}{2}(t^4+u^4)
		+ 2\lagt{8}{3} t^2u^2,
	\label{eq:M4p8}
\end{align}
There are now three independent $\Z_4$-invariant kinematic structures,
$t^4$, $t^2(s^2+u^2)$ and $s^2u^2$, and correspondingly three
for $\Z_{\{2,2\}}$. This is reflected in the six LECs.

Similarly, $\ampl_{10,\{4\}}$ will be a linear combination of $s^5$, $s^3tu$ and
$st^2u^2$, and $\ampl_{10,\{2,2\}}$ will be a linear combination
of $t^5$, $t^3us$ and $tu^2s^2$, since these are the only independent
$\O(p^{10})$ kinematic structures that are invariant under $\Z_{\{4\}}$
and $\Z_{\{2,2\}}$, respectively. The coefficients will be linear combinations
of the LECs of the terms in $\lagr_{10}$ that only contain four $u_\mu$'s. 
These terms, along with the rest of $\lagr_{10}$, have not yet been studied.
The same pattern can be applied to $\O(p^{12})$ and beyond.

Note that the above discussion is fully compatible with section 5
in~\cite{p8lagr} where we have two functions with properties
$B(s,t,u)=B(u,t,s)$ and $C(s,t,u)=C(s,u,t)$. Independent combinations
in $B$ at order $p^{2n}$ are made from $t^{n-2i}(s-u)^{2i}$ and in $C$
from $s^{n-2i}(t-u)^{2i}$.

\subsection{The $\O(p^2)$ 6- and 8-point amplitudes}
\label{sec:M6p2-M8p2}

The leading order in the power counting offers a relatively simple playground
for flavour-ordering, free from splittings and singlets. It is relatively well
explored, and the amplitudes presented here were also calculated
in \cite{Kampf:2013vha} using different methods. 

The 6-point amplitude is given by the diagrams
\begin{equation}
	\tikzineq{
		\draw [thick] (-.5,.5)   -- (0,0) -- (-.5,0);
		\draw [thick] (-.5,-.5)  -- (0,0) -- (.5,-.5);
		\draw [thick] (.5,0)     -- (0,0) -- (.5,.5);
	}{M6p2-0}\qquad
	\tikzineq{
		\draw [thick] (-.5,.5)   -- (0,0) -- (-.5,0);
		\draw [thick] (-.5,-.5)  -- (0,0) -- (1,0) -- (1.5,-.5);
		\draw [thick] (1.5,0)    -- (1,0) -- (1.5,.5);
	}{M6p2-2}\quad.
\end{equation}
Each diagram represents the sum of all distinct labellings of its legs,
as described in section~\ref{sec:flavour}. The amplitude is
\begin{equation}
	\begin{split}
		-4iF^4\ampl_{2,6} =\:&
			s_{12} + s_{23} + s_{34} + s_{45} + s_{56} + s_{61}\\
			&-  \frac{(s_{12} + s_{23})(s_{45} + s_{56})}{s_{123}}
			-  \frac{(s_{23} + s_{34})(s_{56} + s_{61})}{s_{234}}
			-  \frac{(s_{34} + s_{45})(s_{61} + s_{12})}{s_{345}},
	\end{split}
	\label{eq:M6p2-full}
\end{equation}
which suggests the simplified form as defined in \eqref{eq:simplified}
\begin{equation}
	-4iF^4\ampl_{2,6} = \left\{s_{12} - \frac12\frac{(s_{12} + s_{23})(s_{45} + s_{56})}{s_{123}}\right\} + [\Z_6],
	\label{eq:M6p2}
\end{equation}
where $[\Z_6]$ indicates summation over all cyclic permutations. 
Note the factor of $1/2$, which expresses that the second term has twofold
symmetry under rotation, a trait that is shared by the second diagram above.

The 8-point amplitude is given by the diagrams
\begin{equation}
	\tikzineq{
		\draw [thick] (-.5,.5)   -- (0,0) -- (-.5,0);
		\draw [thick] (-.5,-.5)  -- (0,0) -- (.5,-.5);
		\draw [thick] (.5,0)     -- (0,0) -- (.5,.5);
		\draw [thick] (0,.5)     -- (0,0) -- (0,-.5);
	}{M8p2-0}\qquad
	\tikzineq{
		\draw [thick] (-.5,.5)   -- (0,0) -- (-.5,0);
		\draw [thick] (0,.5)     -- (0,0) -- (0,-.5);
		\draw [thick] (-.5,-.5)  -- (0,0) -- (1,0) -- (1.5,-.5);
		\draw [thick] (1.5,0)    -- (1,0) -- (1.5,.5);
	}{M8p2-1}\qquad
	\tikzineq{
		\draw [thick] (-.5,.5)   -- (0,0) -- (-.5,0);
		\draw [thick] (-.5,-.5)  -- (0,0) -- (.5,0) -- (1,0) -- (1.5,-.5);
		\draw [thick] (1.5,0)    -- (1,0) -- (1.5,.5);
		\draw [thick] (.5,.5) -- (.5,0) -- (.5,-.5);
	}{M8p2-2}\qquad
	\tikzineq{
		\draw [thick] (-.5,.5)   -- (0,0) -- (-.5,0);
		\draw [thick] (-.5,-.5)  -- (0,0) -- (.5,0) -- (1,0) -- (1.5,-.5);
		\draw [thick] (1.5,0)    -- (1,0) -- (1.5,.5);
		\draw [thick] (.25,.5) -- (.5,0) -- (.75,.5);
	}{M8p2-3}\quad ,
\end{equation}
and its stripped amplitude is, in a similarly simplified form,
\begin{multline}
	-8iF^6 \ampl_{2,8} = 
		\left\{\frac{4s_{12} + s_{1234}}{2} 
		- \frac{(s_{12}+s_{23})(s_{45}+s_{56}+s_{67}+s_{78} + s_{4567} + s_{5678})}{s_{123}}\right.	\\
		+ \frac12\frac{(s_{12}+s_{23})(s_{1234}+s_{4567})(s_{56}+s_{67})}{s_{123}s_{567}}		\\
		\left.+ \frac{(s_{12}+s_{23})(s_{1234}+s_{45})(s_{67}+s_{78})}{s_{123}s_{678}}\right\} + [\Z_8].
	\label{eq:M8p2}
\end{multline}

The analogous 10- and 12-point amplitudes, the second of which has not been
determined before, are presented in appendices~\ref{app:M10p2} and~\ref{app:M12p2}.

\subsection{The $\O(p^4)$ 6-point amplitude}

The calculation of this amplitude hinges decisively on the use of split-trace
flavour ordering. It was arrived at independently in a different form
by \cite{Low:2019ynd} using recursion relations. Our result agrees with theirs.
The amplitude is given by the four diagrams
\begin{equation}
	\tikzineq{
		\draw [thick] (-.5,.5)   -- (0,0) -- (-.5,0);
		\draw [thick] (-.5,-.5)  -- (0,0) -- (.5,-.5);
		\draw [thick] (.5,0)     -- (0,0) -- (.5,.5);
		\draw (0,0) node [anchor=south] {\ordidx 4};
	}{M6p4-0}\qquad
	\tikzineq{
		\draw [thick, rounded corners=3pt, name path=ad] (-.5, .5) -- (0,0) -- (-.5,-.5);
		\draw [thick, rounded corners=3pt, name path=eh] ( .5,-.5) -- (0,0) -- ( .5, .5);
		
		\path [name path=bf] (-.5, .2) -- (.5,-.2);
		\path [name path=cg] (-.5,-.2) -- (.5, .2);
		
		\draw [thick, name intersections={of=cg and eh}] ( .5, .2) -- (intersection-1);
		\draw [thick, name intersections={of=bf and eh}] ( .5,-.2) -- (intersection-1);
		\draw (0,0) node [anchor=south] {\ordidx 4};
	}{M6p4-1}\qquad
	\tikzineq{
		\draw [thick] (-.5,.5)   -- (0,0) -- (-.5,0);
		\draw [thick] (-.5,-.5)  -- (0,0) -- (1,0) -- (1.5,-.5);
		\draw [thick] (1.5,0)    -- (1,0) -- (1.5,.5);
		\draw (0,0) node [anchor=south] {\ordidx 4};
	}{M6p4-2}\qquad
	\tikzineq{
		\draw [thick, rounded corners=6pt] (-.5,.5) -- (0,0) -- (-.5,0);
		\draw [thick, rounded corners=4pt] (-.5,-.5)  -- (0,0) -- (1,0) -- (1.5,0);
		\draw [thick] (1.5,-.5)    -- (1,0) -- (1.5,.5);
		\draw (0,0) node [anchor=south] {\ordidx 4};
	}{M6p4-3}\quad .
\end{equation}
Note that unlike its $\O(p^2)$ counterpart, the third diagram is not symmetric
due to the asymmetric placement of vertices.
The amplitude has a single-trace and a two-trace part.
The single-trace stripped amplitude is
\begin{multline}
	-iF^6\ampl_{4,\{6\}} = 
		L_3\left\{
			s_{12}\left(s_{12}+s_{34}+s_{45}\right) 
			- \frac{(s_{12} + s_{23})\left(s_{45}^2 + s_{56}^2\right)}{s_{123}}
		\right\} + [\Z_6]	\\
		+ 2L_0\left\{
			s_{12}\left(s_{12} + s_{34} + 2s_{45}\right)
			+ s_{123}\left(s_{612} - s_{61}\right)
			- \frac{(s_{12} + s_{23})\left(s_{45} + s_{56}\right)^2}{s_{123}}
		\right\} + [\Z_6]	\\
	\label{eq:M6p4s6}
\end{multline}
In order to find the simplified form of the two-trace part, it is
extremely helpful to have a closed Mandelstam basis. In terms of the closed
basis $\basis_{\{2,4\}} = \{t_1,\ldots,t_9\}$ of \eqref{eq:24-basis}, it is
\begin{multline}
	-iF^6\ampl_{4,\{2,4\}} = 
		\frac{L_1}{2}\left\{
			t_1\big[t_1 + 2t_2 + t_3 - 3t_5\big]
			+ \frac{(t_2+t_3+t_4)^2\big[t_3 - 2t_5\big]}{2t_1}
		\right\} + [\Z_{\{2,4\}}]	\\
		+ \frac{L_2}{8}\left\{
			t_1\left[t_1 + 2t_2 + \frac{t_3}{2} - 3t_5\right] 
			+ 4t_7^2 - 2t_9^2
			\phantom{\frac{\big[t_3\big]}{2t_1}}
		\right.	\\
		\left.
			+\:\frac{\left[(t_2+t_3+t_4)^2 + 4(t_7+t_8+t_9)^2\right]\big[t_3 - 2t_5\big]}{2t_1}
		\right\} 
		+ [\Z_{\{2,4\}}].
	\label{eq:M6p4s24}
\end{multline}
Note that the summation over cyclic permutations is replaced by summation
over $\Z_{\{2,4\}}$.

\subsection{Further amplitudes}\label{sec:further}

We have computed the $\O(p^6)$ 6-point amplitude, and using the closed
Mandelstam bases presented in appendix~\ref{app:closed-basis}, it is possible to
present its reduced form given in appendix \ref{app:M6p6}. The $\O(p^6)$
divergent part is given explicitly in the supplementary material
\cite{supplementary} as well as the $\O(p^8)$ expression. The $\O(p^2)$ 10-point
is given in appendix \ref{app:M10p2}. Finally the $\O(p^2)$ 12-point amplitudes
is given in appendix \ref{app:M12p2}.

We have also computed several amplitudes whose expressions are too large to
overview. They have been verified by checking their Adler zeroes, and in some
cases by running brute-force Feynman diagram calculations. Beyond these
amplitudes, we have generated the flavour-ordered diagrams of many more
amplitudes using our program \fodge, described in section~\ref{sec:fodge}. Here,
we only summarise the number and general properties of the diagrams to give an
idea of how the complexity scales. The summary is given in
table~\ref{tab:diagr}.

For all entries labelled ``Yes'' in table~\ref{tab:diagr} that are not
included in the main text, the flavour-ordered diagrams are given in the
supplementary material \cite{supplementary}. Most of the amplitudes themselves
are too long to be practically written down, but they can be generated by using
the freely available programs described in section~\ref{sec:fodge}.

\begin{table}[hbtp]
	\centering
	\begin{tabular}{cc|cccccc|p{3cm}}
		\multirow{2}{*}{$\O(p^N)$}	&	 \multirow{2}{*}{$n$}	&	\multicolumn{6}{c|}{Number of diagrams}									&	\multirow{2}{*}{Computed?}	\\
								&						&	$SU(N_f)$	&	$U(N_f)$	&	$SU(3)$	&	$U(3)$	&	$SU(2)$	&	$U(2)$	&							\\
		\hline\hline
		\multirow{6}{*}{$\O(p^2)$}	&	4					&	1		&			&			&			&			&			&	Yes \eqref{eq:M4p2}	\\
								&	6					&	2		&	\multicolumn{5}{c|}{\textit{(same as $SU(N_f)$)}}		&	Yes \eqref{eq:M6p2}	\\
								&	8					&	4		&			&			&			&			&			&	Yes \eqref{eq:M8p2}	\\
								&	10					&	16		&			&			&			&			&			&	Yes \eqref{eq:M10p2}		\\
								&	12					&	73		&			&			&			&			&			&	Yes$^*$ \eqref{eq:M12p2}			\\
								&	14					&	414		&			&			&			&			&			&	No				\\
		\hline
		\multirow{5}{*}{$\O(p^4)$}	&	4					&	2		&			&			&			&	1		&	1		&	Yes \eqref{eq:M4p4}	\\
								&	6					&	4		&	\multicolumn{3}{c}{\textit{(same as $SU(N_f)$)}}
																										&	2		&	2		&	Yes$^{\dag*}$ (\ref{eq:M6p4s6}-\ref{eq:M6p4s24})					\\
								&	8					&	18		&			&			&			&	8		&	8		&	Yes$^{\dag*}$ \mbox{(\ref{eq:M6p6s6}-\ref{eq:M6p6s222})} 		\\
								&	10					&	90		&			&			&			&	43		&	43		&	Yes$^*$ 		\\
								&	12					&	577		&			&			&			&	283		&	283		&	No					\\
		\hline
		\multirow{4}{*}{$\O(p^6)$}	&	4					&	2		&	2		&	2		&	2		&	1		&	1		&	Yes \eqref{eq:M4p6}	\\
								&	6					&	10		&	9		&	9		&	8		&	4		&	3		&	Yes$^{\dag*}$ 		\\
								&	8					&	50		&	45		&	48		&	43		&	18		&	14		&	Yes$^*$ 		\\
								&	10					&	360		&	318		&	348		&	316		&	129		&	98		&	No				\\
		\hline
		\multirow{3}{*}{$\O(p^8)$}	&	4					&	2		&	2		&	2		&	2		&	1		&	1		&	Yes$^*$ \eqref{eq:M4p8}	\\
								&	6					&	11		&	10		&	10		&	9		&	4		&	3		&	Yes$^*$ 		\\
								&	8					&	105		&	85		&	97		&	77		&	34		&	21		&	No
	\end{tabular}
	\caption{Summary of the number of $\O(p^N)$ $n$-point diagrams.
The $SU(N_f)$ column shows the general count, and the $U(N_f)$ column shows
the count without singlet diagrams. The $SU(3)$ and $SU(2)$ columns show the
number of distinct diagrams left when some Lagrangian terms have been
eliminated using the Cayley-Hamilton relation as discussed in
section~\ref{sec:small-Nf}. Note that the distinction for $N_f=2$
assumes we remove the $L_1$ and $L_2$ term and
emerges first at $\O(p^4)$.  For $N_f=3$ it emerges first at $\O(p^6)$.
The distinction between $SU$ and $U$ also emerges first at $\O(p^6)$. 
The rightmost column states whether an amplitude has been computed by us,
and provides references to the explicit amplitudes when possible.
Amplitudes marked with an asterisk have to our knowledge not been calculated
before; the $\O(p^4)$ 6-point amplitude was recently independently reproduced
by~\cite{Low:2019ynd}. Amplitudes marked with a dagger have been verified with
a brute-force Feynman diagram calculation; the remainder rely only on Adler
zeroes for verification.}
\label{tab:diagr}
\end{table}

In the table, we note that the number of diagrams grows more rapidly with $n$
(the number of particles) than with $N$ (the power-counting order). Especially
when $N>n$, the number of new diagrams is very small. This is also reflected
in the computational effort needed: the $\O(p^2)$ 10-point, $\O(p^6)$ 8-point
and $\O(p^8)$ 6-point amplitudes took approximately 10 minutes each to
calculate with \form\ \cite{Vermaseren:2000nd,Kuipers:2012rf},
while the $\O(p^4)$ 10-point amplitude took
almost and hour and the $\O(p^2)$ 12-point amplitude took over 2 days. At
high $N$, the calculation of vertex factors takes significant time, while
at high $n$, the conversion to Mandelstam variables is very time-consuming
due to the large dimension of the kinematic space.

As the table shows, we have calculated all amplitudes with less
than 100 diagrams, excluding $N\geq 10$, where the Lagrangian is not yet known.
If we decide to push the frontier of large $n$ further in the future, we
expect the required computational effort to be severe.

%%%%%%%%%%%%%%%%%%%%%%%%%%%%%%%%%%%%%%%%%%%%%%%%%%%%%%%%%%%%%%%%%%%%%%%%%%%%%%
\section{Conclusions}
\label{sec:conclusions}

In this work we have extended flavour ordering methods to include multiple
traces and higher orders in derivatives. The uniqueness of the method relies
on the extended orthogo\-nality relation \eqref{eq:ortho}. We implemented
the constraints in a diagram generator and used this then to calulate a number
of amplitudes in the NLSM with more legs and derivatives than obtained
previously.

Our methods are fairly constrained in which models they can be applied to,
since they hinge on the existence of flavour structures and the contraction
identity \eqref{eq:contr}. On the other hand, they are readily extended to 
extremely high-order and many-particle amplitudes. They may also
have some applicability to loop diagrams and massive particles under \chpt.
A tentative discussion of these possibilities can be found in 
\cite{masterthesis}.

Flavour-ordering serves as an enhancement of the standard diagrammatic approach,
and as such is rather brute-force in nature. This contrasts with the recursive
approach developed in~\cite{Cheung:2015ota}, in which subtler properties such as
soft limits play a much clearer role. These methods can also be applied to a
wider range of models. The downside is that practical calculations require
algebraic manipulations that are not entirely obvious. 
Flavour-ordering calculations can 
be very extensive, but are mathematically trivial and easily automated. Further
developments of recursion relations in~\cite{Low:2019ynd} have offset the 
algebraic difficulties, but
soft recursion retains the fundamental limitation that recursive calculation of
an $\O(p^m)$ $n$-point amplitude requires $n>m$. Therefore, the $\O(p^6)$ 
6-point can not be reached by such means, and must be supplied as a seed 
amplitude if $\O(p^6)$ amplitudes are to be calculated for more than 6 
particles. For this, our methods seem to be the only viable option other than
brute-force Feynman diagrams.

%%%%%%%%%%%%%%%%%%%%%%%%%%%%%%%%%%%%%%%%%%%%%%%%%%%%%%%%%%%%%%%%%%%%%%%%%%%%
\section*{Acknowledgements}
We thank Malin Sjödahl for correcting an earlier version
of \eqref{eq:ortho}. KK enjoyed kind hospitality at Lund University
while most of this work was realized.
This work is supported in part by the Swedish
Research Council grants contract numbers 2015-04089 and
2016-05996, by the European Research Council (ERC)
under the European Union’s Horizon 2020 research and
innovation programme under grant agreement No 668679,
and the Czech Government projects GACR 18-17224S and LTAUSA17069.

\pagebreak
\appendix

%%%%%%%%%%%%%%%%%%%%%%%%%%%%%%%%%%%%%%%%%%%%%%%%%%%%%%%%%%%%%%%%%%%%%%%%%

\section{The NNLO NLSM Lagrangian}
\label{app:p6lagr}
%%%
%%% To be included in article_VERSION.tex
%%%

The NNLO \chpt\ Lagrangian $\lagr_6^\text{\chpt}$ was first determined in \cite{p6lagr}. It has 21 terms that do not vanish when external fields are removed, but this turns out to be an overcomplete basis for the NLSM. In tandem with the NNNLO \chpt\ Lagrangian $\lagr_8$ in \cite{p8lagr}, the authors of that paper produced a version of $\lagr_6^\text{\chpt}$ where removing external fields yields a \emph{minimal} NLSM Lagrangian with 19 terms. It was not published there, but we present it in table~\ref{tab:p6lagr}. The first 135 terms of the Lagrangian in \cite{p8lagr} constitute a minimal NLSM Lagrangian $\lagr_8$.

The Lagrangian of \cite{p6lagr} is expressed as
\begin{equation}
	\lagr_6^\text{\chpt} = \sum_{i=1}^{112} K_i Y_i,
\end{equation}
where $K_i$ are LECs and $Y_i$ are monomials in the fields. Terms 1--6 and 49--63 remain when external fields are removed. All monomials except $Y_1$, $Y_2$ and $Y_6$ correspond directly to monomials in the minimal NLSM Lagrangian
\begin{equation}
	\lagr_6 = \sum_{i=1}^{19} \lagt{6}{i} \lagm{6}{i},
\end{equation}
where $\lagt{6}{i}$ are LECs and $\lagm{6}{i}$ are monomials. The remaining $Y_i$ can be decomposed in terms of $\lagm{6}{i}$ using the relations described in \cite{p8lagr}, which for the NLSM simplify to
\begin{equation}
	\begin{gathered}
		\nabla_\mu u_\nu = \nabla_\nu u_\mu, \qquad \nabla_\mu u^\mu = 0, \\
		\comm{\nabla_\mu}{\nabla_\nu} u_\rho = \frac14\comm{\comm{u_\mu}{u_\nu}}{u_\rho}.
	\end{gathered}
\end{equation}
This yields the relations
\begin{align}
	Y_1 &= -3\lagm{6}{3} + \lagm{6}{4} + \lagm{6}{{15}} -2\lagm{6}{{16}} + \frac12 \lagm{6}{{17}} + \lagm{6}{{18}} - \frac12 \lagm{6}{{19}},\\
	Y_2 &= -8\lagm{6}{2} + 2\lagm{6}{8} - 2\lagm{6}{9},\\
	Y_6 &= 4\frac{\lagm{6}{2} - \lagm{6}{1}}{3} -2\frac{\lagm{6}{9} - \lagm{6}{8}}{3} - 2\frac{\lagm{6}{{12}} - \lagm{6}{{11}}}{3}.
\end{align}
Furthermore, some factors of 2 appear since \cite{p6lagr} includes higher derivatives in terms of \mbox{$h_{\mu\nu}\equiv\nabla_\mu u_\nu + \nabla_\nu u_\mu$}, which is just $2\nabla_\mu u_\nu$ in the NLSM. 

\begin{table}[hbtp]
	\centering
	\begin{tabular}{|c|ccc|c|}
		\hline
		\multirow{2}{*}{Monomial}	&	\multicolumn{3}{c|}{Number in $\lagr_6$}	& \multirow{2}{*}{Relation to \cite{p6lagr}}		\\
								&	$SU(N_f)$		&	$SU(3)$ 	&	$SU(2)$	&											\\
		\hline
		$\tr{u_\mu\nabla_\nu u_\rho}\tr{u^\mu\nabla^\nu u^\rho}$	&	1	&	1	&	1	&	$K_4 - \frac43 K_6$			\\
		$\tr{u_\mu\nabla_\nu u_\rho}\tr{u^\rho\nabla^\mu u^\nu}$	&	2	&	2	&		&	$-8K_2 + \frac43 K_6$			\\
		$\tr{u_\mu\nabla_\nu u^\mu u_\rho\nabla^\nu u^\rho}$		&	3	&	3	&	2	&	$4K_5 - 3K_1$					\\
		$\tr{u_\mu\nabla_\nu u_\rho u^\mu\nabla^\nu u^\rho}$		&	4	&	4	&	3	&	$4K_3 + K_1$					\\
		$\tr{u_\mu u^\mu}\tr{u_\nu u^\nu}\tr{u_\rho u^\rho}$		&	5	&		&		&	$K_{51}$						\\
		$\tr{u_\mu u^\mu}\tr{u_\nu u_\rho}\tr{u^\nu u^\rho}$		&	6	&		&		&	$K_{56}$						\\
		$\tr{u_\mu u^\nu}\tr{u_\nu u^\rho}\tr{u_\rho u^\mu}$		&	7	&		&		&	$K_{63}$						\\
		$\tr{u_\mu u^\mu}\tr{u_\nu u^\nu u_\rho u^\rho}$			&	8	&	5	&		&	$K_{50} + 2K_2 + \frac23 K_6$	\\
		$\tr{u_\mu u^\mu}\tr{u_\nu u_\rho u^\nu u^\rho}$			&	9	&	6	&		&	$K_{57} - 2K_2 - \frac23 K_6$	\\
		$\tr{u_\mu u^\mu u_\nu}\tr{u^\nu u_\rho u^\rho}$			&	10	&	7	&		&	$K_{53}$						\\
		$\tr{u_\mu u_\nu}\tr{u^\mu u^\nu u_\rho u^\rho}$			&	11	&		&		&	$K_{55} + \frac23 K_6$			\\
		$\tr{u_\mu u_\nu}\tr{u^\mu u_\rho u^\nu u^\rho}$			&	12	&		&		&	$K_{62} - \frac23 K_6$			\\
		$\tr{u_\mu u_\nu u_\rho}\tr{u^\mu u^\nu u^\rho}$			&	13	&		&		&	$K_{59}$						\\
		$\tr{u_\mu u_\nu u_\rho}\tr{u^\mu u^\rho u^\nu}$			&	14	&		&		&	$K_{61}$						\\
		$\tr{u_\mu u^\mu u_\nu u^\nu u_\rho u^\rho}$				&	15	&	8	&	4	&	$K_{49} + K_1$				\\
		$\tr{u_\mu u^\mu u_\nu u_\rho u^\nu u^\rho}$				&	16	&	9	&	5	&	$K_{54} - 2K_1$				\\
		$\tr{u_\mu u^\mu u_\nu u_\rho u^\rho u^\nu}$				&	17	&	10	&		&	$K_{52} + \frac12 K_1$			\\
		$\tr{u_\mu u_\nu u^\mu u_\rho u^\nu u^\rho}$				&	18	&	11	&		&	$K_{60} + K_1$				\\
		$\tr{u_\mu u_\nu u_\rho u^\mu u^\nu u^\rho}$				&	19	&	12	&	6	&	$K_{58} - \frac12 K_1$			\\
		\hline
	\end{tabular}
	\caption{The terms of the NNLO NLSM Lagrangian $\lagr_6$. The numbering of the terms is taken from material produced in tandem with \cite{p8lagr}; the choice of which terms to keep at small $N_f$ (see section~\ref{sec:small-Nf}) is carried over from \cite{p6lagr}. The rightmost column shows how $K_i$ combine to give $\lagt{6}{i}$ when the overcomplete Lagrangian is decomposed.}
	\label{tab:p6lagr}
\end{table}

\subsection{Renormalisation}
\label{app:renorm}	
NNLO \chpt\ was renormalised in \cite{p6lagr-renorm}, based on \cite{p6lagr}. For renormalisation in the NLSM, we transfer those results to the minimal Lagrangian given in table~\ref{tab:p6lagr}. For details on the renormalisation, see \cite{p6lagr-renorm} and sources therein. At NLO, it is performed by splitting the LECs as
\begin{equation}
	L_i = (c\mu)^{d-4}\left[ L_i^r(\mu, d) + \Gamma_i \Lambda\right],\qquad \Lambda = \frac{1}{16\pi^2 (d-4)}.
	\label{eq:p4-renorm}
\end{equation}
The measurable LECs are given by $L_i^r(\mu, d)$ as $d \to 4$, with
\begin{equation}
	\Gamma_0 = \frac{N_f}{48},\qquad \Gamma_1 = \frac{1}{16},\qquad \Gamma_2 = \frac{1}{8}, \qquad \Gamma_3 = \frac{N_f}{24}.
	\label{eq:p4-gamma}
\end{equation}
Likewise, at NNLO the LECs are split as
\begin{equation}
	\lagt{6}{i} = \frac{(c\mu)^{2(d-4)}}{F^2}\left[\lagt{6}{i}^r(\mu, d) - \Gamma^{(2)}_i\Lambda^2 - \left(\Gamma_i^{(1)} + \Gamma_i^{(L)}(\mu,d)\right)\Lambda\right].
	\label{eq:p6-renorm}
\end{equation}
The $\Gamma$'s for the corresponding renormalisation of the $K_i$ are given in \cite{p6lagr-renorm}. Using the rightmost column of table~\ref{tab:p6lagr}, the renormalisation of the minimal $\lagr_6$ is given in table~\ref{tab:renorm}.

%%%%%%%%%%%%%%%%%%%%%%%%%%%%%%%%%%%%%%%%%%%%%%%%%%%%%%%%%%%%%%%%%%%%%%%%%%%%%%%%%%%%%%%%%%%%%%%%%%%%%%%%%%%%%%%%%%%%%%%%%%%%%%%%
% Produced from the .log file (divergence_p6.log) with the following regex replacements:
%
% Put columns in correct order:
% GL \* \(([^\)]*)\)\n\n([^G]*)G1 \* \(([^\)]*)\)\n\n([^G]*)G2 \* \(([^\)]*)\)   ->   $\5$\t&\t$\3$\t%\t$\1$1\t\\\\
% Format i:
%  *\[old0?([0-9]+)\] *= *\n *   ->   \t\t\1\t&\t
% Format math:
% ([0-9]+)/([0-9]+)   ->   \\frac{\1}{\2}
% \*  ->
% Format constants:
% Nf   ->   N_f
% L([0-3])r   ->   L_\1^r
%
%%%%%%%%%%%%%%%%%%%%%%%%%%%%%%%%%%%%%%%%%%%%%%%%%%%%%%%%%%%%%%%%%%%%%%%%%%%%%%%%%%%%%%%%%%%%%%%%%%%%%%%%%%%%%%%%%%%%%%%%%%%%%%%%
\begin{table}[hbtp]
	\centering
	\begin{tabular}{|c|c|c|c|}
		\hline
		$i$	&	$\Gamma^{(2)}_i$	&	$16\pi^2\Gamma^{(1)}_i$	&	$\Gamma^{(L)}_i$		\\
		\hline
		1	&	$  - \frac{5}{72} N_f $	&	$  - \frac{19}{864} N_f $	&	$  - \frac{4}{3} L_3^r - 4 L_0^r $	\\
		2	&	$  - \frac{5}{9} N_f $	&	$ \frac{1}{864} N_f $	&	$  - \frac{16}{3} L_3^r - \frac{8}{3} L_0^r - \frac{8}{3} N_f L_2^r - 8 N_f L_1^r $	\\
		3	&	$  - \frac{5}{16} - \frac{5}{96} N_f^2 $	&	$  - \frac{1}{96} + \frac{35}{6912} N_f^2 $	&	$  - \frac{25}{6} L_2^r - \frac{5}{3} L_1^r - \frac{23}{12} N_f L_3^r - \frac{7}{6} N_f L_0^r $	\\
		4	&	$ \frac{5}{16} + \frac{5}{288} N_f^2 $	&	$ \frac{1}{96} - \frac{17}{6912} N_f^2 $	&	$ \frac{25}{6} L_2^r + \frac{5}{3} L_1^r + \frac{3}{4} N_f L_3^r + \frac{1}{6} N_f L_0^r $	\\
		5	&	$ \frac{1}{64} $	&	$ \frac{5}{256} $	&	$ \frac{1}{4} L_2^r $	\\
		6	&	$  - \frac{1}{32} $	&	$ \frac{3}{128} $	&	$  - \frac{1}{2} L_2^r $	\\
		7	&	$  - \frac{1}{8} $	&	$  - \frac{1}{32} $	&	$  - 2 L_2^r $	\\
		8	&	$ \frac{1}{24} N_f $	&	$ \frac{25}{576} N_f $	&	$ \frac{17}{12} L_3^r + \frac{13}{6} L_0^r + \frac{2}{3} N_f L_2^r - \frac{1}{2} N_f L_1^r $	\\
		9	&	$  - \frac{1}{96} N_f $	&	$  - \frac{5}{1152} N_f $	&	$  - \frac{13}{24} L_3^r - \frac{29}{12} L_0^r + \frac{1}{12} N_f L_2^r $	\\
		10	&	$  - \frac{1}{64} N_f $	&	$  - \frac{5}{256} N_f $	&	$  - \frac{5}{4} L_3^r + L_0^r $	\\
		11	&	$  - \frac{5}{144} N_f $	&	$  - \frac{1}{1728} N_f $	&	$ \frac{2}{3} L_3^r + \frac{4}{3} L_0^r - \frac{2}{3} N_f L_2^r $	\\
		12	&	$  - \frac{13}{144} N_f $	&	$  - \frac{53}{1728} N_f $	&	$  - \frac{7}{6} L_3^r - \frac{19}{3} L_0^r - \frac{1}{3} N_f L_2^r $	\\
		13	&	$  - \frac{1}{192} N_f $	&	$ \frac{65}{2304} N_f $	&	$  - \frac{3}{4} L_3^r + L_0^r $	\\
		14	&	$ \frac{7}{192} N_f $	&	$  - \frac{23}{2304} N_f $	&	$ \frac{5}{4} L_3^r + L_0^r $	\\
		15	&	$ \frac{5}{48} + \frac{1}{144} N_f^2 $	&	$  - \frac{7}{576} - \frac{25}{6912} N_f^2 $	&	$ \frac{5}{6} L_2^r + \frac{5}{3} L_1^r + \frac{1}{6} N_f L_3^r + \frac{1}{3} N_f L_0^r $	\\
		16	&	$  - \frac{5}{24} - \frac{1}{96} N_f^2 $	&	$  - \frac{19}{576} + \frac{5}{1152} N_f^2 $	&	$  - \frac{2}{3} L_2^r - \frac{16}{3} L_1^r - \frac{1}{4} N_f L_3^r - \frac{1}{6} N_f L_0^r $	\\
		17	&	$ \frac{5}{96} - \frac{1}{48} N_f^2 $	&	$ \frac{43}{576} + \frac{49}{13824} N_f^2 $	&	$ \frac{1}{4} L_2^r + \frac{7}{6} L_1^r - \frac{2}{3} N_f L_3^r - \frac{2}{3} N_f L_0^r $	\\
		18	&	$ \frac{5}{48} + \frac{1}{64} N_f^2 $	&	$  - \frac{67}{576} - \frac{7}{1728} N_f^2 $	&	$ \frac{1}{6} L_2^r + 3 L_1^r + \frac{17}{24} N_f L_3^r + \frac{1}{12} N_f L_0^r $	\\
		19	&	$  - \frac{5}{96} - \frac{1}{144} N_f^2 $	&	$ \frac{25}{288} + \frac{5}{4608} N_f^2 $	&	$  - \frac{7}{12} L_2^r - \frac{1}{2} L_1^r - \frac{1}{3} N_f L_3^r $	\\
		\hline
	\end{tabular}
	\caption{The coefficients used to renormalise $\lagr_6$ as per \eqref{eq:p6-renorm}. $L_i^r$ are the renormalised LECs of $\lagr_4$ as per (\ref{eq:p4-renorm}-\ref{eq:p4-gamma}). Note how the highest power of $N_f$ in $\Gamma^{(1,2)}_i$ is 3 minus the number of traces in $\lagm{6}{i}$.}
	\label{tab:renorm}
\end{table}

\subsection{Explicit divergences}
\label{app:div}
In analogy with \eqref{eq:p4-renorm}, we define
\begin{equation}
	\ampl_{4,R} = (c\mu)^{d-4}\left[\ampl_{4,R}^r(\mu,d) + \ampl_{4,R}^{(1)}\Lambda\right],
\end{equation}
where $\ampl_{4,R}$ is some $\O(p^4)$ stripped amplitude, $\ampl_{4,R}^r(\mu,d)$ is the corresponding measurable amplitude expressed in terms of $L_i^r(\mu,d)$, and $\ampl_{6,R}^{(1)}$ is its divergence.

Using this notation and \eqref{eq:p4-gamma}, the divergence of the $\O(p^4)$ 4-point amplitude \eqref{eq:M4p4} is
\begin{equation}
	-iF^4\ampl_{4,\{4\}}^{(1)} = N_f\frac{s^2+t^2+u^2}{12},\qquad -iF^4\ampl_{4,\{2,2\}}^{(1)} = \frac{s^2+t^2+u^2}{2}.
	\label{eq:p4-div}
\end{equation}
These kinematic terms are highly symmetric, more so than the amplitude itself. The divergences of the 6-point amplitude (\ref{eq:M6p4s6}-\ref{eq:M6p4s24}) are
\begin{equation}
	\begin{split}
		-iF^6\ampl_{4,\{6\}}^{(1)} 
			&= \frac{N_f}{12}\left\{
				s_{12}\left(s_{12}+s_{34}+\frac{3s_{45}}{2}+s_{234}\right) - \frac{s_{123}s_{234}}{2}
				\right.\\
			&\hspace{2.5cm}\left.
				-\,\frac{(s_{12}+s_{23})\left(s_{45}^2+s_{56}^2\right) 
					+ s_{12}s_{23}\left(s_{45}+s_{56}\right)}{s_{123}}
			\right\} + [\Z_{\{6\}}]\\
		-iF^6\ampl_{4,\{2,4\}}^{(1)}
			&= \frac{1}{64}\left\{
				3\left(t_1^2 + 2t_1t_2 + \frac{t_1t_3}{2} - 3t_1t_5\right)
				+ 4t_7^2 - 2t_9^2
				\right.\\
			&\left.
				\hspace{1.5cm}+\,\frac{1}{t_1}\left[
					3\left(t_2t_3t_4-2t_2t_3t_5-2t_2t_4t_5-2t_3t_4t_5\right)
					+ 3t_3^2(t_2+t_4)
					\vphantom{\frac{3(t_3 - 2t_5)}{2}}\right.\right.\\
			&\left.\left.
				\hspace{2.5cm}+\,\frac{3(t_3 - 2t_5)}{2}\left(T_{234}^2 + 2T_{789}^2\right)
				\right]
			\right\} + [\Z_{\{2,4\}}].
	\end{split}
\end{equation}
We use the closed basis \eqref{eq:24-basis}, and $T_{ijk} = t_i+t_j+t_k$.

An $\O(p^6)$ analogue of \eqref{eq:p4-div} can be formed based on \eqref{eq:p6-renorm}:
\begin{equation}
	\ampl_{6,R} = \frac{(c\mu)^{2(d-4)}}{F^2}\left[\ampl_{6,R}^r(\mu,d) - \ampl_{6,R}^{(2)}\Lambda^2 - \left(\ampl_{6,R}^{(1)} + \ampl_{6,R}^{(L)}(\mu,d)\right)\Lambda\right],
	\label{eq:p6-div}
\end{equation}
where $\ampl_{6,R}^{(2,L)}$ will gain contributions from both \eqref{eq:p4-renorm} and \eqref{eq:p6-renorm}.

Using this notation and the above renormalisation, the divergences of the $\O(p^6)$ 4-point amplitude are \eqref{eq:M4p6} are
\begin{align}
	-iF^4\ampl_{6,\{4\}}^{(1)} &= \frac{1}{72}\left(2t^3 - s^3-u^3\right) 
		+ \frac{N_f^2}{5184}\left[8t^3 + 35(s^3+u^3)\right],\notag\\
	-iF^4\ampl_{6,\{4\}}^{(2)} &= \frac{5}{12}\left(2t^3 -s^3-u^3\right)
		- \frac{5N_f^2}{72}\left(s^3+u^3\right),\notag\\
	-iF^4\ampl_{6,\{4\}}^{(L)} &= 10\frac{2L_1^r + 5L_2^r}{9}\left(2t^3 -s^3-u^3\right)
		\eqbreak{+}\frac{2N_f L_0^r}{9}\left[2t^3 - 7(s^3+u^3)\right]
		- \frac{N_f L_3^r}{9}\left[2t^3 + 23(s^3+u^3)\right],\\
	-iF^4\ampl_{6,\{2,2\}}^{(1)} &= \frac{N_f}{1296}\left[s^3 + 58(u^3+t^3)\right],\notag\\
	-iF^4\ampl_{6,\{2,2\}}^{(2)} &= -\frac{5N_f}{108}\left[8s^3 + 5(u^3+t^3)\right],\notag\\
	-iF^4\ampl_{6,\{2,2\}}^{(L)} &= -16N_f\frac{3L_1^r + L_2^r}{9}\left(s^3+t^3+u^3\right)
		\eqbreak{+}\frac{2 L_0^r}{9}\left[23(u^3+t^3) - 8s^3\right]
		- \frac{8 L_3^r}{9}\left(u^3+t^3 + 4s^3\right),
\end{align}
with the dependence on $(\mu,d)$ left out for compactness. These expressions do not share the simplicity and symmetry of their $\O(p^4)$ counterparts. The analogous divergences of the $\O(p^6)$ 6-point amplitude (appendix~\ref{app:M6p6}) are given in \cite{supplementary}.

%%%%%%%%%%%%%%%%%%%%%%%%%%%%%%%%%%%%%%%%%%%%%%%%%%%%%%%%%%%%%%%%%%%%%%%%%

\section{The orthogonality of flavour structures}\label{app:ortho}
Here, we prove the orthogonality relation \eqref{eq:ortho} used in section~\ref{sec:unique} to prove the uniqueness of stripped amplitudes. It relies on notation defined in that and previous sections.

Let $\sigma,\rho\in\perm_n$ be two permutations, and $Q,R$ be two $n$-index flavour splittings. We use these to build two flavour structures, and begin by focusing on the trace in $\flav_\sigma(Q)$ that contains $a_{\sigma(n)}$ and the trace in $\flav_\rho(R)$ that containis $a_{\rho(m)}$, where we have picked $m$ such that $\rho(m)=\sigma(n)$. If there are more traces present, we leave them as passive ``spectators'' for the time being. Then, we use \eqref{eq:contr} to contract $a_{\sigma(n)}$ in
\begin{equation}
	\flav_\sigma(Q)\cdot\big[\flav_\rho(R)\big]^* = 
		\left[\tr{X a_{\sigma(n-1)}a_{\rho(m-1)} Y} -
		\frac{1}{N_f}\tr{X a_{\sigma(n-1)}}\tr{a_{\rho(m-1)} Y}\right]\cdot\text{(spectators)},
	\label{eq:init}
\end{equation}
where the product is defined as in \eqref{eq:ortho}.

From here on, we work only to leading order in $N_f$, so we can omit the second term above. (Note that we do not do this because $N_f$ is necessarily large, but because we wish to use power counting of $N_f$ to separate orthogonal flavour structures.) We then move on to contracting $\sigma(n-1)$, followed by $\sigma(n-2)$, and so on. Each time we contract $\sigma(n-i)$, the situation may be one of the following cases:
\begin{enumerate}
	\item $\rho(m-i) = \sigma(n-i)$. We carry on through a special case of the contraction identity \eqref{eq:contr-alt}, and find
	\begin{equation}
		\tr{Xa_{\sigma(n-i)}a_{\sigma(n-i)} Y} = \frac{N_f^2-1}{N_f}\tr{XY}.
	\end{equation}
	This may be repeated as long as there are indices left, and we gain a factor of $N_f$ (plus $\O(N_f^\inv)$, which we ignore) each time. 
	\item $\rho(m-i) \neq \sigma(n-i)$, but $\rho(m') = \sigma(n-i)$ is in the same trace as $\sigma(n-i)$. Here, \eqref{eq:contr-alt} (after some reshuffling of $X$ and $Y$) gives
	\begin{equation}
		\tr{Xa_{\sigma(n-i)} Y a_{\sigma(n-i)}} = \left[\tr{X}\tr{Y} - \frac{1}{N_f}\tr{XY}\right].
	\end{equation}
	the second term is suppressed by a factor of $N_f^\inv$, and the first must eventually take a detour through \eqref{eq:init} before continuing; in any case, this case falls behind case 1 by at least two factors of $N_f$.
	\item $\rho(m') = \sigma(n-i)$ is in a different trace than $\sigma(n-i)$. This forces us to bring in the spectator trace containing $\rho(m')$ and go back to \eqref{eq:init}, so this case falls behind case 1 by at least one factor of $N_f$.
	\item The trace is empty. We gain a factor of $\tr{\1} = N_f$, and if there are no spectator traces left, we are done. Otherwise, we bring in the next pair of spectators and continue from \eqref{eq:init}.
\end{enumerate}
If $Q=R=\{n\}$ and $\sigma \equiv \rho\mod{\Z_R}$, we will only encounter case 1 until we finish with a case 4, and will gain a total factor of $N_f^{n}[1 + \O(N_f^{-2})]$. If $Q=R\neq\{n\}$ on the other hand, we will encounter case 4 at each split, but the leading power of $N_f$ stays the same. 

If $\sigma \not\equiv \rho \mod{\Z_R}$, we must eventually encounter case 2, so this falls behind the $\sigma\equiv\rho\mod{\Z_R}$ case by at least two powers of $N_f$. If $Q\neq R$, we will encounter case 3 (without a corresponding case 4) whenever there is a mismatch in the flavour splits, so we will fall behind the $Q=R$ case by at least one power of $N_f$. This is the reason for the values of $\gamma$ stated below \eqref{eq:ortho}.

Thus, we have proven
\begin{equation}
	\flav_\sigma(Q)\cdot\big[\flav_\rho(R)\big]^* = N_f^n
	\begin{cases}
		1 + \O\left(N_f^{-2}\right)		&	\text{if $Q = R$ and $\sigma\equiv\rho\mod{\Z_R}$,}				\\
		\O\left(N_f^{-\gamma}\right)	&	\text{otherwise ($\gamma\geq 1$)}	
	\end{cases}
\end{equation}
which is \eqref{eq:ortho}.

%%%%%%%%%%%%%%%%%%%%%%%%%%%%%%%%%%%%%%%%%%%%%%%%%%%%%%%%%%%%%%%%%%%%%%%%%%%%%%%%%%%%%%%%%%%%%%%%%%%%%%%%%%%%%%%%%%%%%%%%%%%%%%

\section{The double soft limit}\label{app:dslim}
This appendix provides a derivation of \eqref{eq:dslim}, which is used to calculate the double soft limit of stripped amplitudes. We start by quoting \eqref{eq:dslim-full}, which is proven in \cite{Kampf:2013vha} and gives the double soft limit of the full amplitude:
\begin{multline}
	\lim_{\e\to0}\ampl_{m,n+2}^{aba_1\cdots a_n}(\e p,\e q, p_1,\ldots,p_n) = \\
		-\frac{1}{F^2}\sum_{i=1}^n f^{abc}f^{a_i dc}\frac{p_i\cdot(p-q)}{p_i\cdot(p+q)}
		\ampl_{m,n}^{a_1\cdots a_{(i-1)}da_{(i+1)}\cdots a_n}(p_1,\cdots,p_n).
	\label{eq:dslim-quote}
\end{multline}
In order to find the corresponding expression for a stripped amplitude, we project it out by contracting both sides with $[\flav_\id(R)]^*$ over all flavour indices (see \eqref{eq:ampl} and section~\ref{sec:unique}). On the left-hand side of \eqref{eq:dslim-quote}, this will project out $\lim_{\epsilon\to0}\ampl_{m,R}(\e p,\e q, p_1,\ldots)$. For simplicity, we start with \mbox{$R=\{n+2\}$} before moving on to the general multi-trace case. According to \eqref{eq:ampl}, the right-hand side of \eqref{eq:dslim-quote} has the form (schematically, with kinematic terms omitted)
\begin{equation}
	\sum_{\sigma\in\perm_n/\Z_n}
	f^{abc}f^{a_idc}\tr{{a_{\sigma(1)}}\cdots {a_{\sigma(i-1)}} d {a_{\sigma(i+1)}}\cdots {a_{\sigma(n)}}}
\end{equation}
plus flavour-split structures, but those can be ignored due to \eqref{eq:ortho}. We have omitted the algebra generators for readability; $a_i$ means $t^{a_i}$. The structure constants can be contracted in using \eqref{eq:contr} and $f^{abc}=-i\tr{t^a\comm{t^b}{t^c}}$, leaving
\begin{equation}
	-\tr{{a_{\sigma(1)}}\cdots {a_{\sigma(i-1)}} 
		\comm{\comm{a}{b}}{{a_{\sigma(i)}}}{a_{\sigma(i+1)}}\cdots {a_{\sigma(n)}}}.
\end{equation}
With appendix~\ref{app:ortho} in mind, we immediately see that this is orthogonal to $\flav_\id(n+2)$ unless $\sigma=\id$. The nested commutators expand to
\begin{equation}
	\comm{\comm{a}{b}}{{a_i}} = ab{a_i} - ba{a_i} - {a_i}ab + {a_i}ba.
\end{equation}
Since $a$ comes before $b$ in $\flav_\id(n+2)$, the second and fourth terms vanish under the projection. Also, $ab$ occurs at the beginning (or, equivalently, the end) of the flavour structure, so the first term only contributes when $i=1$, and the third term only contributes when $i=n$. This collapses the sum in \eqref{eq:dslim-quote} to those two cases, leaving
\begin{multline}
	\lim_{\e\to0}\ampl_{m,\{n+2\}}(\e p,\e q, p_1,\ldots,p_n) = \\
		\frac{1}{F^2}\left\{\frac{p_1\cdot(p-q)}{p_1\cdot(p+q)}-\frac{p_n\cdot(p-q)}{p_n\cdot(p+q)}\right\}
		\ampl_{m,\{n\}}(p_1,\cdots,p_n).
	\label{eq:dslim-single}
\end{multline}
If we now move on to general $R$, we see that $a$ and $b$ must reside in the same trace, since the nested commutator on the right-hand side is inside a single trace. This is essentially the condition stated for the validity of \eqref{eq:dslim}, with $(p_n,p,q,p_1)$ mapping to $(p_{i-1},p_i,p_j,p_{j+1})$. The trace they reside in can be treated exactly like the single-trace flavour structure of \eqref{eq:dslim-single}, and all other traces in the flavour structure follow along as ``spectators'', as in a normal application of \eqref{eq:ortho}. The reduction $\{n+2\}\to\{n\}$ in \eqref{eq:dslim-single} then generalises to $R\to R'$ as described below \eqref{eq:dslim}. This generalisation therefore results in \eqref{eq:dslim}, which is thereby proven.

%%%%%%%%%%%%%%%%%%%%%%%%%%%%%%%%%%%%%%%%%%%%%%%%%%%%%%%%%%%%%%%%%%%%%%%%%%%%%%%%%%%%%%%%%%%%%%%%%%%%%%%%

\section{Closed Mandelstam bases}\label{app:closed-basis}
Here, we show the derivation of closed Mandelstam bases for all 6-particle flavour structures as described in section~\ref{sec:mandel}. Note that neither basis is unique, and that better basis choices may exist.

\subsection{The basis for $R=\{2,4\}$}\label{app:24-basis}
This is the only basis other than $\basis_{\{6\}}$ that is needed at $\O(p^4)$. This flavour split permits four different propagator momenta (corresponding to the labellings in \eqref{eq:24-labels}). Since $\Z_{\{2,4\}}$ is Abelian and rather small, it is simple to handle, and some inspired trial-and-error gives the closed basis $\basis_{\{2,4\}} = \{t_1,\ldots,t_9\}$ with elements\footnote{This basis is a slight improvement over the one used in~\cite{masterthesis}. It modifies $t_5$ and $t_6$ so that they map to themselves under $\Z_{\{2,4\}}$.}
\begin{equation}
	\begin{gathered}
		t_1 = s_{123}, \quad 
		t_2 = s_{124}, \quad 
		t_3 = s_{125}, \quad 
		t_4 = s_{126},	\\
		t_5 = s_{45} + s_{56} + \frac{s_{125} - s_{123}}{2},\quad 
		t_6 = s_{45} - s_{56} + \frac{2s_{124} - (s_{124} + s_{125})}{2},		\\
		t_7 = s_{14} + s_{15} + \frac{s_{123} + s_{126}}{2}, \quad 
		t_8 = s_{15} + s_{16} + \frac{s_{123} + s_{124}}{2}, \\
		t_9 = s_{14} + s_{16} + \frac{s_{123} + s_{125}}{2}.
	\end{gathered}
	\label{eq:24-basis}
\end{equation}
Under the action of $\Z_{\{2,4\}}$, they transform as
\begin{equation}
	\begin{split}
		21\,3456:\qquad
		\{t_1,\ldots t_6,\,t_7,t_8,t_9\} &\to \{t_1,\ldots t_6,\,-t_7,-t_8,-t_9\},
		\\
		12\,4563:\qquad
		\{t_1,\ldots t_6,\,t_7,t_8,t_9\} &\to \{t_2,t_3,t_4,t_1,\,t_5,-t_6,\,+t_8,-t_7,-t_9\},		
	\end{split}
\end{equation}
where the first permutation cycles the 2-trace, and the second cycles the 4-trace; together, they generate all of $\Z_{\{2,4\}}$. Note that $\Z_{\{2,4\}}$ does not act as a true permutation on the basis, since some elements change sign. This appears to be unavoidable, but is not a problem --- in fact, any complex phase can be applied without hindering simplification.

\subsection{The basis for $R=\{3,3\}$}\label{app:33-basis}
The group $\Z_{\{3,3\}}$ is generated by the permutations $g_1=231\,456$ and $g_2=456\,123$. The group is not abelian, which makes its effects less predictable. Among all kinematic invariants, only $s_{123}$ maps to itself under both generators, and is also the only squared propagator momentum permitted by this flavour structure. The other 24 invariants de\-compose into a sextuplet and two nonets under the group, and can be mapped out in a variant of a Cayley graph:
\begin{equation}
	\tikzineq{
		\draw[fill=black] ( 90:{1 + 1/sqrt(3)}) node [anchor=south] {\tiny 45};
		\draw[fill=black] (210:{1 + 1/sqrt(3)}) node [anchor=east ] {\tiny 64};
		\draw[fill=black] (330:{1 + 1/sqrt(3)}) node [anchor=west ] {\tiny 56};
		
		\draw[thick, dashed] ( 90:{1 + 1/sqrt(3)}) -- ( 90:{1/sqrt(3)}) node [anchor=south] {\tiny 12};
		\draw[thick, dashed] (210:{1 + 1/sqrt(3)}) -- (210:{1/sqrt(3)}) node [anchor=east ] {\tiny 31};
		\draw[thick, dashed] (330:{1 + 1/sqrt(3)}) -- (330:{1/sqrt(3)}) node [anchor=west ] {\tiny 23};
		
		\draw[thick] ( 90:{1/sqrt(3)}) -- (210:{1/sqrt(3)}) -- (330:{1/sqrt(3)}) -- cycle;
	}{z33-sextet}\qquad
	\tikzineq{
		\draw[thick] (120:1) -- (150:{sqrt(3)}) -- (180:1) -- cycle;
		\draw[thick] (240:1) -- (270:{sqrt(3)}) -- (300:1) -- cycle;
		\draw[thick] (  0:1) -- ( 30:{sqrt(3)}) -- ( 60:1) -- cycle;
		
		\draw[thick, dashed] ( 60:1) -- (120:1);
		\draw[thick, dashed] (180:1) -- (240:1);
		\draw[thick, dashed] (300:1) -- (360:1);
		
		\draw (150:{sqrt(3)}) node[anchor=center] {$\star$} node [anchor=south] {\tiny 14};
		\draw (270:{sqrt(3)}) node[anchor=center] {$\star$} node [anchor=north] {\tiny 36};
		\draw ( 30:{sqrt(3)}) node[anchor=center] {$\star$} node [anchor=south] {\tiny 25};
		
		\draw (120:1) node[anchor=center] {$\bullet$} node [anchor=south] {\tiny 24};
		\draw (240:1) node[anchor=center] {$\bullet$} node [anchor=east ] {\tiny 16};
		\draw (  0:1) node[anchor=center] {$\bullet$} node [anchor=west ] {\tiny 35};
		
		\draw (180:1) node[anchor=center] {$\circ$} node [anchor=east ] {\tiny 34};
		\draw (300:1) node[anchor=center] {$\circ$} node [anchor=west ] {\tiny 26};
		\draw ( 60:1) node[anchor=center] {$\circ$} node [anchor=south] {\tiny 15};
	}{z33-nonet0}\qquad
	\tikzineq[yscale=-1]{
		\draw[thick] (120:1) -- (150:{sqrt(3)}) -- (180:1) -- cycle;
		\draw[thick] (240:1) -- (270:{sqrt(3)}) -- (300:1) -- cycle;
		\draw[thick] (  0:1) -- ( 30:{sqrt(3)}) -- ( 60:1) -- cycle;
		
		\draw[thick, dashed] ( 60:1) -- (120:1);
		\draw[thick, dashed] (180:1) -- (240:1);
		\draw[thick, dashed] (300:1) -- (360:1);
		
		\draw (150:{sqrt(3)}) node [anchor=north] {\tiny 345};
		\draw (270:{sqrt(3)}) node [anchor=south] {\tiny 135};
		\draw ( 30:{sqrt(3)}) node [anchor=north] {\tiny 234};
		
		\draw (120:1) node [anchor=north] {\tiny 145};
		\draw (240:1) node [anchor=east ] {\tiny 134};
		\draw (  0:1) node [anchor=west ] {\tiny 125};
		
		\draw (180:1) node [anchor=east ] {\tiny 245};
		\draw (300:1) node [anchor=west ] {\tiny 124};
		\draw ( 60:1) node [anchor=north] {\tiny 235};
	}{z33-nonet1}
	\label{eq:z33}
\end{equation}
Each node in the graph represents $s_{ij\cdots}$ and is marked with $ij\cdots$. The action of $g_1$ is represented by following the solid-drawn triangles clockwise, and $g_2$ is represented by following the dashed lines.

We must now extract 9 basis elements $t_1,\ldots,t_9$ that are closed under $\Z_{\{3,3\}}$. In the first nonet, we have marked three sets of invariants with $\star$, $\bullet$ and $\circ$. They map to each other as $(\star,\bullet,\circ)\to(\bullet,\circ,\star)$ under $g_1$ and as $(\star,\bullet,\circ)\to(\star,\circ,\bullet)$ under $g_2$, so suitable linear combinations of the elements in each set will be closed under $\Z_{\{3,3\}}$. Similar constructions taken from the sextet and the other nonet turn out not to be linearly independent from these.

Unfortunately, it appears impossible to form a basis that contains the propagator momentum $s_{123}$ as an element, but since there is only one propagator, this is not as much of a problem as it would be under a group that supports more operators. Also, it appears impossible to form real linear combinations without sacrificing either linear independence or closedness. Guided by the fact that $g_1$ has period 3, we instead insert the third root of unity, $\omega = e^{2\pi i/3}$, and find the closed and complete basis $\basis_{\{3,3\}}$ with elements\footnote{The basis presented in~\cite{masterthesis} was not complete. This mistake was not discovered until after its publication, and is corrected here at the cost of losing the propagator.}
\begin{equation}
	\begin{gathered}
		t_1 = -\frac{s_{36} + s_{14} + s_{25}}{3},\quad
		t_2 = -\frac{s_{24} + s_{35} + s_{16}}{3},\quad
		t_3 = -\frac{s_{15} + s_{26} + s_{34}}{3},\\
		t_6 = \frac{s_{36} + \omega s_{14} + \omega^2 s_{25}}{3},\quad
		t_4 = \frac{s_{24} + \omega s_{35} + \omega^2 s_{16}}{3},\quad
		t_5 = \frac{s_{15} + \omega s_{26} + \omega^2 s_{34}}{3},\\
		t_9 = \frac{\omega^2 s_{36} + \omega s_{14} + s_{25}}{3},\quad
		t_7 = \frac{\omega^2 s_{24} + \omega s_{35} + s_{16}}{3},\quad
		t_8 = \frac{\omega^2 s_{15} + \omega s_{26} + s_{34}}{3},
	\end{gathered}	
	\label{eq:33-basis}	 
\end{equation}
In each row above, the first basis element comes from the $\star$ set, the second from the $\bullet$ set, and the third from the $\circ$ set. The propagator momentum is $s_{123}=\frac{3}{2}(t_1+t_2+t_3)$. The basis transforms as
\begin{equation}
	\begin{split}
		g_1:\qquad& \{t_1,t_2,t_3,\,t_4,t_5,t_6,\,t_7,t_8,t_9\} 
			\to \{t_2, t_3, t_1,\,\omega t_5, \omega t_6, \omega t_4,\,\omega^2 t_8, \omega^2 t_9, \omega^2 t_7\},\\
		g_2:\qquad& \{t_1,t_2,t_3,\,t_4,t_5,t_6,\,t_7,t_8,t_9\} 
			\to \{t_1,t_3,t_2,\,t_4,t_6,t_5,\,t_7,t_9,t_8\}.
	\end{split}
\end{equation}
Since stripped amplitudes are real, the complex basis must be compensated for by complex coefficients. Still, $\basis_{\{3,3\}}$ is just as valid as a real basis, and is useable for simplification.

\subsection{The basis for $R=\{2,2,2\}$}\label{app:222-basis}
The group $\Z_{\{2,2,2\}}$ is also non-abelian, and can be tackled similarly to $\Z_{\{3,3\}}$. We choose the generators $g_1=34\,56\,12$, $g_2 = 21\,34\,56$ and $g_3=65\,43\,21$ with the hopes that they be well-behaved, since $\basis_{\{6\}}$ is closed under two of them. This flavour structure permits six propagators that form a sextet under the group. The Cayley graph is
\begin{equation}
	\tikzineq{
		\draw[thick] (-.5,-.5) node[anchor=north] {\tiny 156} 
			-- (-.5,.5) node[anchor=south] {\tiny 126}
			-- ({-(.5 + sqrt(3)/2)},0) node[anchor=east] {\tiny 123} -- cycle;
		\draw[thick] ( .5,-.5) node[anchor=north] {\tiny 125}
			-- ( .5,.5) node[anchor=south] {\tiny 134}
			-- ({ (.5 + sqrt(3)/2)},0) node[anchor=west] {\tiny 124} -- cycle;
		
		\draw[thick, dotted] (-.5,-.5) to [bend right=40] (-.5,.5);
		\draw[thick, dotted] ( .5,-.5) to [bend left =40] ( .5,.5);
		
		\draw[thick, dashed] (-.5,-.5) .. controls ( .5,-.5) and (-.5, .5) .. ( .5, .5);
	}{z222-sextet}
\end{equation}
where $g_1$ and $g_2$ are represented as in \eqref{eq:z33}, and the dotted lines represent the action of $g_3$. The remaining invariants decompose into a triplet, a quadruplet, and a 12-plet:
\begin{equation}
	\tikzineq{
		\draw[thick] (-.5,0) node[anchor=east] {\tiny 12} -- ( .5,0) node[anchor=west] {\tiny 56} 
			-- (0,{sqrt(3)/2}) node [anchor=south] {\tiny 34} -- cycle;
			
		\draw[thick, dotted] (-.5,0) to [bend right=40] ( .5,0);
	}{z222-triplet}\qquad
	\tikzineq{
		\coordinate (top) at (0,{sqrt(3)/2)});
		\draw[thick] (-.5,0) node[anchor=east] {\tiny 136} -- ( .5,0) node[anchor=west] {\tiny 145} 
			-- (top) node [anchor=west] {\tiny 235} -- cycle;
			
		\draw[thick, dashed] (-.5,0) to [bend right=40] ( .5,0);
		\draw[thick, dotted] (top) to [bend right=40] (-.5,0);
		\draw[thick, dashed] (top) -- +(0,.8) node[anchor=south] {\tiny 135};
	}{z222-quadruplet}\qquad
	\tikzineq{
		\draw[thick, dotted] (1,1) .. controls (2,1) .. (3,2);
		\draw[thick, dotted] (0,1) .. controls (1,2) .. (2,2);
		\draw[white, line width=1mm] (1,0) .. controls (2,0) and (1,3) .. (2,3);
		\draw[thick, dotted        ] (1,0) .. controls (2,0) and (1,3) .. (2,3);
		
		\draw[thick] (0,3) -- (1,3) -- (0,2) -- cycle;
		\draw[thick] (2,3) -- (3,2) -- (2,2) -- cycle;
		\draw[thick] (3,1) -- (3,0) -- (2,0) -- cycle;
		\draw[thick] (1,0) -- (0,1) -- (1,1) -- cycle;
		
		\draw[thick, dashed] (0,1) -- (0,2);
		\draw[thick, dashed] (1,3) -- (2,3);
		\draw[thick, dashed] (3,2) -- (3,1);
		\draw[thick, dashed] (2,0) -- (1,0);
		
		\draw (0,3) node[anchor=center] {$\star$} node [anchor=east ] {\tiny 45};
		\draw (2,2) node[anchor=center] {$\star$} node [anchor=north] {\tiny 46};
		\draw (3,0) node[anchor=center] {$\star$} node [anchor=west ] {\tiny 36};
		\draw (1,1) node[anchor=center] {$\star$} node [anchor=south] {\tiny 35};
		
		\draw (0,2) node[anchor=center] {$\bullet$} node [anchor=east ] {\tiny 23};
		\draw (0,1) node[anchor=center] {$\bullet$} node [anchor=east ] {\tiny 13};
		\draw (3,2) node[anchor=center] {$\bullet$} node [anchor=west ] {\tiny 24};
		\draw (3,1) node[anchor=center] {$\bullet$} node [anchor=west ] {\tiny 14};
		
		\draw (1,3) node[anchor=center] {$\circ$} node [anchor=south] {\tiny 61};
		\draw (2,3) node[anchor=center] {$\circ$} node [anchor=south] {\tiny 62};
		\draw (1,0) node[anchor=center] {$\circ$} node [anchor=north] {\tiny 51};
		\draw (2,0) node[anchor=center] {$\circ$} node [anchor=north] {\tiny 52};
		
		\draw[thick, dotted] (0,3) to [bend right=40] (0,2);
		\draw[thick, dotted] (3,0) to [bend right=40] (3,1);
	}{z222-12plet}
\end{equation}
Like in \eqref{eq:z33}, we have marked three closed sets of $s_{ij}$'s. From these, it is possible to construct three linearly independent elements that close the basis without need for the less structured triplet and quadruplet. Thus, $\basis_{\{2,2,2\}}$ has elements
\begin{equation}
	\begin{gathered}
		t_1 = s_{123},\quad t_2 = s_{126},\quad t_3 = s_{156},\quad t_4 = s_{124},\quad t_5 = s_{125},\quad t_6 = s_{134},\\
		t_7 = \frac{s_{61} - s_{62} + s_{52} - s_{51}}{2},\quad
		t_8 = \frac{s_{23} - s_{24} + s_{14} - s_{13}}{2},\quad
		t_9 = \frac{s_{45} - s_{46} + s_{36} - s_{35}}{2},
	\end{gathered}
	\label{eq:222-basis}
\end{equation}
where the factors of $1/2$ remove some large powers of 2 that show up when writing amplitudes in this basis. Unlike in $\basis_{\{3,3\}}$, there was no need to resort to complex numbers. The basis transforms as
\begin{equation}
	\begin{split}
		g_1:\qquad& \{t_1,t_2,t_3,t_4,t_5,t_6,t_7,t_8,t_9\} 
			\to \{t_1,t_2,t_6,t_4,t_5,t_3,-t_7,-t_8,t_9\},\\
		g_2:\qquad& \{t_1,t_2,t_3,t_4,t_5,t_6,t_7,t_8,t_9\} 
			\to \{t_2,t_3,t_1,t_5,t_6,t_4,t_8,t_9,t_7\},\\
		g_3:\qquad& \{t_1,t_2,t_3,t_4,t_5,t_6,t_7,t_8,t_9\} 
			\to \{t_1,t_3,t_2,t_4,t_6,t_5,t_7,t_9,t_8\}.
	\end{split}
\end{equation}
No element is a fixed point, which makes the basis harder to work in.

%%%%%%%%%%%%%%%%%%%%%%%%%%%%%%%%%%%%%%%%%%%%%%%%%%%%%%%%%%%%%%%%%%%%%%%%%%%%%%
\section{Explicit amplitudes}
\label{app:results}

\subsection{The $\O(p^6)$ 6-point amplitude}
\label{app:M6p6}
This amplitude has been simplified using the closed bases of appendix~\ref{app:closed-basis}. The terms were reduced to coset representatives in \fodge\ followed by manual post-processing. Greater simplification might be possible for some terms. The amplitude consists of four stripped amplitudes with flavour split $\{6\}$, $\{2,4\}$, $\{3,3\}$, and $\{2,2,2\}$.

There are three diagrams with a single-trace flavour structure:
\begin{equation}
	\tikzineq{
		\draw [thick] (-.5,.5)   -- (0,0) -- (-.5,0);
		\draw [thick] (-.5,-.5)  -- (0,0) -- (.5,-.5);
		\draw [thick] (.5,0)     -- (0,0) -- (.5,.5);
		\draw (0,0) node [anchor=south] {\ordidx 6};
	}{M6p6-unsplit-0}\qquad
	\tikzineq{
		\draw [thick] (-.5,.5)   -- (0,0) -- (-.5,0);
		\draw [thick] (-.5,-.5)  -- (0,0) -- (1,0) -- (1.5,-.5);
		\draw [thick] (1.5,0)    -- (1,0) -- (1.5,.5);
		\draw (0,0) node [anchor=south] {\ordidx 6};
	}{M6p6-unsplit-1}\qquad
	\tikzineq{
		\draw [thick] (-.5,.5)   -- (0,0) -- (-.5,0);
		\draw [thick] (-.5,-.5)  -- (0,0) -- (1,0) -- (1.5,-.5);
		\draw [thick] (1.5,0)    -- (1,0) -- (1.5,.5);
		\draw (0,0) node [anchor=south] {\ordidx 4};
		\draw (1,0) node [anchor=south] {\ordidx 4};
	}{M6p6-unsplit-2}
\end{equation}
The corresponding stripped amplitude is
\begin{align}
	-iF^8&\ampl_{6,\{6\}} 
		\eqbreak{=} 2(L_3 + 2L_0)^2\left\{2(s_{12}+s_{23})^2(s_{45}+s_{56}) - s_{123}(s_{12}+s_{23})(s_{45}+s_{56})\right\}\notag\\
		&+ 16L_3L_0s_{12}s_{23}(s_{45}+s_{56})\notag\\
		&- 8(L_0^2 + L_3L_0)\frac{(s_{12}+s_{23})^2(s_{45}+s_{56})^2}{s_{123}} - 2L_3^2\frac{(s_{12}^2+s_{23}^2)(s_{45}^2+s_{56}^2)}{s_{123}}\notag\\
		&+\lagt{6}{3}\left\{\left[s_{123}s_{34} 
			- \frac{s_{123}s_{234}}{2}\right](s_{12} + [\Z_6]) - s_{12}s_{23}s_{34} 
			\vphantom{\frac{\left[(s_{12}+s_{23})^3)\right](s_{45}+s_{56})}{3s_{123}}}
			\peqbreak{+} \frac{2(s_{12}+s_{45})^3 + (s_{12}^3+s_{45}^3)}{3} 
			- \frac{\left[(s_{12}+s_{23})^3 + 2(s_{12}^3+s_{23}^3)\right](s_{45}+s_{56})}{3s_{123}}
			\right\}\notag\\
		&+\lagt{6}{4}\left\{s_{123}s_{234}s_{345} 
			+ \frac{s_{123}^2}{2}(s_{234}+s_{345}-2s_{34}) 
			\peqbreak{-} 2s_{123}s_{234}(2s_{12}+s_{23}+2s_{34})
			- \frac{s_{12}s_{45}}{2}(s_{123} + s_{345}) 
			\peqbreak{+} s_{123}\left[s_{12} + s_{56} + 2s_{12}s_{34} + 2s_{34}s_{56} 
				+ 4s_{34}(s_{23}+s_{34}+s_{45})\right] 
			\peqbreak{+} \frac{(s_{12}+s_{45})^3}{2} 
			- \frac{(s_{12}+s_{23})^3(s_{45}+s_{56})}{s_{123}}
			\right\}\notag\\
		&-\lagt{6}{15}\left\{s_{12}s_{45}s_{56}\right\}\notag\\
		&+\lagt{6}{16}\left\{s_{12}(s_{34}+s_{45})(s_{123}+s_{345}) - s_{12}s_{123}s_{345} - s_{12}s_{34}s_{56} 
			- \frac{s_{12}s_{45}}{2}(s_{12} + [\Z_6])\right\}\notag\\
		&+\lagt{6}{17}\left\{s_{12}s_{45}(s_{123} + s_{345} - s_{12} - s_{45})\right\}\notag\\
		&+\lagt{6}{18}\left\{(s_{12}+s_{45} + s_{234}^2)(s_{123} + s_{345}) - s_{12}s_{234}(s_{123} + [\Z_6]) 
			- 4s_{12}s_{123}s_{345} 
			\peqbreak{+} 2s_{12}s_{234}(s_{23} + s_{34} + s_{45} + s_{56} + s_{61}) 
			- 2s_{12}s_{34}(s_{23} + s_{45} + s_{56} + s_{61})
			\right\}\notag\\
		&+\lagt{6}{19}\left\{2s_{123}s_{234}s_{345} + 3s_{234}^2(s_{123} + s_{345})
			- 3s_{12}s_{234}(s_{123} + [\Z_6]) - 6s_{12}s_{123}s_{345} 
			\peqbreak{+} 6s_{12}s_{234}(s_{23} + s_{34} + s_{56} + s_{61}) 
			- 6s_{12}s_{23}s_{34} - 2s_{12}s_{34}s_{56}
			\right\}\notag\\
		&+ [\Z_6].
	\label{eq:M6p6s6}
\end{align}
The ``$+\,[\Z_6]$'' acts on all terms in the amplitude.

There are also three diagrams with a $\{2,4\}$-split flavour structure:
\begin{equation}
	\tikzineq{
		\draw [thick, rounded corners=3pt, name path=ad] (-.5, .5) -- (0,0) -- (-.5,-.5);
		\draw [thick, rounded corners=3pt, name path=eh] ( .5,-.5) -- (0,0) -- ( .5, .5);
		
		\path [name path=bf] (-.5, .2) -- (.5,-.2);
		\path [name path=cg] (-.5,-.2) -- (.5, .2);
		
		\draw [thick, name intersections={of=cg and eh}] ( .5, .2) -- (intersection-1);
		\draw [thick, name intersections={of=bf and eh}] ( .5,-.2) -- (intersection-1);
		\draw (0,0) node [anchor=south] {\ordidx 6};
	}{M6p6-split24-0}\qquad
	\tikzineq{
		\draw [thick, rounded corners=6pt] (-.5,.5) -- (0,0) -- (-.5,0);
		\draw [thick, rounded corners=4pt] (-.5,-.5)  -- (0,0) -- (1,0) -- (1.5,0);
		\draw [thick] (1.5,-.5)    -- (1,0) -- (1.5,.5);
		\draw (0,0) node [anchor=south] {\ordidx 6};
	}{M6p6-split24-1}\qquad
	\tikzineq{
		\draw [thick, rounded corners=6pt] (-.5,.5) -- (0,0) -- (-.5,0);
		\draw [thick, rounded corners=4pt] (-.5,-.5)  -- (0,0) -- (1,0) -- (1.5,0);
		\draw [thick] (1.5,-.5) -- (1,0) -- (1.5,.5);
		\draw (0,0) node [anchor=south] {\ordidx 4};
		\draw (1,0) node [anchor=south] {\ordidx 4};
	}{M6p6-split24-2}
\end{equation}
Using the closed basis \eqref{eq:24-basis}, the stripped amplitude is
{\allowdisplaybreaks
\begin{align}
	-iF^8&\ampl_{6,\{2,4\}} 
		\eqbreak{=} L_0L_1\left\{t_1\left[-t_1^2+t_2^2+2t_3(3t_2+t_3-2t_5) + t_4^2 + 4t_5^2\right] 
			- \frac{T_{234}^2(t_3-2t_5)^2}{t_1}
			\right\}\notag\\
		&+ L_0L_2\left\{
			t_1\left[
				t_1\left(2t_5-\frac{5t_1}{4}\right) 
				+ t_2\left(\frac{t_2}{4} + \frac{3t_2}{2} + 4t_5\right)
				\ppeqbreak{+} 3t_3\left(\frac{t_3}{2} - t_5\right)
				+ \frac{t_4^2}{4} - 3t_5^2 + T_{789}^2\right]
			-\frac{(t_3 - 2t_5)^2(T_{234}^2 + 4T_{789}^2)}{4t_1}
			\right\}\notag\\
		&+ L_3L_1\left\{
			-\frac{t_1^3}{2} + 2t_1^2(t_5-t_6)
			\peqbreak{+} t_1\left[
				2t_2(t_3+2t_5+t_6) + t_3(5t_3 + t_5 -3t_6)
				-\frac{3t_4^2}{2} + t_5^2 + t_6^2\right]
			\peqbreak{+} \frac{1}{t_1}\left[
				- t_2^3(t_2 + t_3 + 2t_4 + 2t_6) 
				+ t_2^2\left(\frac{t_3^2}{2} + t_3(t_5-3t_6) - t_4^2 - 4t_4t_6\right)
				\ppeqbreak{+} t_2(t_3^2t_4 + t_3t_4^2)
				+ 2t_2t_3t_4(t_5 - t_6) - 2t_2t_4^2t_6
				- \frac{t_3^4}{2} 
				\ppeqbreak{+} t_3^3(-t_4 + t_5 + t_6)
				- \frac{t_3^2t_4^2}{2} + (t_5 + t_6)(t_3t_4^2 + 2t_3^2t_4)
				- T_{234}^2(t_5^2 + t_6^2)\right]
			\right\}\notag\\
		&+ L_3L_2\left\{
			-\frac{9t_1^3}{8} + t_1^2\frac{t_5+t_6}{2}
			\peqbreak{+} t_1\left[
				t_2\left(\frac{11t_2}{2} + \frac{t_3}{2} + 3t_5 + \frac{5t_6}{2}\right)
				+ \frac{t_3}{4}\left(5t_3 + t_5 + t_6\right)
				- \frac{3(t_5^2 + t_6^2)}{4} 
				\ppeqbreak{-} \frac{7t_4^2}{8} 
				+ t_7(t_9 - 3t_8) + 5t_8t_9 + \frac{t_8^2 + t_9}{2}
			\right]
			+ t_5(2t_7^2 + t_9^2) + 2t_6t_7t_8
			\peqbreak{+} \frac{1}{t_1}\left[
				\frac{T_{789}^2}{4}\left(
					\left[2t_2 - t_3\right]^2
					+ \left[2t_5 + 2t_6 - t_3\right]^2
					+ 8t_6[t_2-t_5]\right)
				\ppeqbreak{-} \frac{T_{234}^2}{4}(t_5^2 + t_6^2) 
				+ \frac{t_5 + t_6}{4}(2t_3^2t_4 + t_3t_4^2)
				- \frac{t_2^4}{4} 
				- \frac{t_2^3}{4}(t_3 + 2t_4 + 2t_6)
				\ppeqbreak{+} \frac{t_2^2}{8}(t_3^2 + 2t_3t_5 - 6t_3t_6 - 2t_4^2 - 8t_4t_6)
				\ppeqbreak{+} \frac{t_2}{2}\left(
					\frac{t_3^2}{2}[t_4 + 2t_5] 
					+ \frac{t_3}{2}\left[t_4^2 + 2t_4t_5 - 2t_4t_6\right] 
					- t_4^2t_6\right)
				\right]
			\right\}\notag\\
		&+ \lagt{6}{1}\mathcal H(3) + \lagt{6}{2}\mathcal H(-1)\notag\\
		&+ \lagt{6}{8}\left\{
			\frac{t_1}{4}\left[
				t_2^2 - 2t_3^2 + t_4^2 - 2(t_5^2 + 2t_6^2) + 2t_6(t_1 + t_3)\right]
			\right\}\notag\\
		&+ \lagt{6}{9}\left\{
			-t_1\left[
				4t_5^2 + \frac{t_2^2 + t_4^2}{2} + t_2t_3 - t_5(t_1 + 2t_2 + t_3)\right]
			\right\}\notag\\
		&+ \lagt{6}{11}\left\{
			\frac{t_1^3}{16} 
			- \frac{t_1}{16}\left[
				t_2^2 - t_3^2 + t_4^2 + 4(t_8^2 - t_7^2) + 2t_2(t_5+t_6) + 4t_9(t_8+t_7)\right]
			\peqbreak{-} \frac{t_5t_9^2}{8} - \frac{t_6t_7^2}{4}
			\right\}\notag\\
		&+ \lagt{6}{12}\left\{
			\frac{t_1^3}{16}
			- \frac{t_1}{16}\left[(t_2 + t_3)^2 + t_4^2 + 4(t_7+t_8)^2 + 4t_9^2 - 4t_3t_5\right]
			+ t_5\frac{2t_7^2 - t_9^2}{4}
			\right\}\notag\\
		&+ [\Z_{\{2,4\}}],
	\label{eq:M6p6s24}
\end{align}}
where $T_{ijk} = t_i+t_j+t_k$, and
\begin{align}
	\mathcal H(\eta) 
		&= \frac{3t_1^3}{128} - \frac{3t_1^2t_5}{32} 
			\eqbreak{+} t_1\left[
				\frac{9t_2^2}{128} + t_2\frac{3t_3 - 5t_5}{32} + \frac{t_3^2}{32} 
				- \frac{5t_3t_5}{64} + \frac{9t_4^2}{128}
				+ \frac{\eta}{32}(t_7+t_8-t_9)^2 + \frac{\eta}{8}(t_7^2 + t_8^2 - t_9^2)
				\right]
			\eqbreak{-} \eta t_5\frac{2t_7^2 + t_9^2}{16}
			+ \frac{3}{16t_1}\left[
				t_2^3\frac{t_3 - 2t_5}{24} + \frac{t_2^2t_3}{8}(t_3 + t_4 - 2t_5) 
				- \frac{t_2^2t_4t_5}{4}
				\peqbreak{+} \frac{t_2t_3^2}{8}(t_3 + 2t_4 - 2t_5)
				+ \frac{t_2t_3}{8}(t_4^2 - 4t_4t_5)
				+ \frac{t_2t_4^2t_5}{4}
				\peqbreak{+} \frac{t_3^3}{24}(t_3 + 3t_4 + 2t_5)
				+ \frac{t_3^2t_4}{8}(t_4 - 2t_5)
				+ \frac{t_4^4}{24}
				- \frac{t_4^2t_5}{12}(3t_3 + t_4)
				\peqbreak{+} 
				\frac{\eta T_{789}^2}{6}\left(t_3^2 + t_2t_3 - t_2t_5 + t_3t_4 - 2t_3t_5 - 2t_4t_5\right)
				\right]
\end{align}
is used for compactness.

There are two diagrams with a $\{3,3\}$ flavour split:
\begin{equation}
	\tikzineq{
		\draw [thick, rounded corners=3pt, name path=ac] (-.5, .5) -- (0,0) -- (-.5,-.5);
		\draw [thick, rounded corners=3pt, name path=df] ( .5,-.5) -- (0,0) -- ( .5, .5);
		
		\path [name path=be] (-.5, 0) -- (.5,0);
		
		\draw [thick, name intersections={of=be and ac}] (-.5,0) -- (intersection-1);
		\draw [thick, name intersections={of=be and df}] ( .5,0) -- (intersection-1);
		\draw (0,0) node [anchor=south] {\ordidx 6};
	}{M6p6-split33-0}
        \qquad
	\tikzineq{
		\draw [thick] (-.5,.5)   -- (0,0) -- (-.5,0);
		\draw [thick] (-.5,-.5)  -- (0,0) (1,0) -- (1.5,-.5);
		\draw [thick] (1.5,0)    -- (1,0) -- (1.5,.5);
		\draw [thick, dashed] (0,0) -- (1,0);
		\draw (0,0) node [anchor=south] {\ordidx 4};
		\draw (1,0) node [anchor=south] {\ordidx 4};
	}{M6p6-split33-1}
\end{equation}
Using the closed basis \eqref{eq:33-basis}, the stripped amplitude is
\begin{align}
	-iF^8&\ampl_{6,\{3,3\}} 
			\eqbreak{=} 64(L_0+L_3)^2\left\{
				(s_{12}+s_{23})\left(s_{45}^2+s_{45}s_{56}+s_{56}^2\right) 
				+ \left(s_{12}^2+s_{12}s_{23}+s_{23}^2\right)(s_{45}+s_{56})
				\vphantom{\frac{\left(s_{12}^2\right)}{s_{123}}}
				\peqbreak{-} \frac{\left(s_{12}^2+s_{12}s_{23}+s_{23}^2\right)
					\left(s_{45}^2+s_{45}s_{56}+s_{56}^2\right)}{s_{123}}
				\right\}\notag\\
			&+ \lagt{6}{10}\left\{
				\frac{t_1^3}{4} - (t_4^4 + t_7^3) 
				+ \frac{t_1t_2t_3}{2} + t_4t_5t_6 + t_7t_8t_9
				+ \frac{3t_1t_2^2}{2} + 6\omega t_1t_5(t_9-t_8)
				\right\}\notag\\
			&+ \lagt{6}{13}\left\{
				t_1^3 - (t_4^3 + t_7^3)
				- 3\omega t_1t_4t_7
				\right\}\notag\\
			&+ \lagt{6}{14}\left\{
				t_1t_2t_3 - (t_4t_5t_6 + t_7t_8t_9)
				+ 6\omega t_1t_5t_9
				\right\}\notag\\
			&+ [\Z_{\{3,3\}}],
	\label{eq:M6p6s33}
\end{align}
where $\omega=e^{2\pi i/3}$ is a third root of unity. The contribution from the singlet diagram turns out to be simpler to express in the standard basis $\basis_{\{6\}}$ than in the closed basis. 

Lastly, there are two diagrams for the $\{2,2,2\}$ flavour split:
\begin{equation}
	\tikzineq{
		\draw [thick, rounded corners=4pt] (-.5,.5) -- (0,0) -- (-.5,0);
		\draw [thick, rounded corners=4pt] ( .5,.5) -- (0,0) -- ( .5,0);
		\draw [thick, rounded corners=4pt] (-.5,-.5) -- (0,0) -- (.5,-.5);
		\draw (0,0) node [anchor=south] {\ordidx 6};
	}{M6p6-split222-0}\qquad
	\tikzineq{
		\draw [thick, rounded corners=4pt] (-.5,.5) -- (0,0) -- (-.5,0);
		\draw [thick, rounded corners=4pt] (-.5,-.5)  -- (0,0) -- (1,0) -- (1.5,-.5);
		\draw [thick, rounded corners=4pt] (1.5,.5)    -- (1,0) -- (1.5,0);
		\draw (0,0) node [anchor=south] {\ordidx 4};
		\draw (1,0) node [anchor=south] {\ordidx 4};
	}{M6p6-split222-1}
\end{equation}		
Using the closed basis \eqref{eq:222-basis}, the stripped amplitude is
\begin{align}
	-iF^8&\ampl_{6,\{2,2,2\}} 
		\eqbreak{=} L_1^2\left\{
			-\frac{t_1^3}{2} + t_1(2t_2^2 + 6t_2t_4 + t_4^2) 
			\peqbreak{-} \frac{1}{t_1}\left[
				\frac{t_4^4}{2} + (t_2+t_5)^2(t_4^2 + 4t_3t_4 t_3^2) 
				+ 2t_2t_3t_5t_6 + 2t_2t_3^2t_5 
				+ 8t_4^2t_2t_3 + 4t_4^3t_2\right]
			\right\}\notag\\
		&+ L_1L_2\left\{
			t_1t_8^2 + \frac{3t_1t_4^2}{2} 
			+ \frac{t_1t_2}{2}(4t_9 + 5t_4 - 8t_3 - 5t_2) - \frac{5t_1^3}{4}
			\peqbreak{-} \frac{1}{t_1}\left[
				\frac{t_2^2}{2}\left(
					t_2^2 + 2t_3^3 + 3t_4^2 - t_5^2 + 4t_9^2 + 2t_2t_4
					+ 4t_2t_9 + 8t_4t_9 + 8t_3t_4 + 4t_3t_6\right)
				\ppeqbreak{+} t_2t_3(4t_4^2 + 4t_4t_5 + t_5t_6)
				+ 2t_2t_9(t_4^2 - t_5^2) + 2t_2t_9^2(t_4 + t_5)
				\ppeqbreak{+} 2t_2t_4^3 + \frac{t_2t_4^2t_5}{2}
				- t_2t_4t_5^2 
				+ t_4^2t_8^2 + \frac{t_4^4}{4}\right]
			\right\}\notag\\
		&- L_2^2\left\{
			\frac{21t_1^3}{32} + \frac{3t_1t_8^2}{4} - \frac{11t_1t_4^2}{16} 
			+ \frac{t_1t_2}{4}(t_4 + 6t_3 + t_2 - 10t_9)
			\peqbreak{+} \frac{1}{4t_1}\left[
				\frac{t_2^2}{2}\left(
					t_2^2 + 2[t_3+t_4]^2 - t_5^2 + 4[t_8^2 + t_9^2] 
					+ 2t_2[t_4+2t_9] + 8t_4t_9 + 8t_3t_8\right)
				\ppeqbreak{+} t_2t_3\left(2t_4^2 + 2t_4t_5 + t_5t_6 - 4t_5t_8 + 8t_8t_9\right)
				\ppeqbreak{+} t_2t_4\left(t_4^2 + 2t_4t_9 - t_5^2 + 4t_9^2\right)
				- 2t_2t_5^2t_9 + 2t_2t_5\left(t_9^2-t_8^2\right)
				\ppeqbreak{+} t_8^2\left(t_4^2 + 2t_9^2 + 8t_2t_9\right)
				+ \frac{t_4^4}{8}\right]
			\right\}\notag\\
		&+ [\Z_{\{2,2,2\}}]
	\label{eq:M6p6s222}
\end{align}	
This completes the amplitude.

\subsection{The $\O(p^2)$ 10-point amplitude}
\label{app:M10p2}
Due to the absence of flavour splits, $\O(p^2)$ amplitudes are relatively easy to extend to many legs. The 10-point amplitude, which is also computed in \cite{Kampf:2013vha}, is given by the 16 diagrams\footnote{The circular shape is a result of the automatic diagram drawing in \fodge. The external legs are evenly distributed around a circle, and the location of each vertex is generated from the mean locations of all legs and vertices connected to it.}
\begin{center}
	\tikzsetfigurename{M10p2-}
	%% M10p2_tikz.tex is autogenerated by FODGE and should have accompanied this file.
	%% Generate using $ fodge 2 10 -t -r 0.8 -o path/to/this/directory/
	%\input{M10p2_tikz.tex}
	
	%Rather than inputting, I've copied the entire contents of M10p2_tikz.tex
	%and modified it somewhat.
	%This is for easier portability 
	%(but makes it harder to regenerate the diagrams)
	
%%% [0] O(p^2) 10-point diagram
\tikzineq[>=stealth]{
	\draw[thick] (-0.000, 0.000) -- (0.282, -0.855);
	\draw[thick] (-0.000, 0.000) -- (0.731, -0.525);
	\draw[thick] (-0.000, 0.000) -- (0.900, 0.005);
	\draw[thick] (-0.000, 0.000) -- (0.725, 0.533);
	\draw[thick] (-0.000, 0.000) -- (0.274, 0.857);
	\draw[thick] (-0.000, 0.000) -- (-0.282, 0.855);
	\draw[thick] (-0.000, 0.000) -- (-0.731, 0.525);
	\draw[thick] (-0.000, 0.000) -- (-0.900, -0.005);
	\draw[thick] (-0.000, 0.000) -- (-0.725, -0.533);
	\draw[thick] (-0.000, 0.000) -- (-0.274, -0.857);
}{M10p2-0}\fodgespace
%
%%% [1] O(p^2) 10-point diagram
\tikzineq[>=stealth]{
	\draw[thick] (-0.471, 0.000) -- (-0.728, 0.529);
	\draw[thick] (-0.471, 0.000) -- (-0.900, 0.000);
	\draw[thick] (-0.471, 0.000) -- (-0.728, -0.529);
	\draw[thick] (0.262, -0.000) -- (-0.471, 0.000);
	\draw[thick] (0.262, -0.000) -- (-0.278, -0.856);
	\draw[thick] (0.262, -0.000) -- (0.278, -0.856);
	\draw[thick] (0.262, -0.000) -- (0.728, -0.529);
	\draw[thick] (0.262, -0.000) -- (0.900, -0.000);
	\draw[thick] (0.262, -0.000) -- (0.728, 0.529);
	\draw[thick] (0.262, -0.000) -- (0.278, 0.856);
	\draw[thick] (0.262, -0.000) -- (-0.278, 0.856);
}{M10p2-1}\fodgespace
%
%%% [2] O(p^2) 10-point diagram
\tikzineq[>=stealth]{
	\draw[thick] (0.416, -0.000) -- (0.278, -0.856);
	\draw[thick] (0.416, -0.000) -- (0.728, -0.529);
	\draw[thick] (0.416, -0.000) -- (0.900, -0.000);
	\draw[thick] (0.416, -0.000) -- (0.728, 0.529);
	\draw[thick] (0.416, -0.000) -- (0.278, 0.856);
	\draw[thick] (-0.416, 0.000) -- (0.416, -0.000);
	\draw[thick] (-0.416, 0.000) -- (-0.278, 0.856);
	\draw[thick] (-0.416, 0.000) -- (-0.728, 0.529);
	\draw[thick] (-0.416, 0.000) -- (-0.900, 0.000);
	\draw[thick] (-0.416, 0.000) -- (-0.728, -0.529);
	\draw[thick] (-0.416, 0.000) -- (-0.278, -0.856);
}{M10p2-2}\fodgespace
%
%%% [3] O(p^2) 10-point diagram
\tikzineq[>=stealth]{
	\draw[thick] (0.434, -0.185) -- (0.463, -0.772);
	\draw[thick] (0.434, -0.185) -- (0.828, -0.353);
	\draw[thick] (0.434, -0.185) -- (0.877, 0.202);
	\draw[thick] (0.134, 0.315) -- (0.434, -0.185);
	\draw[thick] (0.134, 0.315) -- (0.591, 0.679);
	\draw[thick] (0.134, 0.315) -- (0.079, 0.896);
	\draw[thick] (-0.405, -0.093) -- (-0.463, 0.772);
	\draw[thick] (-0.405, -0.093) -- (-0.828, 0.353);
	\draw[thick] (-0.405, -0.093) -- (-0.877, -0.202);
	\draw[thick] (-0.405, -0.093) -- (-0.591, -0.679);
	\draw[thick] (-0.405, -0.093) -- (-0.079, -0.896);
	\draw[thick] (0.134, 0.315) -- (-0.405, -0.093);
}{M10p2-3}\fodgespace
%
%%% [4] O(p^2) 10-point diagram
\tikzineq[>=stealth]{
	\draw[thick] (0.277, -0.381) -- (0.000, -0.900);
	\draw[thick] (0.277, -0.381) -- (0.529, -0.728);
	\draw[thick] (0.277, -0.381) -- (0.856, -0.278);
	\draw[thick] (-0.396, -0.000) -- (0.277, -0.381);
	\draw[thick] (0.277, 0.381) -- (0.856, 0.278);
	\draw[thick] (0.277, 0.381) -- (0.529, 0.728);
	\draw[thick] (0.277, 0.381) -- (-0.000, 0.900);
	\draw[thick] (-0.396, -0.000) -- (0.277, 0.381);
	\draw[thick] (-0.396, -0.000) -- (-0.529, 0.728);
	\draw[thick] (-0.396, -0.000) -- (-0.856, 0.278);
	\draw[thick] (-0.396, -0.000) -- (-0.856, -0.278);
	\draw[thick] (-0.396, -0.000) -- (-0.529, -0.728);
}{M10p2-4}\fodgespace
%
%%% [5] O(p^2) 10-point diagram
\tikzineq[>=stealth]{
	\draw[thick] (-0.448, 0.146) -- (-0.529, 0.728);
	\draw[thick] (-0.448, 0.146) -- (-0.856, 0.278);
	\draw[thick] (-0.448, 0.146) -- (-0.856, -0.278);
	\draw[thick] (-0.000, -0.208) -- (-0.448, 0.146);
	\draw[thick] (-0.000, -0.208) -- (-0.529, -0.728);
	\draw[thick] (-0.000, -0.208) -- (-0.000, -0.900);
	\draw[thick] (-0.000, -0.208) -- (0.529, -0.728);
	\draw[thick] (0.448, 0.146) -- (0.856, -0.278);
	\draw[thick] (0.448, 0.146) -- (0.856, 0.278);
	\draw[thick] (0.448, 0.146) -- (0.529, 0.728);
	\draw[thick] (-0.000, -0.208) -- (0.448, 0.146);
	\draw[thick] (-0.000, -0.208) -- (0.000, 0.900);
}{M10p2-5}\fodgespace
%
%%% [6] O(p^2) 10-point diagram
\tikzineq[>=stealth]{
	\draw[thick] (0.471, -0.000) -- (0.728, -0.529);
	\draw[thick] (0.471, -0.000) -- (0.900, -0.000);
	\draw[thick] (0.471, -0.000) -- (0.728, 0.529);
	\draw[thick] (-0.000, -0.000) -- (0.471, -0.000);
	\draw[thick] (-0.000, -0.000) -- (0.278, 0.856);
	\draw[thick] (-0.000, -0.000) -- (-0.278, 0.856);
	\draw[thick] (-0.471, 0.000) -- (-0.728, 0.529);
	\draw[thick] (-0.471, 0.000) -- (-0.900, 0.000);
	\draw[thick] (-0.471, 0.000) -- (-0.728, -0.529);
	\draw[thick] (-0.000, -0.000) -- (-0.471, 0.000);
	\draw[thick] (-0.000, -0.000) -- (-0.278, -0.856);
	\draw[thick] (-0.000, -0.000) -- (0.278, -0.856);
}{M10p2-6}\fodgespace
%
%%% [7] O(p^2) 10-point diagram
\tikzineq[>=stealth]{
	\draw[thick] (0.134, -0.315) -- (0.079, -0.896);
	\draw[thick] (0.134, -0.315) -- (0.591, -0.679);
	\draw[thick] (0.434, 0.185) -- (0.877, -0.202);
	\draw[thick] (0.434, 0.185) -- (0.828, 0.353);
	\draw[thick] (0.434, 0.185) -- (0.463, 0.772);
	\draw[thick] (0.134, -0.315) -- (0.434, 0.185);
	\draw[thick] (-0.405, 0.093) -- (0.134, -0.315);
	\draw[thick] (-0.405, 0.093) -- (-0.079, 0.896);
	\draw[thick] (-0.405, 0.093) -- (-0.591, 0.679);
	\draw[thick] (-0.405, 0.093) -- (-0.877, 0.202);
	\draw[thick] (-0.405, 0.093) -- (-0.828, -0.353);
	\draw[thick] (-0.405, 0.093) -- (-0.463, -0.772);
}{M10p2-7}\fodgespace
%
%%% [8] O(p^2) 10-point diagram
\tikzineq[>=stealth]{
	\draw[thick] (0.111, -0.000) -- (0.278, -0.856);
	\draw[thick] (0.471, -0.000) -- (0.728, -0.529);
	\draw[thick] (0.471, -0.000) -- (0.900, -0.000);
	\draw[thick] (0.471, -0.000) -- (0.728, 0.529);
	\draw[thick] (0.111, -0.000) -- (0.471, -0.000);
	\draw[thick] (0.111, -0.000) -- (0.278, 0.856);
	\draw[thick] (-0.416, 0.000) -- (0.111, -0.000);
	\draw[thick] (-0.416, 0.000) -- (-0.278, 0.856);
	\draw[thick] (-0.416, 0.000) -- (-0.728, 0.529);
	\draw[thick] (-0.416, 0.000) -- (-0.900, 0.000);
	\draw[thick] (-0.416, 0.000) -- (-0.728, -0.529);
	\draw[thick] (-0.416, 0.000) -- (-0.278, -0.856);
}{M10p2-8}\fodgespace
%
%%% [9] O(p^2) 10-point diagram
\tikzineq[>=stealth]{
	\draw[thick] (0.471, -0.000) -- (0.728, -0.529);
	\draw[thick] (0.471, -0.000) -- (0.900, -0.000);
	\draw[thick] (0.471, -0.000) -- (0.728, 0.529);
	\draw[thick] (0.000, 0.342) -- (0.471, -0.000);
	\draw[thick] (0.000, 0.342) -- (0.278, 0.856);
	\draw[thick] (0.000, 0.342) -- (-0.278, 0.856);
	\draw[thick] (-0.000, -0.342) -- (0.000, 0.342);
	\draw[thick] (-0.471, 0.000) -- (-0.728, 0.529);
	\draw[thick] (-0.471, 0.000) -- (-0.900, 0.000);
	\draw[thick] (-0.471, 0.000) -- (-0.728, -0.529);
	\draw[thick] (-0.000, -0.342) -- (-0.471, 0.000);
	\draw[thick] (-0.000, -0.342) -- (-0.278, -0.856);
	\draw[thick] (-0.000, -0.342) -- (0.278, -0.856);
}{M10p2-9}\fodgespace
%
%%% [10] O(p^2) 10-point diagram
\tikzineq[>=stealth]{
	\draw[thick] (0.424, -0.205) -- (0.426, -0.793);
	\draw[thick] (0.424, -0.205) -- (0.811, -0.391);
	\draw[thick] (0.424, -0.205) -- (0.886, 0.160);
	\draw[thick] (0.149, 0.308) -- (0.424, -0.205);
	\draw[thick] (0.149, 0.308) -- (0.622, 0.650);
	\draw[thick] (0.149, 0.308) -- (0.121, 0.892);
	\draw[thick] (-0.109, -0.020) -- (0.149, 0.308);
	\draw[thick] (-0.109, -0.020) -- (-0.426, 0.793);
	\draw[thick] (-0.464, -0.084) -- (-0.811, 0.391);
	\draw[thick] (-0.464, -0.084) -- (-0.886, -0.160);
	\draw[thick] (-0.464, -0.084) -- (-0.622, -0.650);
	\draw[thick] (-0.109, -0.020) -- (-0.464, -0.084);
	\draw[thick] (-0.109, -0.020) -- (-0.121, -0.892);
}{M10p2-10}\fodgespace
%
%%% [11] O(p^2) 10-point diagram
\tikzineq[>=stealth]{
	\draw[thick] (0.381, -0.277) -- (0.278, -0.856);
	\draw[thick] (0.381, -0.277) -- (0.728, -0.529);
	\draw[thick] (0.381, -0.277) -- (0.900, -0.000);
	\draw[thick] (0.201, 0.277) -- (0.381, -0.277);
	\draw[thick] (0.201, 0.277) -- (0.728, 0.529);
	\draw[thick] (0.201, 0.277) -- (0.278, 0.856);
	\draw[thick] (-0.201, 0.277) -- (0.201, 0.277);
	\draw[thick] (-0.201, 0.277) -- (-0.278, 0.856);
	\draw[thick] (-0.201, 0.277) -- (-0.728, 0.529);
	\draw[thick] (-0.381, -0.277) -- (-0.900, 0.000);
	\draw[thick] (-0.381, -0.277) -- (-0.728, -0.529);
	\draw[thick] (-0.381, -0.277) -- (-0.278, -0.856);
	\draw[thick] (-0.201, 0.277) -- (-0.381, -0.277);
}{M10p2-11}\fodgespace
%
%%% [12] O(p^2) 10-point diagram
\tikzineq[>=stealth]{
	\draw[thick] (-0.000, 0.471) -- (0.529, 0.728);
	\draw[thick] (-0.000, 0.471) -- (-0.000, 0.900);
	\draw[thick] (-0.000, 0.471) -- (-0.529, 0.728);
	\draw[thick] (0.000, -0.180) -- (-0.000, 0.471);
	\draw[thick] (-0.448, -0.146) -- (-0.856, 0.278);
	\draw[thick] (-0.448, -0.146) -- (-0.856, -0.278);
	\draw[thick] (-0.448, -0.146) -- (-0.529, -0.728);
	\draw[thick] (0.000, -0.180) -- (-0.448, -0.146);
	\draw[thick] (0.000, -0.180) -- (0.000, -0.900);
	\draw[thick] (0.448, -0.146) -- (0.000, -0.180);
	\draw[thick] (0.448, -0.146) -- (0.529, -0.728);
	\draw[thick] (0.448, -0.146) -- (0.856, -0.278);
	\draw[thick] (0.448, -0.146) -- (0.856, 0.278);
}{M10p2-12}\fodgespace
%
%%% [13] O(p^2) 10-point diagram
\tikzineq[>=stealth]{
	\draw[thick] (0.464, -0.084) -- (0.622, -0.650);
	\draw[thick] (0.464, -0.084) -- (0.886, -0.160);
	\draw[thick] (0.464, -0.084) -- (0.811, 0.391);
	\draw[thick] (0.109, -0.020) -- (0.464, -0.084);
	\draw[thick] (0.109, -0.020) -- (0.426, 0.793);
	\draw[thick] (-0.149, 0.308) -- (-0.121, 0.892);
	\draw[thick] (-0.149, 0.308) -- (-0.622, 0.650);
	\draw[thick] (-0.424, -0.205) -- (-0.886, 0.160);
	\draw[thick] (-0.424, -0.205) -- (-0.811, -0.391);
	\draw[thick] (-0.424, -0.205) -- (-0.426, -0.793);
	\draw[thick] (-0.149, 0.308) -- (-0.424, -0.205);
	\draw[thick] (0.109, -0.020) -- (-0.149, 0.308);
	\draw[thick] (0.109, -0.020) -- (0.121, -0.892);
}{M10p2-13}\fodgespace
%
%%% [14] O(p^2) 10-point diagram
\tikzineq[>=stealth]{
	\draw[thick] (-0.000, 0.342) -- (0.278, 0.856);
	\draw[thick] (-0.000, 0.342) -- (-0.278, 0.856);
	\draw[thick] (-0.471, -0.000) -- (-0.728, 0.529);
	\draw[thick] (-0.471, -0.000) -- (-0.900, -0.000);
	\draw[thick] (-0.471, -0.000) -- (-0.728, -0.529);
	\draw[thick] (-0.000, 0.342) -- (-0.471, -0.000);
	\draw[thick] (0.000, -0.342) -- (-0.000, 0.342);
	\draw[thick] (0.000, -0.342) -- (-0.278, -0.856);
	\draw[thick] (0.000, -0.342) -- (0.278, -0.856);
	\draw[thick] (0.471, 0.000) -- (0.000, -0.342);
	\draw[thick] (0.471, 0.000) -- (0.728, -0.529);
	\draw[thick] (0.471, 0.000) -- (0.900, 0.000);
	\draw[thick] (0.471, 0.000) -- (0.728, 0.529);
}{M10p2-14}\fodgespace
%
%%% [15] O(p^2) 10-point diagram
\tikzineq[>=stealth]{
	\draw[thick] (0.471, -0.000) -- (0.728, -0.529);
	\draw[thick] (0.471, -0.000) -- (0.900, -0.000);
	\draw[thick] (0.471, -0.000) -- (0.728, 0.529);
	\draw[thick] (0.111, -0.000) -- (0.471, -0.000);
	\draw[thick] (0.111, -0.000) -- (0.278, 0.856);
	\draw[thick] (-0.111, 0.000) -- (-0.278, 0.856);
	\draw[thick] (-0.471, 0.000) -- (-0.728, 0.529);
	\draw[thick] (-0.471, 0.000) -- (-0.900, 0.000);
	\draw[thick] (-0.471, 0.000) -- (-0.728, -0.529);
	\draw[thick] (-0.111, 0.000) -- (-0.471, 0.000);
	\draw[thick] (-0.111, 0.000) -- (-0.278, -0.856);
	\draw[thick] (0.111, -0.000) -- (-0.111, 0.000);
	\draw[thick] (0.111, -0.000) -- (0.278, -0.856);
}{M10p2-15}\fodgespace
\end{center}
and has the stripped amplitude
{\addtolength{\peqindent}{-2cm}\addtolength{\ppeqindent}{-2cm}
\begin{align}
	-16&iF^8\ampl_{2,\{10\}} = 
		5s_{12} + 2s_{1234} 
		\eqbreak{-} \frac{(s_{12}+s_{23}+s_{34}+s_{45}+s_\ol{14}+s_\ol{25})
			(s_{67}+s_{78}+s_{89}+s_{9\A}+s_\ol{69}s_\ol{7\A})}{2s_\ol{15}}
		\eqbreak{-} \frac{s_{12}+s_{23}}{s_{123}}\left\{
			2\left(s_{45}+s_{56}+s_{67}+s_{78}+s_{89}+s_{9\A}\right)
			+ 2\left(s_\ol{\A3} + s_\ol{14}\right) 
			\mvphantom
			\peqbreak{-} 
			\frac{(s_{67}+s_{78})(s_{45}+s_{9\A}+s_\ol{\A3}+s_\ol{14}+s_\ol{58}+s_\ol{69})}{2s_{678}}
			\peqbreak{-} 
			\frac{(s_{78}+s_{89})(s_{45}+s_{56} + s_\ol{14}+s_\ol{47}+s_\ol{7\A} + s_\ol{69})}{s_{789}}
			\peqbreak{-} 
			\frac{(s_{89}+s_{9\A})(s_{45}+s_{56}+s_{67} + s_\ol{14}+s_\ol{47}+s_\ol{7\A})}{s_{89\A}}
			\peqbreak{+} \frac{s_{9\A} + s_\ol{\A3}}{s_\ol{48}}\left[
				\frac{(s_{67}+s_{78})(s_{45}+s_\ol{58})}{2s_{678}}
				- \left(s_{45}+s_{56}+s_{67}+s_{78} + s_\ol{47}+s_\ol{58}\right)
			\right]
			\peqbreak{+} \frac{s_\ol{\A3}+s_\ol{14}}{s_\ol{59}}\left[
				\frac{(s_{67}+s_{78})(s_\ol{58}+s_\ol{69})}{2s_{678}} 
				- \left(s_{56}+s_{67}+s_{78}+s_{89} + s_\ol{58}+s_\ol{69}\right)
			\right]
			\peqbreak{+}\frac{s_\ol{14} + s_{45}}{s_\ol{15}}\left[
				- \left(s_{67}+s_{78}+s_{89}+s_{9\A} + s_\ol{69}+s_\ol{7\A}\right)
				\mvphantom
				\ppeqbreak{+} \frac{(s_{67}+s_{78})(s_\ol{69}+s_{9\A})}{2s_{678}}
				+ \frac{(s_{78}+s_{89})(s_\ol{69}+s_\ol{7\A})}{s_{789}}
				+ \frac{(s_{89}+s_{9\A})(s_{67}+s_\ol{7\A})}{s_{89\A}}
			\right]
			\peqbreak{+} 
			s_\ol{47}+s_\ol{58}+s_\ol{69}+s_\ol{7\A} 
			+ \frac{(s_\ol{7\A}+s_\ol{\A3})(s_{45}+s_{56})(s_{78}+s_{89})}{s_{456}s_{789}}
		\right\} + [\Z_{10}].
	\label{eq:M10p2}
\end{align}}
To avoid problems with multi-digit indices, we switch to hexadecimal and write $\A$ instead of $10$. To abbreviate long index lists, we write $\ol{ij}$ for $i(i+1)\cdots(j-1)j$. Indices wrap around cyclically; $\ol{\A3}$ means $\A123$.

\subsection{The $\O(p^2)$ 12-point amplitude}
\label{app:M12p2}
This is a novel amplitude, and takes the most time to compute of all amplitudes presented in this work. It consists of 73 diagrams:
\begin{center}
	%% M12p2_tikz.tex is autogenerated by FODGE and should have accompanied this file.
	%% Generate using $ fodge 2 12 -t -r 0.8 -o path/to/this/directory/
	\tikzsetfigurename{M12p2-}
	%\input{M12p2_tikz.tex}
	
	%Rather than inputting, I've copied the entire contents of M12p2_tikz.tex
	%and modified it somewhat.
	%This is for easier portability 
	%(but makes it harder to regenerate the diagrams)
	
%%% [0] O(p^2) 12-point diagram
\tikzineq[>=stealth]{
	\draw[thick] (-0.000, 0.000) -- (0.222, -0.975);
	\draw[thick] (-0.000, 0.000) -- (0.679, -0.734);
	\draw[thick] (-0.000, 0.000) -- (0.955, -0.296);
	\draw[thick] (-0.000, 0.000) -- (0.975, 0.222);
	\draw[thick] (-0.000, 0.000) -- (0.734, 0.679);
	\draw[thick] (-0.000, 0.000) -- (0.296, 0.955);
	\draw[thick] (-0.000, 0.000) -- (-0.222, 0.975);
	\draw[thick] (-0.000, 0.000) -- (-0.679, 0.734);
	\draw[thick] (-0.000, 0.000) -- (-0.955, 0.296);
	\draw[thick] (-0.000, 0.000) -- (-0.975, -0.222);
	\draw[thick] (-0.000, 0.000) -- (-0.734, -0.679);
	\draw[thick] (-0.000, 0.000) -- (-0.296, -0.955);
}{M12p2-0}\fodgespace
%
%%% [1] O(p^2) 12-point diagram
\tikzineq[>=stealth]{
	\draw[thick] (-0.546, 0.000) -- (-0.866, 0.500);
	\draw[thick] (-0.546, 0.000) -- (-1.000, 0.000);
	\draw[thick] (-0.546, 0.000) -- (-0.866, -0.500);
	\draw[thick] (0.248, -0.000) -- (-0.546, 0.000);
	\draw[thick] (0.248, -0.000) -- (-0.500, -0.866);
	\draw[thick] (0.248, -0.000) -- (-0.000, -1.000);
	\draw[thick] (0.248, -0.000) -- (0.500, -0.866);
	\draw[thick] (0.248, -0.000) -- (0.866, -0.500);
	\draw[thick] (0.248, -0.000) -- (1.000, -0.000);
	\draw[thick] (0.248, -0.000) -- (0.866, 0.500);
	\draw[thick] (0.248, -0.000) -- (0.500, 0.866);
	\draw[thick] (0.248, -0.000) -- (0.000, 1.000);
	\draw[thick] (0.248, -0.000) -- (-0.500, 0.866);
}{M12p2-1}\fodgespace
%
%%% [2] O(p^2) 12-point diagram
\tikzineq[>=stealth]{
	\draw[thick] (-0.415, -0.000) -- (-0.000, 1.000);
	\draw[thick] (-0.415, -0.000) -- (-0.500, 0.866);
	\draw[thick] (-0.415, -0.000) -- (-0.866, 0.500);
	\draw[thick] (-0.415, -0.000) -- (-1.000, -0.000);
	\draw[thick] (-0.415, -0.000) -- (-0.866, -0.500);
	\draw[thick] (-0.415, -0.000) -- (-0.500, -0.866);
	\draw[thick] (-0.415, -0.000) -- (0.000, -1.000);
	\draw[thick] (0.533, 0.000) -- (-0.415, -0.000);
	\draw[thick] (0.533, 0.000) -- (0.500, -0.866);
	\draw[thick] (0.533, 0.000) -- (0.866, -0.500);
	\draw[thick] (0.533, 0.000) -- (1.000, 0.000);
	\draw[thick] (0.533, 0.000) -- (0.866, 0.500);
	\draw[thick] (0.533, 0.000) -- (0.500, 0.866);
}{M12p2-2}\fodgespace
%
%%% [3] O(p^2) 12-point diagram
\tikzineq[>=stealth]{
	\draw[thick] (0.511, -0.192) -- (0.635, -0.773);
	\draw[thick] (0.511, -0.192) -- (0.936, -0.352);
	\draw[thick] (0.511, -0.192) -- (0.987, 0.163);
	\draw[thick] (0.225, 0.314) -- (0.511, -0.192);
	\draw[thick] (0.225, 0.314) -- (0.773, 0.635);
	\draw[thick] (0.225, 0.314) -- (0.352, 0.936);
	\draw[thick] (-0.409, -0.068) -- (-0.163, 0.987);
	\draw[thick] (-0.409, -0.068) -- (-0.635, 0.773);
	\draw[thick] (-0.409, -0.068) -- (-0.936, 0.352);
	\draw[thick] (-0.409, -0.068) -- (-0.987, -0.163);
	\draw[thick] (-0.409, -0.068) -- (-0.773, -0.635);
	\draw[thick] (-0.409, -0.068) -- (-0.352, -0.936);
	\draw[thick] (-0.409, -0.068) -- (0.163, -0.987);
	\draw[thick] (0.225, 0.314) -- (-0.409, -0.068);
}{M12p2-3}\fodgespace
%
%%% [4] O(p^2) 12-point diagram
\tikzineq[>=stealth]{
	\draw[thick] (0.386, -0.386) -- (0.259, -0.966);
	\draw[thick] (0.386, -0.386) -- (0.707, -0.707);
	\draw[thick] (0.386, -0.386) -- (0.966, -0.259);
	\draw[thick] (-0.429, 0.000) -- (0.386, -0.386);
	\draw[thick] (0.386, 0.386) -- (0.966, 0.259);
	\draw[thick] (0.386, 0.386) -- (0.707, 0.707);
	\draw[thick] (0.386, 0.386) -- (0.259, 0.966);
	\draw[thick] (-0.429, 0.000) -- (0.386, 0.386);
	\draw[thick] (-0.429, 0.000) -- (-0.259, 0.966);
	\draw[thick] (-0.429, 0.000) -- (-0.707, 0.707);
	\draw[thick] (-0.429, 0.000) -- (-0.966, 0.259);
	\draw[thick] (-0.429, 0.000) -- (-0.966, -0.259);
	\draw[thick] (-0.429, 0.000) -- (-0.707, -0.707);
	\draw[thick] (-0.429, 0.000) -- (-0.259, -0.966);
}{M12p2-4}\fodgespace
%
%%% [5] O(p^2) 12-point diagram
\tikzineq[>=stealth]{
	\draw[thick] (0.473, 0.273) -- (1.000, -0.000);
	\draw[thick] (0.473, 0.273) -- (0.866, 0.500);
	\draw[thick] (0.473, 0.273) -- (0.500, 0.866);
	\draw[thick] (-0.000, -0.304) -- (0.473, 0.273);
	\draw[thick] (-0.000, -0.304) -- (0.000, 1.000);
	\draw[thick] (-0.473, 0.273) -- (-0.500, 0.866);
	\draw[thick] (-0.473, 0.273) -- (-0.866, 0.500);
	\draw[thick] (-0.473, 0.273) -- (-1.000, 0.000);
	\draw[thick] (-0.000, -0.304) -- (-0.473, 0.273);
	\draw[thick] (-0.000, -0.304) -- (-0.866, -0.500);
	\draw[thick] (-0.000, -0.304) -- (-0.500, -0.866);
	\draw[thick] (-0.000, -0.304) -- (-0.000, -1.000);
	\draw[thick] (-0.000, -0.304) -- (0.500, -0.866);
	\draw[thick] (-0.000, -0.304) -- (0.866, -0.500);
}{M12p2-5}\fodgespace
%
%%% [6] O(p^2) 12-point diagram
\tikzineq[>=stealth]{
	\draw[thick] (-0.528, 0.141) -- (-0.707, 0.707);
	\draw[thick] (-0.528, 0.141) -- (-0.966, 0.259);
	\draw[thick] (-0.528, 0.141) -- (-0.966, -0.259);
	\draw[thick] (-0.000, -0.157) -- (-0.528, 0.141);
	\draw[thick] (-0.000, -0.157) -- (-0.707, -0.707);
	\draw[thick] (-0.000, -0.157) -- (-0.259, -0.966);
	\draw[thick] (-0.000, -0.157) -- (0.259, -0.966);
	\draw[thick] (-0.000, -0.157) -- (0.707, -0.707);
	\draw[thick] (0.528, 0.141) -- (0.966, -0.259);
	\draw[thick] (0.528, 0.141) -- (0.966, 0.259);
	\draw[thick] (0.528, 0.141) -- (0.707, 0.707);
	\draw[thick] (-0.000, -0.157) -- (0.528, 0.141);
	\draw[thick] (-0.000, -0.157) -- (0.259, 0.966);
	\draw[thick] (-0.000, -0.157) -- (-0.259, 0.966);
}{M12p2-6}\fodgespace
%
%%% [7] O(p^2) 12-point diagram
\tikzineq[>=stealth]{
	\draw[thick] (-0.323, -0.440) -- (-0.916, -0.402);
	\draw[thick] (-0.323, -0.440) -- (-0.592, -0.806);
	\draw[thick] (-0.323, -0.440) -- (-0.109, -0.994);
	\draw[thick] (-0.299, 0.373) -- (-0.323, -0.440);
	\draw[thick] (0.530, -0.058) -- (0.402, -0.916);
	\draw[thick] (0.530, -0.058) -- (0.806, -0.592);
	\draw[thick] (0.530, -0.058) -- (0.994, -0.109);
	\draw[thick] (0.530, -0.058) -- (0.916, 0.402);
	\draw[thick] (0.530, -0.058) -- (0.592, 0.806);
	\draw[thick] (-0.299, 0.373) -- (0.530, -0.058);
	\draw[thick] (-0.299, 0.373) -- (0.109, 0.994);
	\draw[thick] (-0.299, 0.373) -- (-0.402, 0.916);
	\draw[thick] (-0.299, 0.373) -- (-0.806, 0.592);
	\draw[thick] (-0.299, 0.373) -- (-0.994, 0.109);
}{M12p2-7}\fodgespace
%
%%% [8] O(p^2) 12-point diagram
\tikzineq[>=stealth]{
	\draw[thick] (0.546, -0.000) -- (0.866, -0.500);
	\draw[thick] (0.546, -0.000) -- (1.000, -0.000);
	\draw[thick] (0.546, -0.000) -- (0.866, 0.500);
	\draw[thick] (-0.000, 0.000) -- (0.546, -0.000);
	\draw[thick] (-0.000, 0.000) -- (0.500, 0.866);
	\draw[thick] (-0.000, 0.000) -- (0.000, 1.000);
	\draw[thick] (-0.000, 0.000) -- (-0.500, 0.866);
	\draw[thick] (-0.546, 0.000) -- (-0.866, 0.500);
	\draw[thick] (-0.546, 0.000) -- (-1.000, 0.000);
	\draw[thick] (-0.546, 0.000) -- (-0.866, -0.500);
	\draw[thick] (-0.000, 0.000) -- (-0.546, 0.000);
	\draw[thick] (-0.000, 0.000) -- (-0.500, -0.866);
	\draw[thick] (-0.000, 0.000) -- (-0.000, -1.000);
	\draw[thick] (-0.000, 0.000) -- (0.500, -0.866);
}{M12p2-8}\fodgespace
%
%%% [9] O(p^2) 12-point diagram
\tikzineq[>=stealth]{
	\draw[thick] (-0.523, 0.105) -- (-0.319, 0.948);
	\draw[thick] (-0.523, 0.105) -- (-0.750, 0.661);
	\draw[thick] (-0.523, 0.105) -- (-0.980, 0.198);
	\draw[thick] (-0.523, 0.105) -- (-0.948, -0.319);
	\draw[thick] (-0.523, 0.105) -- (-0.661, -0.750);
	\draw[thick] (0.153, -0.230) -- (-0.523, 0.105);
	\draw[thick] (0.153, -0.230) -- (-0.198, -0.980);
	\draw[thick] (0.153, -0.230) -- (0.319, -0.948);
	\draw[thick] (0.153, -0.230) -- (0.750, -0.661);
	\draw[thick] (0.518, 0.174) -- (0.980, -0.198);
	\draw[thick] (0.518, 0.174) -- (0.948, 0.319);
	\draw[thick] (0.518, 0.174) -- (0.661, 0.750);
	\draw[thick] (0.153, -0.230) -- (0.518, 0.174);
	\draw[thick] (0.153, -0.230) -- (0.198, 0.980);
}{M12p2-9}\fodgespace
%
%%% [10] O(p^2) 12-point diagram
\tikzineq[>=stealth]{
	\draw[thick] (-0.533, 0.000) -- (-0.500, 0.866);
	\draw[thick] (-0.533, 0.000) -- (-0.866, 0.500);
	\draw[thick] (-0.533, 0.000) -- (-1.000, 0.000);
	\draw[thick] (-0.533, 0.000) -- (-0.866, -0.500);
	\draw[thick] (-0.533, 0.000) -- (-0.500, -0.866);
	\draw[thick] (0.143, -0.000) -- (-0.533, 0.000);
	\draw[thick] (0.143, -0.000) -- (-0.000, -1.000);
	\draw[thick] (0.143, -0.000) -- (0.500, -0.866);
	\draw[thick] (0.546, -0.000) -- (0.866, -0.500);
	\draw[thick] (0.546, -0.000) -- (1.000, -0.000);
	\draw[thick] (0.546, -0.000) -- (0.866, 0.500);
	\draw[thick] (0.143, -0.000) -- (0.546, -0.000);
	\draw[thick] (0.143, -0.000) -- (0.500, 0.866);
	\draw[thick] (0.143, -0.000) -- (0.000, 1.000);
}{M12p2-10}\fodgespace
%
%%% [11] O(p^2) 12-point diagram
\tikzineq[>=stealth]{
	\draw[thick] (-0.518, 0.174) -- (-0.661, 0.750);
	\draw[thick] (-0.518, 0.174) -- (-0.948, 0.319);
	\draw[thick] (-0.518, 0.174) -- (-0.980, -0.198);
	\draw[thick] (-0.153, -0.230) -- (-0.518, 0.174);
	\draw[thick] (-0.153, -0.230) -- (-0.750, -0.661);
	\draw[thick] (-0.153, -0.230) -- (-0.319, -0.948);
	\draw[thick] (-0.153, -0.230) -- (0.198, -0.980);
	\draw[thick] (0.523, 0.105) -- (0.661, -0.750);
	\draw[thick] (0.523, 0.105) -- (0.948, -0.319);
	\draw[thick] (0.523, 0.105) -- (0.980, 0.198);
	\draw[thick] (0.523, 0.105) -- (0.750, 0.661);
	\draw[thick] (0.523, 0.105) -- (0.319, 0.948);
	\draw[thick] (-0.153, -0.230) -- (0.523, 0.105);
	\draw[thick] (-0.153, -0.230) -- (-0.198, 0.980);
}{M12p2-11}\fodgespace
%
%%% [12] O(p^2) 12-point diagram
\tikzineq[>=stealth]{
	\draw[thick] (-0.546, 0.000) -- (-0.866, 0.500);
	\draw[thick] (-0.546, 0.000) -- (-1.000, 0.000);
	\draw[thick] (-0.546, 0.000) -- (-0.866, -0.500);
	\draw[thick] (-0.200, 0.000) -- (-0.546, 0.000);
	\draw[thick] (-0.200, 0.000) -- (-0.500, -0.866);
	\draw[thick] (0.415, -0.000) -- (-0.000, -1.000);
	\draw[thick] (0.415, -0.000) -- (0.500, -0.866);
	\draw[thick] (0.415, -0.000) -- (0.866, -0.500);
	\draw[thick] (0.415, -0.000) -- (1.000, -0.000);
	\draw[thick] (0.415, -0.000) -- (0.866, 0.500);
	\draw[thick] (0.415, -0.000) -- (0.500, 0.866);
	\draw[thick] (0.415, -0.000) -- (0.000, 1.000);
	\draw[thick] (-0.200, 0.000) -- (0.415, -0.000);
	\draw[thick] (-0.200, 0.000) -- (-0.500, 0.866);
}{M12p2-12}\fodgespace
%
%%% [13] O(p^2) 12-point diagram
\tikzineq[>=stealth]{
	\draw[thick] (-0.409, 0.068) -- (0.163, 0.987);
	\draw[thick] (-0.409, 0.068) -- (-0.352, 0.936);
	\draw[thick] (-0.409, 0.068) -- (-0.773, 0.635);
	\draw[thick] (-0.409, 0.068) -- (-0.987, 0.163);
	\draw[thick] (-0.409, 0.068) -- (-0.936, -0.352);
	\draw[thick] (-0.409, 0.068) -- (-0.635, -0.773);
	\draw[thick] (-0.409, 0.068) -- (-0.163, -0.987);
	\draw[thick] (0.225, -0.314) -- (-0.409, 0.068);
	\draw[thick] (0.225, -0.314) -- (0.352, -0.936);
	\draw[thick] (0.225, -0.314) -- (0.773, -0.635);
	\draw[thick] (0.511, 0.192) -- (0.225, -0.314);
	\draw[thick] (0.511, 0.192) -- (0.987, -0.163);
	\draw[thick] (0.511, 0.192) -- (0.936, 0.352);
	\draw[thick] (0.511, 0.192) -- (0.635, 0.773);
}{M12p2-13}\fodgespace
%
%%% [14] O(p^2) 12-point diagram
\tikzineq[>=stealth]{
	\draw[thick] (0.323, -0.440) -- (0.109, -0.994);
	\draw[thick] (0.323, -0.440) -- (0.592, -0.806);
	\draw[thick] (0.323, -0.440) -- (0.916, -0.402);
	\draw[thick] (0.299, 0.373) -- (0.323, -0.440);
	\draw[thick] (0.299, 0.373) -- (0.994, 0.109);
	\draw[thick] (0.299, 0.373) -- (0.806, 0.592);
	\draw[thick] (0.299, 0.373) -- (0.402, 0.916);
	\draw[thick] (0.299, 0.373) -- (-0.109, 0.994);
	\draw[thick] (-0.530, -0.058) -- (0.299, 0.373);
	\draw[thick] (-0.530, -0.058) -- (-0.592, 0.806);
	\draw[thick] (-0.530, -0.058) -- (-0.916, 0.402);
	\draw[thick] (-0.530, -0.058) -- (-0.994, -0.109);
	\draw[thick] (-0.530, -0.058) -- (-0.806, -0.592);
	\draw[thick] (-0.530, -0.058) -- (-0.402, -0.916);
}{M12p2-14}\fodgespace
%
%%% [15] O(p^2) 12-point diagram
\tikzineq[>=stealth]{
	\draw[thick] (0.515, -0.138) -- (0.259, -0.966);
	\draw[thick] (0.515, -0.138) -- (0.707, -0.707);
	\draw[thick] (0.515, -0.138) -- (0.966, -0.259);
	\draw[thick] (0.515, -0.138) -- (0.966, 0.259);
	\draw[thick] (0.515, -0.138) -- (0.707, 0.707);
	\draw[thick] (0.000, 0.386) -- (0.515, -0.138);
	\draw[thick] (0.000, 0.386) -- (0.259, 0.966);
	\draw[thick] (0.000, 0.386) -- (-0.259, 0.966);
	\draw[thick] (-0.515, -0.138) -- (0.000, 0.386);
	\draw[thick] (-0.515, -0.138) -- (-0.707, 0.707);
	\draw[thick] (-0.515, -0.138) -- (-0.966, 0.259);
	\draw[thick] (-0.515, -0.138) -- (-0.966, -0.259);
	\draw[thick] (-0.515, -0.138) -- (-0.707, -0.707);
	\draw[thick] (-0.515, -0.138) -- (-0.259, -0.966);
}{M12p2-15}\fodgespace
%
%%% [16] O(p^2) 12-point diagram
\tikzineq[>=stealth]{
	\draw[thick] (0.000, 0.000) -- (0.000, -1.000);
	\draw[thick] (0.533, 0.000) -- (0.500, -0.866);
	\draw[thick] (0.533, 0.000) -- (0.866, -0.500);
	\draw[thick] (0.533, 0.000) -- (1.000, 0.000);
	\draw[thick] (0.533, 0.000) -- (0.866, 0.500);
	\draw[thick] (0.533, 0.000) -- (0.500, 0.866);
	\draw[thick] (0.000, 0.000) -- (0.533, 0.000);
	\draw[thick] (0.000, 0.000) -- (0.000, 1.000);
	\draw[thick] (-0.533, 0.000) -- (0.000, 0.000);
	\draw[thick] (-0.533, 0.000) -- (-0.500, 0.866);
	\draw[thick] (-0.533, 0.000) -- (-0.866, 0.500);
	\draw[thick] (-0.533, 0.000) -- (-1.000, 0.000);
	\draw[thick] (-0.533, 0.000) -- (-0.866, -0.500);
	\draw[thick] (-0.533, 0.000) -- (-0.500, -0.866);
}{M12p2-16}\fodgespace
%
%%% [17] O(p^2) 12-point diagram
\tikzineq[>=stealth]{
	\draw[thick] (0.528, 0.141) -- (0.966, -0.259);
	\draw[thick] (0.528, 0.141) -- (0.966, 0.259);
	\draw[thick] (0.528, 0.141) -- (0.707, 0.707);
	\draw[thick] (0.000, 0.386) -- (0.528, 0.141);
	\draw[thick] (0.000, 0.386) -- (0.259, 0.966);
	\draw[thick] (0.000, 0.386) -- (-0.259, 0.966);
	\draw[thick] (-0.000, -0.478) -- (0.000, 0.386);
	\draw[thick] (-0.528, 0.141) -- (-0.707, 0.707);
	\draw[thick] (-0.528, 0.141) -- (-0.966, 0.259);
	\draw[thick] (-0.528, 0.141) -- (-0.966, -0.259);
	\draw[thick] (-0.000, -0.478) -- (-0.528, 0.141);
	\draw[thick] (-0.000, -0.478) -- (-0.707, -0.707);
	\draw[thick] (-0.000, -0.478) -- (-0.259, -0.966);
	\draw[thick] (-0.000, -0.478) -- (0.259, -0.966);
	\draw[thick] (-0.000, -0.478) -- (0.707, -0.707);
}{M12p2-17}\fodgespace
%
%%% [18] O(p^2) 12-point diagram
\tikzineq[>=stealth]{
	\draw[thick] (0.541, -0.074) -- (0.790, -0.613);
	\draw[thick] (0.541, -0.074) -- (0.991, -0.136);
	\draw[thick] (0.541, -0.074) -- (0.926, 0.377);
	\draw[thick] (0.150, 0.356) -- (0.541, -0.074);
	\draw[thick] (0.150, 0.356) -- (0.613, 0.790);
	\draw[thick] (0.150, 0.356) -- (0.136, 0.991);
	\draw[thick] (-0.107, -0.254) -- (-0.377, 0.926);
	\draw[thick] (-0.541, 0.074) -- (-0.790, 0.613);
	\draw[thick] (-0.541, 0.074) -- (-0.991, 0.136);
	\draw[thick] (-0.541, 0.074) -- (-0.926, -0.377);
	\draw[thick] (-0.107, -0.254) -- (-0.541, 0.074);
	\draw[thick] (-0.107, -0.254) -- (-0.613, -0.790);
	\draw[thick] (-0.107, -0.254) -- (-0.136, -0.991);
	\draw[thick] (-0.107, -0.254) -- (0.377, -0.926);
	\draw[thick] (0.150, 0.356) -- (-0.107, -0.254);
}{M12p2-18}\fodgespace
%
%%% [19] O(p^2) 12-point diagram
\tikzineq[>=stealth]{
	\draw[thick] (0.501, -0.217) -- (0.596, -0.803);
	\draw[thick] (0.501, -0.217) -- (0.918, -0.398);
	\draw[thick] (0.501, -0.217) -- (0.993, 0.114);
	\draw[thick] (0.240, 0.303) -- (0.501, -0.217);
	\draw[thick] (0.240, 0.303) -- (0.803, 0.596);
	\draw[thick] (0.240, 0.303) -- (0.398, 0.918);
	\draw[thick] (-0.142, -0.016) -- (0.240, 0.303);
	\draw[thick] (-0.142, -0.016) -- (-0.114, 0.993);
	\draw[thick] (-0.142, -0.016) -- (-0.596, 0.803);
	\draw[thick] (-0.543, -0.063) -- (-0.918, 0.398);
	\draw[thick] (-0.543, -0.063) -- (-0.993, -0.114);
	\draw[thick] (-0.543, -0.063) -- (-0.803, -0.596);
	\draw[thick] (-0.142, -0.016) -- (-0.543, -0.063);
	\draw[thick] (-0.142, -0.016) -- (-0.398, -0.918);
	\draw[thick] (-0.142, -0.016) -- (0.114, -0.993);
}{M12p2-19}\fodgespace
%
%%% [20] O(p^2) 12-point diagram
\tikzineq[>=stealth]{
	\draw[thick] (0.100, -0.373) -- (0.000, -1.000);
	\draw[thick] (0.100, -0.373) -- (0.500, -0.866);
	\draw[thick] (0.546, 0.000) -- (0.866, -0.500);
	\draw[thick] (0.546, 0.000) -- (1.000, 0.000);
	\draw[thick] (0.546, 0.000) -- (0.866, 0.500);
	\draw[thick] (0.100, 0.373) -- (0.546, 0.000);
	\draw[thick] (0.100, 0.373) -- (0.500, 0.866);
	\draw[thick] (0.100, 0.373) -- (-0.000, 1.000);
	\draw[thick] (0.100, -0.373) -- (0.100, 0.373);
	\draw[thick] (-0.533, -0.000) -- (0.100, -0.373);
	\draw[thick] (-0.533, -0.000) -- (-0.500, 0.866);
	\draw[thick] (-0.533, -0.000) -- (-0.866, 0.500);
	\draw[thick] (-0.533, -0.000) -- (-1.000, -0.000);
	\draw[thick] (-0.533, -0.000) -- (-0.866, -0.500);
	\draw[thick] (-0.533, -0.000) -- (-0.500, -0.866);
}{M12p2-20}\fodgespace
%
%%% [21] O(p^2) 12-point diagram
\tikzineq[>=stealth]{
	\draw[thick] (-0.437, 0.328) -- (-0.393, 0.920);
	\draw[thick] (-0.437, 0.328) -- (-0.800, 0.600);
	\draw[thick] (-0.437, 0.328) -- (-0.993, 0.120);
	\draw[thick] (-0.304, -0.239) -- (-0.437, 0.328);
	\draw[thick] (-0.304, -0.239) -- (-0.920, -0.393);
	\draw[thick] (-0.304, -0.239) -- (-0.600, -0.800);
	\draw[thick] (0.170, -0.217) -- (-0.304, -0.239);
	\draw[thick] (0.170, -0.217) -- (-0.120, -0.993);
	\draw[thick] (0.170, -0.217) -- (0.393, -0.920);
	\draw[thick] (0.170, -0.217) -- (0.800, -0.600);
	\draw[thick] (0.503, 0.215) -- (0.993, -0.120);
	\draw[thick] (0.503, 0.215) -- (0.920, 0.393);
	\draw[thick] (0.503, 0.215) -- (0.600, 0.800);
	\draw[thick] (0.170, -0.217) -- (0.503, 0.215);
	\draw[thick] (0.170, -0.217) -- (0.120, 0.993);
}{M12p2-21}\fodgespace
%
%%% [22] O(p^2) 12-point diagram
\tikzineq[>=stealth]{
	\draw[thick] (0.503, -0.213) -- (0.603, -0.798);
	\draw[thick] (0.503, -0.213) -- (0.921, -0.390);
	\draw[thick] (0.503, -0.213) -- (0.992, 0.123);
	\draw[thick] (0.237, 0.305) -- (0.503, -0.213);
	\draw[thick] (0.237, 0.305) -- (0.798, 0.603);
	\draw[thick] (0.237, 0.305) -- (0.390, 0.921);
	\draw[thick] (-0.000, -0.000) -- (0.237, 0.305);
	\draw[thick] (-0.000, -0.000) -- (-0.123, 0.992);
	\draw[thick] (-0.529, -0.066) -- (-0.603, 0.798);
	\draw[thick] (-0.529, -0.066) -- (-0.921, 0.390);
	\draw[thick] (-0.529, -0.066) -- (-0.992, -0.123);
	\draw[thick] (-0.529, -0.066) -- (-0.798, -0.603);
	\draw[thick] (-0.529, -0.066) -- (-0.390, -0.921);
	\draw[thick] (-0.000, -0.000) -- (-0.529, -0.066);
	\draw[thick] (-0.000, -0.000) -- (0.123, -0.992);
}{M12p2-22}\fodgespace
%
%%% [23] O(p^2) 12-point diagram
\tikzineq[>=stealth]{
	\draw[thick] (0.141, -0.528) -- (-0.259, -0.966);
	\draw[thick] (0.141, -0.528) -- (0.259, -0.966);
	\draw[thick] (0.141, -0.528) -- (0.707, -0.707);
	\draw[thick] (-0.478, -0.000) -- (0.141, -0.528);
	\draw[thick] (0.528, 0.141) -- (0.966, -0.259);
	\draw[thick] (0.528, 0.141) -- (0.966, 0.259);
	\draw[thick] (0.528, 0.141) -- (0.707, 0.707);
	\draw[thick] (-0.000, 0.386) -- (0.528, 0.141);
	\draw[thick] (-0.000, 0.386) -- (0.259, 0.966);
	\draw[thick] (-0.000, 0.386) -- (-0.259, 0.966);
	\draw[thick] (-0.478, -0.000) -- (-0.000, 0.386);
	\draw[thick] (-0.478, -0.000) -- (-0.707, 0.707);
	\draw[thick] (-0.478, -0.000) -- (-0.966, 0.259);
	\draw[thick] (-0.478, -0.000) -- (-0.966, -0.259);
	\draw[thick] (-0.478, -0.000) -- (-0.707, -0.707);
}{M12p2-23}\fodgespace
%
%%% [24] O(p^2) 12-point diagram
\tikzineq[>=stealth]{
	\draw[thick] (0.415, -0.355) -- (0.334, -0.943);
	\draw[thick] (0.415, -0.355) -- (0.760, -0.649);
	\draw[thick] (0.415, -0.355) -- (0.983, -0.182);
	\draw[thick] (0.318, 0.219) -- (0.415, -0.355);
	\draw[thick] (0.318, 0.219) -- (0.943, 0.334);
	\draw[thick] (0.318, 0.219) -- (0.649, 0.760);
	\draw[thick] (-0.030, 0.385) -- (0.318, 0.219);
	\draw[thick] (-0.030, 0.385) -- (0.182, 0.983);
	\draw[thick] (-0.030, 0.385) -- (-0.334, 0.943);
	\draw[thick] (-0.503, -0.178) -- (-0.760, 0.649);
	\draw[thick] (-0.503, -0.178) -- (-0.983, 0.182);
	\draw[thick] (-0.503, -0.178) -- (-0.943, -0.334);
	\draw[thick] (-0.503, -0.178) -- (-0.649, -0.760);
	\draw[thick] (-0.503, -0.178) -- (-0.182, -0.983);
	\draw[thick] (-0.030, 0.385) -- (-0.503, -0.178);
}{M12p2-24}\fodgespace
%
%%% [25] O(p^2) 12-point diagram
\tikzineq[>=stealth]{
	\draw[thick] (-0.546, 0.000) -- (-0.866, 0.500);
	\draw[thick] (-0.546, 0.000) -- (-1.000, 0.000);
	\draw[thick] (-0.546, 0.000) -- (-0.866, -0.500);
	\draw[thick] (-0.000, -0.390) -- (-0.546, 0.000);
	\draw[thick] (-0.000, -0.390) -- (-0.500, -0.866);
	\draw[thick] (-0.000, -0.390) -- (-0.000, -1.000);
	\draw[thick] (-0.000, -0.390) -- (0.500, -0.866);
	\draw[thick] (0.546, -0.000) -- (0.866, -0.500);
	\draw[thick] (0.546, -0.000) -- (1.000, -0.000);
	\draw[thick] (0.546, -0.000) -- (0.866, 0.500);
	\draw[thick] (-0.000, -0.390) -- (0.546, -0.000);
	\draw[thick] (0.000, 0.546) -- (0.500, 0.866);
	\draw[thick] (0.000, 0.546) -- (0.000, 1.000);
	\draw[thick] (0.000, 0.546) -- (-0.500, 0.866);
	\draw[thick] (-0.000, -0.390) -- (0.000, 0.546);
}{M12p2-25}\fodgespace
%
%%% [26] O(p^2) 12-point diagram
\tikzineq[>=stealth]{
	\draw[thick] (-0.545, 0.037) -- (-0.830, 0.558);
	\draw[thick] (-0.545, 0.037) -- (-0.998, 0.068);
	\draw[thick] (-0.545, 0.037) -- (-0.898, -0.440);
	\draw[thick] (-0.152, -0.133) -- (-0.545, 0.037);
	\draw[thick] (-0.152, -0.133) -- (-0.558, -0.830);
	\draw[thick] (-0.152, -0.133) -- (-0.068, -0.998);
	\draw[thick] (0.454, -0.305) -- (0.440, -0.898);
	\draw[thick] (0.454, -0.305) -- (0.830, -0.558);
	\draw[thick] (0.454, -0.305) -- (0.998, -0.068);
	\draw[thick] (-0.152, -0.133) -- (0.454, -0.305);
	\draw[thick] (0.305, 0.454) -- (0.898, 0.440);
	\draw[thick] (0.305, 0.454) -- (0.558, 0.830);
	\draw[thick] (0.305, 0.454) -- (0.068, 0.998);
	\draw[thick] (-0.152, -0.133) -- (0.305, 0.454);
	\draw[thick] (-0.152, -0.133) -- (-0.440, 0.898);
}{M12p2-26}\fodgespace
%
%%% [27] O(p^2) 12-point diagram
\tikzineq[>=stealth]{
	\draw[thick] (-0.454, -0.305) -- (-0.998, -0.068);
	\draw[thick] (-0.454, -0.305) -- (-0.830, -0.558);
	\draw[thick] (-0.454, -0.305) -- (-0.440, -0.898);
	\draw[thick] (0.152, -0.133) -- (-0.454, -0.305);
	\draw[thick] (0.152, -0.133) -- (0.068, -0.998);
	\draw[thick] (0.152, -0.133) -- (0.558, -0.830);
	\draw[thick] (0.545, 0.037) -- (0.898, -0.440);
	\draw[thick] (0.545, 0.037) -- (0.998, 0.068);
	\draw[thick] (0.545, 0.037) -- (0.830, 0.558);
	\draw[thick] (0.152, -0.133) -- (0.545, 0.037);
	\draw[thick] (0.152, -0.133) -- (0.440, 0.898);
	\draw[thick] (-0.305, 0.454) -- (-0.068, 0.998);
	\draw[thick] (-0.305, 0.454) -- (-0.558, 0.830);
	\draw[thick] (-0.305, 0.454) -- (-0.898, 0.440);
	\draw[thick] (0.152, -0.133) -- (-0.305, 0.454);
}{M12p2-27}\fodgespace
%
%%% [28] O(p^2) 12-point diagram
\tikzineq[>=stealth]{
	\draw[thick] (0.000, -0.200) -- (0.000, -1.000);
	\draw[thick] (0.473, -0.273) -- (0.500, -0.866);
	\draw[thick] (0.473, -0.273) -- (0.866, -0.500);
	\draw[thick] (0.473, -0.273) -- (1.000, 0.000);
	\draw[thick] (0.000, -0.200) -- (0.473, -0.273);
	\draw[thick] (0.273, 0.473) -- (0.866, 0.500);
	\draw[thick] (0.273, 0.473) -- (0.500, 0.866);
	\draw[thick] (0.273, 0.473) -- (-0.000, 1.000);
	\draw[thick] (0.000, -0.200) -- (0.273, 0.473);
	\draw[thick] (-0.533, -0.000) -- (0.000, -0.200);
	\draw[thick] (-0.533, -0.000) -- (-0.500, 0.866);
	\draw[thick] (-0.533, -0.000) -- (-0.866, 0.500);
	\draw[thick] (-0.533, -0.000) -- (-1.000, -0.000);
	\draw[thick] (-0.533, -0.000) -- (-0.866, -0.500);
	\draw[thick] (-0.533, -0.000) -- (-0.500, -0.866);
}{M12p2-28}\fodgespace
%
%%% [29] O(p^2) 12-point diagram
\tikzineq[>=stealth]{
	\draw[thick] (0.000, -0.386) -- (-0.259, -0.966);
	\draw[thick] (0.000, -0.386) -- (0.259, -0.966);
	\draw[thick] (0.528, -0.141) -- (0.707, -0.707);
	\draw[thick] (0.528, -0.141) -- (0.966, -0.259);
	\draw[thick] (0.528, -0.141) -- (0.966, 0.259);
	\draw[thick] (0.000, -0.386) -- (0.528, -0.141);
	\draw[thick] (-0.478, -0.000) -- (0.000, -0.386);
	\draw[thick] (0.141, 0.528) -- (0.707, 0.707);
	\draw[thick] (0.141, 0.528) -- (0.259, 0.966);
	\draw[thick] (0.141, 0.528) -- (-0.259, 0.966);
	\draw[thick] (-0.478, -0.000) -- (0.141, 0.528);
	\draw[thick] (-0.478, -0.000) -- (-0.707, 0.707);
	\draw[thick] (-0.478, -0.000) -- (-0.966, 0.259);
	\draw[thick] (-0.478, -0.000) -- (-0.966, -0.259);
	\draw[thick] (-0.478, -0.000) -- (-0.707, -0.707);
}{M12p2-29}\fodgespace
%
%%% [30] O(p^2) 12-point diagram
\tikzineq[>=stealth]{
	\draw[thick] (-0.473, -0.273) -- (-1.000, 0.000);
	\draw[thick] (-0.473, -0.273) -- (-0.866, -0.500);
	\draw[thick] (-0.473, -0.273) -- (-0.500, -0.866);
	\draw[thick] (-0.000, -0.200) -- (-0.473, -0.273);
	\draw[thick] (-0.000, -0.200) -- (-0.000, -1.000);
	\draw[thick] (0.533, -0.000) -- (0.500, -0.866);
	\draw[thick] (0.533, -0.000) -- (0.866, -0.500);
	\draw[thick] (0.533, -0.000) -- (1.000, -0.000);
	\draw[thick] (0.533, -0.000) -- (0.866, 0.500);
	\draw[thick] (0.533, -0.000) -- (0.500, 0.866);
	\draw[thick] (-0.000, -0.200) -- (0.533, -0.000);
	\draw[thick] (-0.273, 0.473) -- (-0.000, -0.200);
	\draw[thick] (-0.273, 0.473) -- (0.000, 1.000);
	\draw[thick] (-0.273, 0.473) -- (-0.500, 0.866);
	\draw[thick] (-0.273, 0.473) -- (-0.866, 0.500);
}{M12p2-30}\fodgespace
%
%%% [31] O(p^2) 12-point diagram
\tikzineq[>=stealth]{
	\draw[thick] (-0.310, -0.450) -- (-0.903, -0.430);
	\draw[thick] (-0.310, -0.450) -- (-0.567, -0.824);
	\draw[thick] (-0.310, -0.450) -- (-0.079, -0.997);
	\draw[thick] (-0.310, 0.364) -- (-0.310, -0.450);
	\draw[thick] (0.199, -0.016) -- (0.430, -0.903);
	\draw[thick] (0.545, -0.043) -- (0.824, -0.567);
	\draw[thick] (0.545, -0.043) -- (0.997, -0.079);
	\draw[thick] (0.545, -0.043) -- (0.903, 0.430);
	\draw[thick] (0.199, -0.016) -- (0.545, -0.043);
	\draw[thick] (0.199, -0.016) -- (0.567, 0.824);
	\draw[thick] (-0.310, 0.364) -- (0.199, -0.016);
	\draw[thick] (-0.310, 0.364) -- (0.079, 0.997);
	\draw[thick] (-0.310, 0.364) -- (-0.430, 0.903);
	\draw[thick] (-0.310, 0.364) -- (-0.824, 0.567);
	\draw[thick] (-0.310, 0.364) -- (-0.997, 0.079);
}{M12p2-31}\fodgespace
%
%%% [32] O(p^2) 12-point diagram
\tikzineq[>=stealth]{
	\draw[thick] (0.395, -0.377) -- (0.281, -0.960);
	\draw[thick] (0.395, -0.377) -- (0.723, -0.691);
	\draw[thick] (0.395, -0.377) -- (0.972, -0.236);
	\draw[thick] (-0.000, 0.000) -- (0.395, -0.377);
	\draw[thick] (-0.000, 0.000) -- (0.960, 0.281);
	\draw[thick] (0.129, 0.531) -- (0.691, 0.723);
	\draw[thick] (0.129, 0.531) -- (0.236, 0.972);
	\draw[thick] (0.129, 0.531) -- (-0.281, 0.960);
	\draw[thick] (-0.000, 0.000) -- (0.129, 0.531);
	\draw[thick] (-0.000, 0.000) -- (-0.723, 0.691);
	\draw[thick] (-0.524, -0.154) -- (-0.972, 0.236);
	\draw[thick] (-0.524, -0.154) -- (-0.960, -0.281);
	\draw[thick] (-0.524, -0.154) -- (-0.691, -0.723);
	\draw[thick] (-0.000, 0.000) -- (-0.524, -0.154);
	\draw[thick] (-0.000, 0.000) -- (-0.236, -0.972);
}{M12p2-32}\fodgespace
%
%%% [33] O(p^2) 12-point diagram
\tikzineq[>=stealth]{
	\draw[thick] (0.199, 0.016) -- (0.567, -0.824);
	\draw[thick] (0.545, 0.043) -- (0.903, -0.430);
	\draw[thick] (0.545, 0.043) -- (0.997, 0.079);
	\draw[thick] (0.545, 0.043) -- (0.824, 0.567);
	\draw[thick] (0.199, 0.016) -- (0.545, 0.043);
	\draw[thick] (0.199, 0.016) -- (0.430, 0.903);
	\draw[thick] (-0.310, -0.364) -- (0.199, 0.016);
	\draw[thick] (-0.310, 0.450) -- (-0.079, 0.997);
	\draw[thick] (-0.310, 0.450) -- (-0.567, 0.824);
	\draw[thick] (-0.310, 0.450) -- (-0.903, 0.430);
	\draw[thick] (-0.310, -0.364) -- (-0.310, 0.450);
	\draw[thick] (-0.310, -0.364) -- (-0.997, -0.079);
	\draw[thick] (-0.310, -0.364) -- (-0.824, -0.567);
	\draw[thick] (-0.310, -0.364) -- (-0.430, -0.903);
	\draw[thick] (-0.310, -0.364) -- (0.079, -0.997);
}{M12p2-33}\fodgespace
%
%%% [34] O(p^2) 12-point diagram
\tikzineq[>=stealth]{
	\draw[thick] (-0.503, 0.215) -- (-0.600, 0.800);
	\draw[thick] (-0.503, 0.215) -- (-0.920, 0.393);
	\draw[thick] (-0.503, 0.215) -- (-0.993, -0.120);
	\draw[thick] (-0.170, -0.217) -- (-0.503, 0.215);
	\draw[thick] (-0.170, -0.217) -- (-0.800, -0.600);
	\draw[thick] (-0.170, -0.217) -- (-0.393, -0.920);
	\draw[thick] (-0.170, -0.217) -- (0.120, -0.993);
	\draw[thick] (0.304, -0.239) -- (0.600, -0.800);
	\draw[thick] (0.304, -0.239) -- (0.920, -0.393);
	\draw[thick] (0.437, 0.328) -- (0.993, 0.120);
	\draw[thick] (0.437, 0.328) -- (0.800, 0.600);
	\draw[thick] (0.437, 0.328) -- (0.393, 0.920);
	\draw[thick] (0.304, -0.239) -- (0.437, 0.328);
	\draw[thick] (-0.170, -0.217) -- (0.304, -0.239);
	\draw[thick] (-0.170, -0.217) -- (-0.120, 0.993);
}{M12p2-34}\fodgespace
%
%%% [35] O(p^2) 12-point diagram
\tikzineq[>=stealth]{
	\draw[thick] (0.273, -0.473) -- (0.000, -1.000);
	\draw[thick] (0.273, -0.473) -- (0.500, -0.866);
	\draw[thick] (0.273, -0.473) -- (0.866, -0.500);
	\draw[thick] (0.200, 0.000) -- (0.273, -0.473);
	\draw[thick] (0.200, 0.000) -- (1.000, 0.000);
	\draw[thick] (0.273, 0.473) -- (0.866, 0.500);
	\draw[thick] (0.273, 0.473) -- (0.500, 0.866);
	\draw[thick] (0.273, 0.473) -- (-0.000, 1.000);
	\draw[thick] (0.200, 0.000) -- (0.273, 0.473);
	\draw[thick] (-0.533, -0.000) -- (-0.500, 0.866);
	\draw[thick] (-0.533, -0.000) -- (-0.866, 0.500);
	\draw[thick] (-0.533, -0.000) -- (-1.000, -0.000);
	\draw[thick] (-0.533, -0.000) -- (-0.866, -0.500);
	\draw[thick] (-0.533, -0.000) -- (-0.500, -0.866);
	\draw[thick] (0.200, 0.000) -- (-0.533, -0.000);
}{M12p2-35}\fodgespace
%
%%% [36] O(p^2) 12-point diagram
\tikzineq[>=stealth]{
	\draw[thick] (0.528, 0.141) -- (0.966, -0.259);
	\draw[thick] (0.528, 0.141) -- (0.966, 0.259);
	\draw[thick] (0.528, 0.141) -- (0.707, 0.707);
	\draw[thick] (-0.000, -0.478) -- (0.528, 0.141);
	\draw[thick] (0.000, 0.386) -- (0.259, 0.966);
	\draw[thick] (0.000, 0.386) -- (-0.259, 0.966);
	\draw[thick] (-0.528, 0.141) -- (-0.707, 0.707);
	\draw[thick] (-0.528, 0.141) -- (-0.966, 0.259);
	\draw[thick] (-0.528, 0.141) -- (-0.966, -0.259);
	\draw[thick] (0.000, 0.386) -- (-0.528, 0.141);
	\draw[thick] (-0.000, -0.478) -- (0.000, 0.386);
	\draw[thick] (-0.000, -0.478) -- (-0.707, -0.707);
	\draw[thick] (-0.000, -0.478) -- (-0.259, -0.966);
	\draw[thick] (-0.000, -0.478) -- (0.259, -0.966);
	\draw[thick] (-0.000, -0.478) -- (0.707, -0.707);
}{M12p2-36}\fodgespace
%
%%% [37] O(p^2) 12-point diagram
\tikzineq[>=stealth]{
	\draw[thick] (-0.197, 0.036) -- (-0.338, 0.941);
	\draw[thick] (-0.538, 0.097) -- (-0.763, 0.646);
	\draw[thick] (-0.538, 0.097) -- (-0.984, 0.178);
	\draw[thick] (-0.538, 0.097) -- (-0.941, -0.338);
	\draw[thick] (-0.197, 0.036) -- (-0.538, 0.097);
	\draw[thick] (-0.197, 0.036) -- (-0.646, -0.763);
	\draw[thick] (0.157, -0.227) -- (-0.197, 0.036);
	\draw[thick] (0.157, -0.227) -- (-0.178, -0.984);
	\draw[thick] (0.157, -0.227) -- (0.338, -0.941);
	\draw[thick] (0.157, -0.227) -- (0.763, -0.646);
	\draw[thick] (0.514, 0.185) -- (0.984, -0.178);
	\draw[thick] (0.514, 0.185) -- (0.941, 0.338);
	\draw[thick] (0.514, 0.185) -- (0.646, 0.763);
	\draw[thick] (0.157, -0.227) -- (0.514, 0.185);
	\draw[thick] (0.157, -0.227) -- (0.178, 0.984);
}{M12p2-37}\fodgespace
%
%%% [38] O(p^2) 12-point diagram
\tikzineq[>=stealth]{
	\draw[thick] (0.197, 0.036) -- (0.646, -0.763);
	\draw[thick] (0.538, 0.097) -- (0.941, -0.338);
	\draw[thick] (0.538, 0.097) -- (0.984, 0.178);
	\draw[thick] (0.538, 0.097) -- (0.763, 0.646);
	\draw[thick] (0.197, 0.036) -- (0.538, 0.097);
	\draw[thick] (0.197, 0.036) -- (0.338, 0.941);
	\draw[thick] (-0.157, -0.227) -- (0.197, 0.036);
	\draw[thick] (-0.157, -0.227) -- (-0.178, 0.984);
	\draw[thick] (-0.514, 0.185) -- (-0.646, 0.763);
	\draw[thick] (-0.514, 0.185) -- (-0.941, 0.338);
	\draw[thick] (-0.514, 0.185) -- (-0.984, -0.178);
	\draw[thick] (-0.157, -0.227) -- (-0.514, 0.185);
	\draw[thick] (-0.157, -0.227) -- (-0.763, -0.646);
	\draw[thick] (-0.157, -0.227) -- (-0.338, -0.941);
	\draw[thick] (-0.157, -0.227) -- (0.178, -0.984);
}{M12p2-38}\fodgespace
%
%%% [39] O(p^2) 12-point diagram
\tikzineq[>=stealth]{
	\draw[thick] (0.543, -0.063) -- (0.803, -0.596);
	\draw[thick] (0.543, -0.063) -- (0.993, -0.114);
	\draw[thick] (0.543, -0.063) -- (0.918, 0.398);
	\draw[thick] (0.142, -0.016) -- (0.543, -0.063);
	\draw[thick] (0.142, -0.016) -- (0.596, 0.803);
	\draw[thick] (0.142, -0.016) -- (0.114, 0.993);
	\draw[thick] (-0.240, 0.303) -- (-0.398, 0.918);
	\draw[thick] (-0.240, 0.303) -- (-0.803, 0.596);
	\draw[thick] (-0.501, -0.217) -- (-0.993, 0.114);
	\draw[thick] (-0.501, -0.217) -- (-0.918, -0.398);
	\draw[thick] (-0.501, -0.217) -- (-0.596, -0.803);
	\draw[thick] (-0.240, 0.303) -- (-0.501, -0.217);
	\draw[thick] (0.142, -0.016) -- (-0.240, 0.303);
	\draw[thick] (0.142, -0.016) -- (-0.114, -0.993);
	\draw[thick] (0.142, -0.016) -- (0.398, -0.918);
}{M12p2-39}\fodgespace
%
%%% [40] O(p^2) 12-point diagram
\tikzineq[>=stealth]{
	\draw[thick] (-0.503, 0.178) -- (-0.182, 0.983);
	\draw[thick] (-0.503, 0.178) -- (-0.649, 0.760);
	\draw[thick] (-0.503, 0.178) -- (-0.943, 0.334);
	\draw[thick] (-0.503, 0.178) -- (-0.983, -0.182);
	\draw[thick] (-0.503, 0.178) -- (-0.760, -0.649);
	\draw[thick] (-0.030, -0.385) -- (-0.503, 0.178);
	\draw[thick] (-0.030, -0.385) -- (-0.334, -0.943);
	\draw[thick] (-0.030, -0.385) -- (0.182, -0.983);
	\draw[thick] (0.318, -0.219) -- (-0.030, -0.385);
	\draw[thick] (0.318, -0.219) -- (0.649, -0.760);
	\draw[thick] (0.318, -0.219) -- (0.943, -0.334);
	\draw[thick] (0.415, 0.355) -- (0.318, -0.219);
	\draw[thick] (0.415, 0.355) -- (0.983, 0.182);
	\draw[thick] (0.415, 0.355) -- (0.760, 0.649);
	\draw[thick] (0.415, 0.355) -- (0.334, 0.943);
}{M12p2-40}\fodgespace
%
%%% [41] O(p^2) 12-point diagram
\tikzineq[>=stealth]{
	\draw[thick] (0.150, -0.356) -- (0.136, -0.991);
	\draw[thick] (0.150, -0.356) -- (0.613, -0.790);
	\draw[thick] (0.541, 0.074) -- (0.926, -0.377);
	\draw[thick] (0.541, 0.074) -- (0.991, 0.136);
	\draw[thick] (0.541, 0.074) -- (0.790, 0.613);
	\draw[thick] (0.150, -0.356) -- (0.541, 0.074);
	\draw[thick] (-0.107, 0.254) -- (0.150, -0.356);
	\draw[thick] (-0.107, 0.254) -- (0.377, 0.926);
	\draw[thick] (-0.107, 0.254) -- (-0.136, 0.991);
	\draw[thick] (-0.107, 0.254) -- (-0.613, 0.790);
	\draw[thick] (-0.541, -0.074) -- (-0.926, 0.377);
	\draw[thick] (-0.541, -0.074) -- (-0.991, -0.136);
	\draw[thick] (-0.541, -0.074) -- (-0.790, -0.613);
	\draw[thick] (-0.107, 0.254) -- (-0.541, -0.074);
	\draw[thick] (-0.107, 0.254) -- (-0.377, -0.926);
}{M12p2-41}\fodgespace
%
%%% [42] O(p^2) 12-point diagram
\tikzineq[>=stealth]{
	\draw[thick] (0.200, -0.000) -- (0.500, -0.866);
	\draw[thick] (0.546, -0.000) -- (0.866, -0.500);
	\draw[thick] (0.546, -0.000) -- (1.000, -0.000);
	\draw[thick] (0.546, -0.000) -- (0.866, 0.500);
	\draw[thick] (0.200, -0.000) -- (0.546, -0.000);
	\draw[thick] (0.200, -0.000) -- (0.500, 0.866);
	\draw[thick] (-0.143, 0.000) -- (0.200, -0.000);
	\draw[thick] (-0.143, 0.000) -- (0.000, 1.000);
	\draw[thick] (-0.143, 0.000) -- (-0.500, 0.866);
	\draw[thick] (-0.546, 0.000) -- (-0.866, 0.500);
	\draw[thick] (-0.546, 0.000) -- (-1.000, 0.000);
	\draw[thick] (-0.546, 0.000) -- (-0.866, -0.500);
	\draw[thick] (-0.143, 0.000) -- (-0.546, 0.000);
	\draw[thick] (-0.143, 0.000) -- (-0.500, -0.866);
	\draw[thick] (-0.143, 0.000) -- (-0.000, -1.000);
}{M12p2-42}\fodgespace
%
%%% [43] O(p^2) 12-point diagram
\tikzineq[>=stealth]{
	\draw[thick] (-0.013, -0.386) -- (-0.292, -0.956);
	\draw[thick] (-0.013, -0.386) -- (0.225, -0.974);
	\draw[thick] (0.195, 0.045) -- (0.682, -0.731);
	\draw[thick] (0.532, 0.123) -- (0.956, -0.292);
	\draw[thick] (0.532, 0.123) -- (0.974, 0.225);
	\draw[thick] (0.532, 0.123) -- (0.731, 0.682);
	\draw[thick] (0.195, 0.045) -- (0.532, 0.123);
	\draw[thick] (0.195, 0.045) -- (0.292, 0.956);
	\draw[thick] (-0.013, -0.386) -- (0.195, 0.045);
	\draw[thick] (-0.510, 0.156) -- (-0.013, -0.386);
	\draw[thick] (-0.510, 0.156) -- (-0.225, 0.974);
	\draw[thick] (-0.510, 0.156) -- (-0.682, 0.731);
	\draw[thick] (-0.510, 0.156) -- (-0.956, 0.292);
	\draw[thick] (-0.510, 0.156) -- (-0.974, -0.225);
	\draw[thick] (-0.510, 0.156) -- (-0.731, -0.682);
}{M12p2-43}\fodgespace
%
%%% [44] O(p^2) 12-point diagram
\tikzineq[>=stealth]{
	\draw[thick] (0.000, -0.000) -- (-0.123, -0.992);
	\draw[thick] (0.237, -0.305) -- (0.390, -0.921);
	\draw[thick] (0.237, -0.305) -- (0.798, -0.603);
	\draw[thick] (0.503, 0.213) -- (0.992, -0.123);
	\draw[thick] (0.503, 0.213) -- (0.921, 0.390);
	\draw[thick] (0.503, 0.213) -- (0.603, 0.798);
	\draw[thick] (0.237, -0.305) -- (0.503, 0.213);
	\draw[thick] (0.000, -0.000) -- (0.237, -0.305);
	\draw[thick] (0.000, -0.000) -- (0.123, 0.992);
	\draw[thick] (-0.529, 0.066) -- (0.000, -0.000);
	\draw[thick] (-0.529, 0.066) -- (-0.390, 0.921);
	\draw[thick] (-0.529, 0.066) -- (-0.798, 0.603);
	\draw[thick] (-0.529, 0.066) -- (-0.992, 0.123);
	\draw[thick] (-0.529, 0.066) -- (-0.921, -0.390);
	\draw[thick] (-0.529, 0.066) -- (-0.603, -0.798);
}{M12p2-44}\fodgespace
%
%%% [45] O(p^2) 12-point diagram
\tikzineq[>=stealth]{
	\draw[thick] (0.000, 0.000) -- (0.000, -1.000);
	\draw[thick] (0.200, 0.000) -- (0.500, -0.866);
	\draw[thick] (0.546, 0.000) -- (0.866, -0.500);
	\draw[thick] (0.546, 0.000) -- (1.000, 0.000);
	\draw[thick] (0.546, 0.000) -- (0.866, 0.500);
	\draw[thick] (0.200, 0.000) -- (0.546, 0.000);
	\draw[thick] (0.200, 0.000) -- (0.500, 0.866);
	\draw[thick] (0.000, 0.000) -- (0.200, 0.000);
	\draw[thick] (0.000, 0.000) -- (0.000, 1.000);
	\draw[thick] (-0.533, 0.000) -- (0.000, 0.000);
	\draw[thick] (-0.533, 0.000) -- (-0.500, 0.866);
	\draw[thick] (-0.533, 0.000) -- (-0.866, 0.500);
	\draw[thick] (-0.533, 0.000) -- (-1.000, 0.000);
	\draw[thick] (-0.533, 0.000) -- (-0.866, -0.500);
	\draw[thick] (-0.533, 0.000) -- (-0.500, -0.866);
}{M12p2-45}\fodgespace
%
%%% [46] O(p^2) 12-point diagram
\tikzineq[>=stealth]{
	\draw[thick] (0.100, -0.373) -- (0.000, -1.000);
	\draw[thick] (0.100, -0.373) -- (0.500, -0.866);
	\draw[thick] (0.546, 0.000) -- (0.866, -0.500);
	\draw[thick] (0.546, 0.000) -- (1.000, 0.000);
	\draw[thick] (0.546, 0.000) -- (0.866, 0.500);
	\draw[thick] (0.100, -0.373) -- (0.546, 0.000);
	\draw[thick] (0.100, 0.373) -- (0.100, -0.373);
	\draw[thick] (0.100, 0.373) -- (0.500, 0.866);
	\draw[thick] (0.100, 0.373) -- (-0.000, 1.000);
	\draw[thick] (-0.533, -0.000) -- (-0.500, 0.866);
	\draw[thick] (-0.533, -0.000) -- (-0.866, 0.500);
	\draw[thick] (-0.533, -0.000) -- (-1.000, -0.000);
	\draw[thick] (-0.533, -0.000) -- (-0.866, -0.500);
	\draw[thick] (-0.533, -0.000) -- (-0.500, -0.866);
	\draw[thick] (0.100, 0.373) -- (-0.533, -0.000);
}{M12p2-46}\fodgespace
%
%%% [47] O(p^2) 12-point diagram
\tikzineq[>=stealth]{
	\draw[thick] (0.195, -0.045) -- (0.292, -0.956);
	\draw[thick] (0.532, -0.123) -- (0.731, -0.682);
	\draw[thick] (0.532, -0.123) -- (0.974, -0.225);
	\draw[thick] (0.532, -0.123) -- (0.956, 0.292);
	\draw[thick] (0.195, -0.045) -- (0.532, -0.123);
	\draw[thick] (0.195, -0.045) -- (0.682, 0.731);
	\draw[thick] (-0.013, 0.386) -- (0.195, -0.045);
	\draw[thick] (-0.013, 0.386) -- (0.225, 0.974);
	\draw[thick] (-0.013, 0.386) -- (-0.292, 0.956);
	\draw[thick] (-0.510, -0.156) -- (-0.731, 0.682);
	\draw[thick] (-0.510, -0.156) -- (-0.974, 0.225);
	\draw[thick] (-0.510, -0.156) -- (-0.956, -0.292);
	\draw[thick] (-0.510, -0.156) -- (-0.682, -0.731);
	\draw[thick] (-0.510, -0.156) -- (-0.225, -0.974);
	\draw[thick] (-0.013, 0.386) -- (-0.510, -0.156);
}{M12p2-47}\fodgespace
%
%%% [48] O(p^2) 12-point diagram
\tikzineq[>=stealth]{
	\draw[thick] (0.528, -0.141) -- (0.707, -0.707);
	\draw[thick] (0.528, -0.141) -- (0.966, -0.259);
	\draw[thick] (0.528, -0.141) -- (0.966, 0.259);
	\draw[thick] (0.193, 0.335) -- (0.528, -0.141);
	\draw[thick] (0.193, 0.335) -- (0.707, 0.707);
	\draw[thick] (0.193, 0.335) -- (0.259, 0.966);
	\draw[thick] (-0.193, 0.335) -- (0.193, 0.335);
	\draw[thick] (-0.193, 0.335) -- (-0.259, 0.966);
	\draw[thick] (-0.193, 0.335) -- (-0.707, 0.707);
	\draw[thick] (-0.528, -0.141) -- (-0.966, 0.259);
	\draw[thick] (-0.528, -0.141) -- (-0.966, -0.259);
	\draw[thick] (-0.528, -0.141) -- (-0.707, -0.707);
	\draw[thick] (-0.000, -0.386) -- (-0.528, -0.141);
	\draw[thick] (-0.000, -0.386) -- (-0.259, -0.966);
	\draw[thick] (-0.000, -0.386) -- (0.259, -0.966);
	\draw[thick] (-0.193, 0.335) -- (-0.000, -0.386);
}{M12p2-48}\fodgespace
%
%%% [49] O(p^2) 12-point diagram
\tikzineq[>=stealth]{
	\draw[thick] (0.542, -0.071) -- (0.793, -0.609);
	\draw[thick] (0.542, -0.071) -- (0.991, -0.131);
	\draw[thick] (0.542, -0.071) -- (0.924, 0.383);
	\draw[thick] (0.148, 0.357) -- (0.542, -0.071);
	\draw[thick] (0.148, 0.357) -- (0.609, 0.793);
	\draw[thick] (0.148, 0.357) -- (0.131, 0.991);
	\draw[thick] (0.077, -0.185) -- (0.148, 0.357);
	\draw[thick] (-0.433, 0.333) -- (-0.383, 0.924);
	\draw[thick] (-0.433, 0.333) -- (-0.793, 0.609);
	\draw[thick] (-0.433, 0.333) -- (-0.991, 0.131);
	\draw[thick] (0.077, -0.185) -- (-0.433, 0.333);
	\draw[thick] (-0.333, -0.433) -- (-0.924, -0.383);
	\draw[thick] (-0.333, -0.433) -- (-0.609, -0.793);
	\draw[thick] (-0.333, -0.433) -- (-0.131, -0.991);
	\draw[thick] (0.077, -0.185) -- (-0.333, -0.433);
	\draw[thick] (0.077, -0.185) -- (0.383, -0.924);
}{M12p2-49}\fodgespace
%
%%% [50] O(p^2) 12-point diagram
\tikzineq[>=stealth]{
	\draw[thick] (0.433, 0.333) -- (0.991, 0.131);
	\draw[thick] (0.433, 0.333) -- (0.793, 0.609);
	\draw[thick] (0.433, 0.333) -- (0.383, 0.924);
	\draw[thick] (-0.148, 0.357) -- (0.433, 0.333);
	\draw[thick] (-0.148, 0.357) -- (-0.131, 0.991);
	\draw[thick] (-0.148, 0.357) -- (-0.609, 0.793);
	\draw[thick] (-0.077, -0.185) -- (-0.148, 0.357);
	\draw[thick] (-0.542, -0.071) -- (-0.924, 0.383);
	\draw[thick] (-0.542, -0.071) -- (-0.991, -0.131);
	\draw[thick] (-0.542, -0.071) -- (-0.793, -0.609);
	\draw[thick] (-0.077, -0.185) -- (-0.542, -0.071);
	\draw[thick] (-0.077, -0.185) -- (-0.383, -0.924);
	\draw[thick] (0.333, -0.433) -- (0.131, -0.991);
	\draw[thick] (0.333, -0.433) -- (0.609, -0.793);
	\draw[thick] (0.333, -0.433) -- (0.924, -0.383);
	\draw[thick] (-0.077, -0.185) -- (0.333, -0.433);
}{M12p2-50}\fodgespace
%
%%% [51] O(p^2) 12-point diagram
\tikzineq[>=stealth]{
	\draw[thick] (0.546, -0.000) -- (0.866, -0.500);
	\draw[thick] (0.546, -0.000) -- (1.000, -0.000);
	\draw[thick] (0.546, -0.000) -- (0.866, 0.500);
	\draw[thick] (0.100, 0.373) -- (0.546, -0.000);
	\draw[thick] (0.100, 0.373) -- (0.500, 0.866);
	\draw[thick] (0.100, 0.373) -- (0.000, 1.000);
	\draw[thick] (0.100, -0.373) -- (0.100, 0.373);
	\draw[thick] (-0.200, 0.000) -- (-0.500, 0.866);
	\draw[thick] (-0.546, 0.000) -- (-0.866, 0.500);
	\draw[thick] (-0.546, 0.000) -- (-1.000, 0.000);
	\draw[thick] (-0.546, 0.000) -- (-0.866, -0.500);
	\draw[thick] (-0.200, 0.000) -- (-0.546, 0.000);
	\draw[thick] (-0.200, 0.000) -- (-0.500, -0.866);
	\draw[thick] (0.100, -0.373) -- (-0.200, 0.000);
	\draw[thick] (0.100, -0.373) -- (-0.000, -1.000);
	\draw[thick] (0.100, -0.373) -- (0.500, -0.866);
}{M12p2-51}\fodgespace
%
%%% [52] O(p^2) 12-point diagram
\tikzineq[>=stealth]{
	\draw[thick] (0.535, -0.113) -- (0.744, -0.668);
	\draw[thick] (0.535, -0.113) -- (0.979, -0.206);
	\draw[thick] (0.535, -0.113) -- (0.950, 0.311);
	\draw[thick] (0.175, 0.345) -- (0.535, -0.113);
	\draw[thick] (0.175, 0.345) -- (0.668, 0.744);
	\draw[thick] (0.175, 0.345) -- (0.206, 0.979);
	\draw[thick] (-0.000, -0.000) -- (0.175, 0.345);
	\draw[thick] (-0.000, -0.000) -- (-0.311, 0.950);
	\draw[thick] (-0.535, 0.113) -- (-0.744, 0.668);
	\draw[thick] (-0.535, 0.113) -- (-0.979, 0.206);
	\draw[thick] (-0.535, 0.113) -- (-0.950, -0.311);
	\draw[thick] (-0.175, -0.345) -- (-0.535, 0.113);
	\draw[thick] (-0.175, -0.345) -- (-0.668, -0.744);
	\draw[thick] (-0.175, -0.345) -- (-0.206, -0.979);
	\draw[thick] (-0.000, -0.000) -- (-0.175, -0.345);
	\draw[thick] (-0.000, -0.000) -- (0.311, -0.950);
}{M12p2-52}\fodgespace
%
%%% [53] O(p^2) 12-point diagram
\tikzineq[>=stealth]{
	\draw[thick] (0.071, 0.542) -- (0.609, 0.793);
	\draw[thick] (0.071, 0.542) -- (0.131, 0.991);
	\draw[thick] (0.071, 0.542) -- (-0.383, 0.924);
	\draw[thick] (0.077, -0.185) -- (0.071, 0.542);
	\draw[thick] (-0.542, 0.071) -- (-0.793, 0.609);
	\draw[thick] (-0.542, 0.071) -- (-0.991, 0.131);
	\draw[thick] (-0.542, 0.071) -- (-0.924, -0.383);
	\draw[thick] (-0.148, -0.357) -- (-0.542, 0.071);
	\draw[thick] (-0.148, -0.357) -- (-0.609, -0.793);
	\draw[thick] (-0.148, -0.357) -- (-0.131, -0.991);
	\draw[thick] (0.077, -0.185) -- (-0.148, -0.357);
	\draw[thick] (0.077, -0.185) -- (0.383, -0.924);
	\draw[thick] (0.542, -0.071) -- (0.077, -0.185);
	\draw[thick] (0.542, -0.071) -- (0.793, -0.609);
	\draw[thick] (0.542, -0.071) -- (0.991, -0.131);
	\draw[thick] (0.542, -0.071) -- (0.924, 0.383);
}{M12p2-53}\fodgespace
%
%%% [54] O(p^2) 12-point diagram
\tikzineq[>=stealth]{
	\draw[thick] (0.528, -0.141) -- (0.707, -0.707);
	\draw[thick] (0.528, -0.141) -- (0.966, -0.259);
	\draw[thick] (0.528, -0.141) -- (0.966, 0.259);
	\draw[thick] (0.193, 0.335) -- (0.528, -0.141);
	\draw[thick] (0.193, 0.335) -- (0.707, 0.707);
	\draw[thick] (0.193, 0.335) -- (0.259, 0.966);
	\draw[thick] (-0.000, -0.386) -- (0.193, 0.335);
	\draw[thick] (-0.193, 0.335) -- (-0.259, 0.966);
	\draw[thick] (-0.193, 0.335) -- (-0.707, 0.707);
	\draw[thick] (-0.528, -0.141) -- (-0.966, 0.259);
	\draw[thick] (-0.528, -0.141) -- (-0.966, -0.259);
	\draw[thick] (-0.528, -0.141) -- (-0.707, -0.707);
	\draw[thick] (-0.193, 0.335) -- (-0.528, -0.141);
	\draw[thick] (-0.000, -0.386) -- (-0.193, 0.335);
	\draw[thick] (-0.000, -0.386) -- (-0.259, -0.966);
	\draw[thick] (-0.000, -0.386) -- (0.259, -0.966);
}{M12p2-54}\fodgespace
%
%%% [55] O(p^2) 12-point diagram
\tikzineq[>=stealth]{
	\draw[thick] (0.500, -0.220) -- (0.592, -0.806);
	\draw[thick] (0.500, -0.220) -- (0.916, -0.402);
	\draw[thick] (0.500, -0.220) -- (0.994, 0.110);
	\draw[thick] (0.242, 0.302) -- (0.500, -0.220);
	\draw[thick] (0.242, 0.302) -- (0.806, 0.592);
	\draw[thick] (0.242, 0.302) -- (0.402, 0.916);
	\draw[thick] (-0.000, -0.000) -- (0.242, 0.302);
	\draw[thick] (-0.000, -0.000) -- (-0.110, 0.994);
	\draw[thick] (-0.199, -0.022) -- (-0.592, 0.806);
	\draw[thick] (-0.543, -0.060) -- (-0.916, 0.402);
	\draw[thick] (-0.543, -0.060) -- (-0.994, -0.110);
	\draw[thick] (-0.543, -0.060) -- (-0.806, -0.592);
	\draw[thick] (-0.199, -0.022) -- (-0.543, -0.060);
	\draw[thick] (-0.199, -0.022) -- (-0.402, -0.916);
	\draw[thick] (-0.000, -0.000) -- (-0.199, -0.022);
	\draw[thick] (-0.000, -0.000) -- (0.110, -0.994);
}{M12p2-55}\fodgespace
%
%%% [56] O(p^2) 12-point diagram
\tikzineq[>=stealth]{
	\draw[thick] (0.473, -0.273) -- (0.500, -0.866);
	\draw[thick] (0.473, -0.273) -- (0.866, -0.500);
	\draw[thick] (0.473, -0.273) -- (1.000, -0.000);
	\draw[thick] (0.273, 0.273) -- (0.473, -0.273);
	\draw[thick] (0.273, 0.273) -- (0.866, 0.500);
	\draw[thick] (0.273, 0.273) -- (0.500, 0.866);
	\draw[thick] (-0.000, -0.000) -- (0.273, 0.273);
	\draw[thick] (-0.000, -0.000) -- (0.000, 1.000);
	\draw[thick] (-0.273, 0.273) -- (-0.500, 0.866);
	\draw[thick] (-0.273, 0.273) -- (-0.866, 0.500);
	\draw[thick] (-0.473, -0.273) -- (-1.000, 0.000);
	\draw[thick] (-0.473, -0.273) -- (-0.866, -0.500);
	\draw[thick] (-0.473, -0.273) -- (-0.500, -0.866);
	\draw[thick] (-0.273, 0.273) -- (-0.473, -0.273);
	\draw[thick] (-0.000, -0.000) -- (-0.273, 0.273);
	\draw[thick] (-0.000, -0.000) -- (-0.000, -1.000);
}{M12p2-56}\fodgespace
%
%%% [57] O(p^2) 12-point diagram
\tikzineq[>=stealth]{
	\draw[thick] (0.386, -0.386) -- (0.259, -0.966);
	\draw[thick] (0.386, -0.386) -- (0.707, -0.707);
	\draw[thick] (0.386, -0.386) -- (0.966, -0.259);
	\draw[thick] (0.335, 0.193) -- (0.386, -0.386);
	\draw[thick] (0.335, 0.193) -- (0.966, 0.259);
	\draw[thick] (0.335, 0.193) -- (0.707, 0.707);
	\draw[thick] (0.000, 0.386) -- (0.335, 0.193);
	\draw[thick] (0.000, 0.386) -- (0.259, 0.966);
	\draw[thick] (0.000, 0.386) -- (-0.259, 0.966);
	\draw[thick] (-0.193, -0.052) -- (0.000, 0.386);
	\draw[thick] (-0.193, -0.052) -- (-0.707, 0.707);
	\draw[thick] (-0.528, -0.141) -- (-0.966, 0.259);
	\draw[thick] (-0.528, -0.141) -- (-0.966, -0.259);
	\draw[thick] (-0.528, -0.141) -- (-0.707, -0.707);
	\draw[thick] (-0.193, -0.052) -- (-0.528, -0.141);
	\draw[thick] (-0.193, -0.052) -- (-0.259, -0.966);
}{M12p2-57}\fodgespace
%
%%% [58] O(p^2) 12-point diagram
\tikzineq[>=stealth]{
	\draw[thick] (0.370, -0.402) -- (0.219, -0.976);
	\draw[thick] (0.370, -0.402) -- (0.678, -0.735);
	\draw[thick] (0.370, -0.402) -- (0.955, -0.298);
	\draw[thick] (0.342, 0.179) -- (0.370, -0.402);
	\draw[thick] (0.342, 0.179) -- (0.976, 0.219);
	\draw[thick] (0.342, 0.179) -- (0.735, 0.678);
	\draw[thick] (0.016, 0.386) -- (0.342, 0.179);
	\draw[thick] (0.016, 0.386) -- (0.298, 0.955);
	\draw[thick] (0.016, 0.386) -- (-0.219, 0.976);
	\draw[thick] (-0.326, 0.207) -- (-0.678, 0.735);
	\draw[thick] (-0.326, 0.207) -- (-0.955, 0.298);
	\draw[thick] (-0.402, -0.370) -- (-0.976, -0.219);
	\draw[thick] (-0.402, -0.370) -- (-0.735, -0.678);
	\draw[thick] (-0.402, -0.370) -- (-0.298, -0.955);
	\draw[thick] (-0.326, 0.207) -- (-0.402, -0.370);
	\draw[thick] (0.016, 0.386) -- (-0.326, 0.207);
}{M12p2-58}\fodgespace
%
%%% [59] O(p^2) 12-point diagram
\tikzineq[>=stealth]{
	\draw[thick] (-0.071, 0.542) -- (0.383, 0.924);
	\draw[thick] (-0.071, 0.542) -- (-0.131, 0.991);
	\draw[thick] (-0.071, 0.542) -- (-0.609, 0.793);
	\draw[thick] (-0.077, -0.185) -- (-0.071, 0.542);
	\draw[thick] (-0.542, -0.071) -- (-0.924, 0.383);
	\draw[thick] (-0.542, -0.071) -- (-0.991, -0.131);
	\draw[thick] (-0.542, -0.071) -- (-0.793, -0.609);
	\draw[thick] (-0.077, -0.185) -- (-0.542, -0.071);
	\draw[thick] (-0.077, -0.185) -- (-0.383, -0.924);
	\draw[thick] (0.148, -0.357) -- (-0.077, -0.185);
	\draw[thick] (0.148, -0.357) -- (0.131, -0.991);
	\draw[thick] (0.148, -0.357) -- (0.609, -0.793);
	\draw[thick] (0.542, 0.071) -- (0.924, -0.383);
	\draw[thick] (0.542, 0.071) -- (0.991, 0.131);
	\draw[thick] (0.542, 0.071) -- (0.793, 0.609);
	\draw[thick] (0.148, -0.357) -- (0.542, 0.071);
}{M12p2-59}\fodgespace
%
%%% [60] O(p^2) 12-point diagram
\tikzineq[>=stealth]{
	\draw[thick] (0.366, -0.406) -- (0.208, -0.978);
	\draw[thick] (0.366, -0.406) -- (0.669, -0.743);
	\draw[thick] (0.366, -0.406) -- (0.951, -0.309);
	\draw[thick] (0.000, 0.000) -- (0.366, -0.406);
	\draw[thick] (0.406, 0.366) -- (0.978, 0.208);
	\draw[thick] (0.406, 0.366) -- (0.743, 0.669);
	\draw[thick] (0.406, 0.366) -- (0.309, 0.951);
	\draw[thick] (0.000, 0.000) -- (0.406, 0.366);
	\draw[thick] (-0.366, 0.406) -- (-0.208, 0.978);
	\draw[thick] (-0.366, 0.406) -- (-0.669, 0.743);
	\draw[thick] (-0.366, 0.406) -- (-0.951, 0.309);
	\draw[thick] (0.000, 0.000) -- (-0.366, 0.406);
	\draw[thick] (-0.406, -0.366) -- (0.000, 0.000);
	\draw[thick] (-0.406, -0.366) -- (-0.978, -0.208);
	\draw[thick] (-0.406, -0.366) -- (-0.743, -0.669);
	\draw[thick] (-0.406, -0.366) -- (-0.309, -0.951);
}{M12p2-60}\fodgespace
%
%%% [61] O(p^2) 12-point diagram
\tikzineq[>=stealth]{
	\draw[thick] (0.546, -0.000) -- (0.866, -0.500);
	\draw[thick] (0.546, -0.000) -- (1.000, -0.000);
	\draw[thick] (0.546, -0.000) -- (0.866, 0.500);
	\draw[thick] (0.200, -0.000) -- (0.546, -0.000);
	\draw[thick] (0.200, -0.000) -- (0.500, 0.866);
	\draw[thick] (0.000, 0.200) -- (0.000, 1.000);
	\draw[thick] (-0.473, 0.273) -- (-0.500, 0.866);
	\draw[thick] (-0.473, 0.273) -- (-0.866, 0.500);
	\draw[thick] (-0.473, 0.273) -- (-1.000, 0.000);
	\draw[thick] (0.000, 0.200) -- (-0.473, 0.273);
	\draw[thick] (-0.273, -0.473) -- (-0.866, -0.500);
	\draw[thick] (-0.273, -0.473) -- (-0.500, -0.866);
	\draw[thick] (-0.273, -0.473) -- (-0.000, -1.000);
	\draw[thick] (0.000, 0.200) -- (-0.273, -0.473);
	\draw[thick] (0.200, -0.000) -- (0.000, 0.200);
	\draw[thick] (0.200, -0.000) -- (0.500, -0.866);
}{M12p2-61}\fodgespace
%
%%% [62] O(p^2) 12-point diagram
\tikzineq[>=stealth]{
	\draw[thick] (0.333, 0.433) -- (0.924, 0.383);
	\draw[thick] (0.333, 0.433) -- (0.609, 0.793);
	\draw[thick] (0.333, 0.433) -- (0.131, 0.991);
	\draw[thick] (-0.077, 0.185) -- (0.333, 0.433);
	\draw[thick] (-0.077, 0.185) -- (-0.383, 0.924);
	\draw[thick] (-0.542, 0.071) -- (-0.793, 0.609);
	\draw[thick] (-0.542, 0.071) -- (-0.991, 0.131);
	\draw[thick] (-0.542, 0.071) -- (-0.924, -0.383);
	\draw[thick] (-0.077, 0.185) -- (-0.542, 0.071);
	\draw[thick] (-0.148, -0.357) -- (-0.077, 0.185);
	\draw[thick] (-0.148, -0.357) -- (-0.609, -0.793);
	\draw[thick] (-0.148, -0.357) -- (-0.131, -0.991);
	\draw[thick] (0.433, -0.333) -- (0.383, -0.924);
	\draw[thick] (0.433, -0.333) -- (0.793, -0.609);
	\draw[thick] (0.433, -0.333) -- (0.991, -0.131);
	\draw[thick] (-0.148, -0.357) -- (0.433, -0.333);
}{M12p2-62}\fodgespace
%
%%% [63] O(p^2) 12-point diagram
\tikzineq[>=stealth]{
	\draw[thick] (0.433, 0.333) -- (0.991, 0.131);
	\draw[thick] (0.433, 0.333) -- (0.793, 0.609);
	\draw[thick] (0.433, 0.333) -- (0.383, 0.924);
	\draw[thick] (-0.077, -0.185) -- (0.433, 0.333);
	\draw[thick] (-0.148, 0.357) -- (-0.131, 0.991);
	\draw[thick] (-0.148, 0.357) -- (-0.609, 0.793);
	\draw[thick] (-0.542, -0.071) -- (-0.924, 0.383);
	\draw[thick] (-0.542, -0.071) -- (-0.991, -0.131);
	\draw[thick] (-0.542, -0.071) -- (-0.793, -0.609);
	\draw[thick] (-0.148, 0.357) -- (-0.542, -0.071);
	\draw[thick] (-0.077, -0.185) -- (-0.148, 0.357);
	\draw[thick] (-0.077, -0.185) -- (-0.383, -0.924);
	\draw[thick] (0.333, -0.433) -- (-0.077, -0.185);
	\draw[thick] (0.333, -0.433) -- (0.131, -0.991);
	\draw[thick] (0.333, -0.433) -- (0.609, -0.793);
	\draw[thick] (0.333, -0.433) -- (0.924, -0.383);
}{M12p2-63}\fodgespace
%
%%% [64] O(p^2) 12-point diagram
\tikzineq[>=stealth]{
	\draw[thick] (0.273, -0.473) -- (0.000, -1.000);
	\draw[thick] (0.273, -0.473) -- (0.500, -0.866);
	\draw[thick] (0.273, -0.473) -- (0.866, -0.500);
	\draw[thick] (-0.000, 0.200) -- (0.273, -0.473);
	\draw[thick] (0.473, 0.273) -- (1.000, 0.000);
	\draw[thick] (0.473, 0.273) -- (0.866, 0.500);
	\draw[thick] (0.473, 0.273) -- (0.500, 0.866);
	\draw[thick] (-0.000, 0.200) -- (0.473, 0.273);
	\draw[thick] (-0.000, 0.200) -- (-0.000, 1.000);
	\draw[thick] (-0.200, -0.000) -- (-0.000, 0.200);
	\draw[thick] (-0.200, -0.000) -- (-0.500, 0.866);
	\draw[thick] (-0.546, -0.000) -- (-0.866, 0.500);
	\draw[thick] (-0.546, -0.000) -- (-1.000, -0.000);
	\draw[thick] (-0.546, -0.000) -- (-0.866, -0.500);
	\draw[thick] (-0.200, -0.000) -- (-0.546, -0.000);
	\draw[thick] (-0.200, -0.000) -- (-0.500, -0.866);
}{M12p2-64}\fodgespace
%
%%% [65] O(p^2) 12-point diagram
\tikzineq[>=stealth]{
	\draw[thick] (0.546, -0.000) -- (0.866, -0.500);
	\draw[thick] (0.546, -0.000) -- (1.000, -0.000);
	\draw[thick] (0.546, -0.000) -- (0.866, 0.500);
	\draw[thick] (0.200, -0.000) -- (0.546, -0.000);
	\draw[thick] (0.200, -0.000) -- (0.500, 0.866);
	\draw[thick] (-0.273, 0.473) -- (0.000, 1.000);
	\draw[thick] (-0.273, 0.473) -- (-0.500, 0.866);
	\draw[thick] (-0.273, 0.473) -- (-0.866, 0.500);
	\draw[thick] (-0.200, 0.000) -- (-0.273, 0.473);
	\draw[thick] (-0.200, 0.000) -- (-1.000, 0.000);
	\draw[thick] (-0.273, -0.473) -- (-0.866, -0.500);
	\draw[thick] (-0.273, -0.473) -- (-0.500, -0.866);
	\draw[thick] (-0.273, -0.473) -- (-0.000, -1.000);
	\draw[thick] (-0.200, 0.000) -- (-0.273, -0.473);
	\draw[thick] (0.200, -0.000) -- (-0.200, 0.000);
	\draw[thick] (0.200, -0.000) -- (0.500, -0.866);
}{M12p2-65}\fodgespace
%
%%% [66] O(p^2) 12-point diagram
\tikzineq[>=stealth]{
	\draw[thick] (0.528, -0.141) -- (0.707, -0.707);
	\draw[thick] (0.528, -0.141) -- (0.966, -0.259);
	\draw[thick] (0.528, -0.141) -- (0.966, 0.259);
	\draw[thick] (0.193, -0.052) -- (0.528, -0.141);
	\draw[thick] (0.193, -0.052) -- (0.707, 0.707);
	\draw[thick] (0.000, 0.386) -- (0.259, 0.966);
	\draw[thick] (0.000, 0.386) -- (-0.259, 0.966);
	\draw[thick] (-0.335, 0.193) -- (-0.707, 0.707);
	\draw[thick] (-0.335, 0.193) -- (-0.966, 0.259);
	\draw[thick] (-0.386, -0.386) -- (-0.966, -0.259);
	\draw[thick] (-0.386, -0.386) -- (-0.707, -0.707);
	\draw[thick] (-0.386, -0.386) -- (-0.259, -0.966);
	\draw[thick] (-0.335, 0.193) -- (-0.386, -0.386);
	\draw[thick] (0.000, 0.386) -- (-0.335, 0.193);
	\draw[thick] (0.193, -0.052) -- (0.000, 0.386);
	\draw[thick] (0.193, -0.052) -- (0.259, -0.966);
}{M12p2-66}\fodgespace
%
%%% [67] O(p^2) 12-point diagram
\tikzineq[>=stealth]{
	\draw[thick] (-0.000, 0.386) -- (0.259, 0.966);
	\draw[thick] (-0.000, 0.386) -- (-0.259, 0.966);
	\draw[thick] (-0.528, 0.141) -- (-0.707, 0.707);
	\draw[thick] (-0.528, 0.141) -- (-0.966, 0.259);
	\draw[thick] (-0.528, 0.141) -- (-0.966, -0.259);
	\draw[thick] (-0.000, 0.386) -- (-0.528, 0.141);
	\draw[thick] (-0.193, -0.335) -- (-0.000, 0.386);
	\draw[thick] (-0.193, -0.335) -- (-0.707, -0.707);
	\draw[thick] (-0.193, -0.335) -- (-0.259, -0.966);
	\draw[thick] (0.193, -0.335) -- (-0.193, -0.335);
	\draw[thick] (0.193, -0.335) -- (0.259, -0.966);
	\draw[thick] (0.193, -0.335) -- (0.707, -0.707);
	\draw[thick] (0.528, 0.141) -- (0.966, -0.259);
	\draw[thick] (0.528, 0.141) -- (0.966, 0.259);
	\draw[thick] (0.528, 0.141) -- (0.707, 0.707);
	\draw[thick] (0.193, -0.335) -- (0.528, 0.141);
}{M12p2-67}\fodgespace
%
%%% [68] O(p^2) 12-point diagram
\tikzineq[>=stealth]{
	\draw[thick] (0.528, -0.141) -- (0.707, -0.707);
	\draw[thick] (0.528, -0.141) -- (0.966, -0.259);
	\draw[thick] (0.528, -0.141) -- (0.966, 0.259);
	\draw[thick] (0.193, -0.052) -- (0.528, -0.141);
	\draw[thick] (0.193, -0.052) -- (0.707, 0.707);
	\draw[thick] (0.000, 0.386) -- (0.259, 0.966);
	\draw[thick] (0.000, 0.386) -- (-0.259, 0.966);
	\draw[thick] (-0.193, -0.052) -- (-0.707, 0.707);
	\draw[thick] (-0.528, -0.141) -- (-0.966, 0.259);
	\draw[thick] (-0.528, -0.141) -- (-0.966, -0.259);
	\draw[thick] (-0.528, -0.141) -- (-0.707, -0.707);
	\draw[thick] (-0.193, -0.052) -- (-0.528, -0.141);
	\draw[thick] (-0.193, -0.052) -- (-0.259, -0.966);
	\draw[thick] (0.000, 0.386) -- (-0.193, -0.052);
	\draw[thick] (0.193, -0.052) -- (0.000, 0.386);
	\draw[thick] (0.193, -0.052) -- (0.259, -0.966);
}{M12p2-68}\fodgespace
%
%%% [69] O(p^2) 12-point diagram
\tikzineq[>=stealth]{
	\draw[thick] (0.199, -0.022) -- (0.402, -0.916);
	\draw[thick] (0.543, -0.060) -- (0.806, -0.592);
	\draw[thick] (0.543, -0.060) -- (0.994, -0.110);
	\draw[thick] (0.543, -0.060) -- (0.916, 0.402);
	\draw[thick] (0.199, -0.022) -- (0.543, -0.060);
	\draw[thick] (0.199, -0.022) -- (0.592, 0.806);
	\draw[thick] (-0.000, 0.000) -- (0.199, -0.022);
	\draw[thick] (-0.000, 0.000) -- (0.110, 0.994);
	\draw[thick] (-0.242, 0.302) -- (-0.402, 0.916);
	\draw[thick] (-0.242, 0.302) -- (-0.806, 0.592);
	\draw[thick] (-0.500, -0.220) -- (-0.994, 0.110);
	\draw[thick] (-0.500, -0.220) -- (-0.916, -0.402);
	\draw[thick] (-0.500, -0.220) -- (-0.592, -0.806);
	\draw[thick] (-0.242, 0.302) -- (-0.500, -0.220);
	\draw[thick] (-0.000, 0.000) -- (-0.242, 0.302);
	\draw[thick] (-0.000, 0.000) -- (-0.110, -0.994);
}{M12p2-69}\fodgespace
%
%%% [70] O(p^2) 12-point diagram
\tikzineq[>=stealth]{
	\draw[thick] (0.100, -0.373) -- (0.000, -1.000);
	\draw[thick] (0.100, -0.373) -- (0.500, -0.866);
	\draw[thick] (0.546, 0.000) -- (0.866, -0.500);
	\draw[thick] (0.546, 0.000) -- (1.000, 0.000);
	\draw[thick] (0.546, 0.000) -- (0.866, 0.500);
	\draw[thick] (0.100, -0.373) -- (0.546, 0.000);
	\draw[thick] (0.100, 0.373) -- (0.100, -0.373);
	\draw[thick] (0.100, 0.373) -- (0.500, 0.866);
	\draw[thick] (0.100, 0.373) -- (-0.000, 1.000);
	\draw[thick] (-0.200, -0.000) -- (0.100, 0.373);
	\draw[thick] (-0.200, -0.000) -- (-0.500, 0.866);
	\draw[thick] (-0.546, -0.000) -- (-0.866, 0.500);
	\draw[thick] (-0.546, -0.000) -- (-1.000, -0.000);
	\draw[thick] (-0.546, -0.000) -- (-0.866, -0.500);
	\draw[thick] (-0.200, -0.000) -- (-0.546, -0.000);
	\draw[thick] (-0.200, -0.000) -- (-0.500, -0.866);
}{M12p2-70}\fodgespace
%
%%% [71] O(p^2) 12-point diagram
\tikzineq[>=stealth]{
	\draw[thick] (0.175, -0.345) -- (0.206, -0.979);
	\draw[thick] (0.175, -0.345) -- (0.668, -0.744);
	\draw[thick] (0.535, 0.113) -- (0.950, -0.311);
	\draw[thick] (0.535, 0.113) -- (0.979, 0.206);
	\draw[thick] (0.535, 0.113) -- (0.744, 0.668);
	\draw[thick] (0.175, -0.345) -- (0.535, 0.113);
	\draw[thick] (-0.000, 0.000) -- (0.175, -0.345);
	\draw[thick] (-0.000, 0.000) -- (0.311, 0.950);
	\draw[thick] (-0.175, 0.345) -- (-0.206, 0.979);
	\draw[thick] (-0.175, 0.345) -- (-0.668, 0.744);
	\draw[thick] (-0.535, -0.113) -- (-0.950, 0.311);
	\draw[thick] (-0.535, -0.113) -- (-0.979, -0.206);
	\draw[thick] (-0.535, -0.113) -- (-0.744, -0.668);
	\draw[thick] (-0.175, 0.345) -- (-0.535, -0.113);
	\draw[thick] (-0.000, 0.000) -- (-0.175, 0.345);
	\draw[thick] (-0.000, 0.000) -- (-0.311, -0.950);
}{M12p2-71}\fodgespace
%
%%% [72] O(p^2) 12-point diagram
\tikzineq[>=stealth]{
	\draw[thick] (0.546, -0.000) -- (0.866, -0.500);
	\draw[thick] (0.546, -0.000) -- (1.000, -0.000);
	\draw[thick] (0.546, -0.000) -- (0.866, 0.500);
	\draw[thick] (0.200, -0.000) -- (0.546, -0.000);
	\draw[thick] (0.200, -0.000) -- (0.500, 0.866);
	\draw[thick] (-0.000, -0.000) -- (0.000, 1.000);
	\draw[thick] (-0.200, 0.000) -- (-0.500, 0.866);
	\draw[thick] (-0.546, 0.000) -- (-0.866, 0.500);
	\draw[thick] (-0.546, 0.000) -- (-1.000, 0.000);
	\draw[thick] (-0.546, 0.000) -- (-0.866, -0.500);
	\draw[thick] (-0.200, 0.000) -- (-0.546, 0.000);
	\draw[thick] (-0.200, 0.000) -- (-0.500, -0.866);
	\draw[thick] (-0.000, -0.000) -- (-0.200, 0.000);
	\draw[thick] (-0.000, -0.000) -- (-0.000, -1.000);
	\draw[thick] (0.200, -0.000) -- (-0.000, -0.000);
	\draw[thick] (0.200, -0.000) -- (0.500, -0.866);
}{M12p2-72}\fodgespace
\end{center}
and has the stripped amplitude
{\addtolength{\peqindent}{-2cm}\addtolength{\ppeqindent}{-2cm}\addtolength{\pppeqindent}{-2cm}
\allowdisplaybreaks
\begin{align}
	-32&iF^{10}\ampl_{2,\{12\}} = 14s_{12} + 5s_{1234} + 2s_{123456}	
		\eqbreak{+}\frac{s_{12}+\ldots+s_{45} + s_\ol{14}+s_\ol{25}}{s_\ol{15}}\left\{
			-\frac{\left(s_{67}+\ldots+s_{9\A} + s_\ol{69}+s_\ol{7\A}\right)(s_\ol{69}+s_\ol{7\A})}{s_\ol{6\A}}
			\peqbreak{\hspace{4cm}\times}
			\left[2(s_{67}+\ldots+s_{\B\C}) + (s_\ol{69}+\ldots+s_\ol{9\C}) + 2(s_\ol{16} + s_\ol{6\B})\right]
			\peqbreak{\hspace{2cm}+}
			\frac{\left(s_{78}+\ldots+s_{\A\B} + s_\ol{7\A}+s_\ol{8\B}\right)(s_\ol{16}+s_\ol{6\B})}{2s_\ol{7\B}}
		\right\}
		\eqbreak{+} \frac{s_{12} + s_{23}}{s_{123}}\left\{
			\mvphantom
			-\left[5\left(s_{45}+\ldots+s_{\B\C} + s_\ol{\C3}+s_\ol{14}\right) + 2\left(s_\ol{47}+\ldots+s_\ol{9\C} + s_\ol{49}+\ldots+s_\ol{7\C}\right)\right]
			\peqbreak{+} \frac{s_{45} + s_{56}}{s_{456}}\left[	
				\mvphantom
				2\left(s_{78}+\ldots+s_{\B\C} + s_\ol{\C3}+s_\ol{47} + s_\ol{16}\right)
				+ s_\ol{7\A}+s_\ol{8\B}+s_\ol{9\C} + s_\ol{49}
				\ppeqbreak{+} \frac{s_{78} + s_{89}}{s_{789}}\left(
					-\left[s_{\A\B}+s_{\B\C} + s_\ol{7\A}+s_\ol{\C3} + s_\ol{16}+s_\ol{49}\right] 
					+ \frac{s_{\A\B} + s_{\B\C}}{2s_{\A\B\C}}
					\pppeqbreak{+} \frac{(s_\ol{7\A} + s_{\A\B})(s_\ol{16}+s_\ol{\C3})}{s_\ol{7\B}}
					+ \frac{(s_\ol{7\A} + s_\ol{49})(s_{\B\C} + s_{\C3})}{s_\ol{\B3}}
				\right)
				\ppeqbreak{+} \frac{s_{89} + s_{9\A}}{s_{89\A}}\left(
					-\left[s_{\B\C} + s_\ol{47}+s_\ol{7\A}+s_\ol{8\B}+s_\ol{\C3} + s_\ol{16}\right]
					+ \frac{(s_\ol{7\A} + s_\ol{8\B})(s_\ol{\C3} + s_\ol{16})}{s_\ol{7\B}}
					\pppeqbreak{+} \frac{(s_\ol{8\B} + s_\ol{\B\C})(s_\ol{47} + s_\ol{16})}{s_\ol{8\C}}
					+ \frac{(s_\ol{\B\C} + s_\ol{\C3})(s_\ol{47} + s_\ol{7\A})}{s_\ol{\B3}}
				\right)
				\ppeqbreak{+} \frac{s_{9\A} + s_{\A\B}}{s_{9\A\B}}\left(
					-\left[s_\ol{47}+s_{78} + s_\ol{9\C}+s_\ol{\C3} + s_\ol{16}\right]
					+ \frac{(s_\ol{47}+s_{78})(s_\ol{9\C}+s_\ol{\C3})}{s_\ol{48}}
					\pppeqbreak{+} \frac{(s_{78}+s_\ol{8\B})(s_\ol{\C3} + s_\ol{16})}{s_\ol{7\B}}
					+ \frac{(s_\ol{8\B} + s_\ol{9\C})(s_\ol{47} + s_\ol{16})}{s_\ol{8\C}}
				\right)
				\ppeqbreak{+} \frac{s_\ol{47} + s_{78}}{s_\ol{48}}\left(
					\frac{(s_\ol{49}+s_{9\A})(s_{\B\C}+s_\ol{\C3})}{s_\ol{\B3}}
					- \left[s_{9\A}+s_{\A\B}+s_{\B\C} + s_\ol{9\C}+s_\ol{\C3} + s_\ol{49}\right]
				\right)
				\ppeqbreak{-} 
				\frac{(s_\ol{\C3} + s_\ol{16})(s_{78}+\ldots+s_{\A\B} + s_\ol{7\A}+s_\ol{8\B})}{s_\ol{7\B}}
				\ppeqbreak{-} 
				\frac{(s_\ol{47} + s_\ol{16})(s_{89}+\ldots+s_{\B\C} + s_{8\B}+s_\ol{9\C})}{s_{8\C}}
				\ppeqbreak{-} 
				\frac{(s_{\B\C} + s_\ol{\C3})(s_{78}+s_{89}+s_{9\A} + s_\ol{47}+s_\ol{7\A} + s_\ol{49})}
					{s_\ol{\B3}}
			\right]
			\peqbreak{+} \frac{s_{56} + s_{67}}{s_{567}}\left[
				\mvphantom
				2\left(s_{89}+\ldots+s_{\B\C} + s_\ol{\C3}+s_\ol{14} + s_\ol{47}+s_\ol{58}\right) 
				+ s_\ol{8\B}+s_\ol{9\C} + s_\ol{49}+s_\ol{5\A}
				\ppeqbreak{+}\frac{s_{9\A} + s_{\A\B}}{s_{9\A\B}}\left(
					\frac{(s_\ol{47}+s_\ol{58})(s_\ol{9\C}+s_\ol{\C3})}{s_\ol{48}}
					- 1
				\right)
				\ppeqbreak{+}\frac{s_\ol{47}+s_\ol{58}}{s_\ol{48}}\left(
					\frac{(s_\ol{49}+s_{9\A})(s_{\B\C}+s_\ol{\C3})}{s_\ol{\B3}}
					- \left(s_{9\A}+s_{\A\B}+s_{\B\C} + s_\ol{9\C}+s_\ol{\C3} + s_\ol{49}\right)
				\right)
				\ppeqbreak{+}\frac{s_\ol{58}+s_{89}}{s_\ol{59}}\left(
					\frac{(s_{\B\C}+s_\ol{\C3})(s_\ol{49}+s_\ol{5\A})}{s_\ol{\B3}}
					+ \frac{(s_\ol{\C3}+s_\ol{14})(s_\ol{5\A}+s_{\A\B})}{s_\ol{\C4}}
					\pppeqbreak{-} 
					\mvphantom
					\left(s_{\A\B}+s_{\B\C} + s_\ol{\C3}+s_\ol{14} + s_\ol{49}+s_\ol{5\A}\right)
				\right)
				\ppeqbreak{-} 
				\frac{(s_\ol{14}+s_\ol{47})(s_{89}+\ldots+s_{\B\C} + s_\ol{8\B}+s_\ol{9\C})}{s_\ol{8\C}}
				\ppeqbreak{-} 
				\frac{(s_{\B\C}+s_\ol{\C3})(s_{89}+s_{9\A} + s_\ol{47}+s_\ol{58} + s_\ol{49}+s_\ol{5\A})}
					{s_\ol{\B3}}
				\ppeqbreak{-} 
				\frac{(s_\ol{\C3}+s_\ol{14})(s_{89}+s_{9\A}+s_{\A\B} + s_\ol{58}+s_\ol{8\B} + s_\ol{5\A})}
					{s_\ol{\C4}}
			\right]
			\peqbreak{+} \frac{s_{67} + s_{78}}{s_{678}}\left[
				2\left(s_{45} + s_{9\A}+s_{\A\B}+s_{\B\C} + s_\ol{\C3}+s_\ol{14} + s_\ol{58}+s_\ol{69}\right)
				+ s_\ol{9\C} + s_\ol{49}+s_\ol{5\A}+s_\ol{6\B}
				\ppeqbreak{+}\frac{s_{45} + s_\ol{14}}{s_\ol{15}}\left(
					\frac{(s_\ol{69}+s_{9\A})(s_\ol{6\B}+s_{\B\C})}{s_\ol{6\A}}
					- \left(s_{9\A}+s_{\A\B}+s_{\B\C} + s_\ol{69}+s_\ol{9\C} + s_\ol{6\B}\right)
				\right)
				\ppeqbreak{+}\frac{s_{45} + s_\ol{58}}{s_\ol{48}}\left(
					\frac{(s_\ol{49}+s_{9\A})(s_{\B\C}+s_\ol{\C3})}{s_\ol{\B3}}
					- \left(s_{9\A}+s_{\A\B}+s_{\B\C} + s_\ol{9\C}+s_\ol{\C3} + s_\ol{49}\right)
				\right)
				\ppeqbreak{+}\frac{s_\ol{58}+s_\ol{69}}{s_\ol{59}}\left(
					\frac{(s_{\B\C}+s_\ol{\C3})(s_\ol{49}+s_\ol{5\A})}{s_\ol{\B3}}
					+ \frac{(s_\ol{\C3}+s_\ol{14})(s_\ol{5\A}+s_{\A\B})}{s_\ol{\C4}}
					\pppeqbreak{-}
					\mvphantom
					\left(s_{\A\B}+s_{\B\C} + s_\ol{\C3}+s_\ol{14} + s_\ol{49}+s_\ol{5\A}\right)
				\right)
				\ppeqbreak{+}\frac{s_\ol{69}+s_{9\A}}{s_\ol{6\A}}\left(
					\frac{(s_{45}+s_\ol{5\A})(s_{\B\C}+s_\ol{\C3})}{s_\ol{\B3}}
					+ \frac{(s_\ol{\C3}+s_\ol{14})(s_\ol{5\A}+s_\ol{6\B})}{s_\ol{\C4}}
					\pppeqbreak{-}
					\mvphantom
					\left(s_{\B\C} + s_\ol{\C3}+s_\ol{14} + s_\ol{5\A}+s_\ol{6\B}\right)
				\right)
				\ppeqbreak{-}
				\frac{(s_{\B\C}+s_\ol{\C3})(s_{45}+s_{\B\C} + s_\ol{58}+s_\ol{69} + s_\ol{49}+s_\ol{5\A})}
					{s_\ol{\B3}}
				\ppeqbreak{-}
				\frac{(s_\ol{\C3}+s_\ol{14})(s_{9\A}+s_{\A\B} + s_\ol{58}+s_\ol{69} + s_\ol{5\A}+s_\ol{6\B})}
					{s_\ol{\C4}}
			\right]
			\peqbreak{+} \frac{s_{78} + s_{89}}{s_{789}}\left[
				\mvphantom
				\frac12(s_{45}+s_{56} + s_\ol{14}+s_\ol{69})
				+ 2(s_\ol{16}+s_\ol{49}+s_\ol{5\A}+s_\ol{6\B})
				\ppeqbreak{+}\frac{s_\ol{14}+s_{45}}{s_\ol{15}}\left(
					\frac{(s_\ol{6\B}+s_{\B\C})(s_\ol{69}+s_\ol{7\A})}{s_\ol{6\A}}
					+ \frac{(s_\ol{7\A}+s_{\A\B})(s_\ol{16}+s_\ol{6\B})}{2s_\ol{7\B}}
					\pppeqbreak{-}
					\mvphantom
					\left(s_{\A\B}+s_{\B\C} + s_\ol{69}+s_\ol{7\A} + s_\ol{16}+s_\ol{6\B}\right)
				\right)
				\ppeqbreak{+}\frac{s_{56}+s_\ol{69}}{s_\ol{59}}\left(
					\frac{(s_{\B\C}+s_\ol{\C3})(s_\ol{49}+s_\ol{5\A})}{2s_\ol{\B3}}
					+ \frac{(s_\ol{\C3}+s_\ol{14})(s_\ol{5\A}+s_{\A\B})}{s_\ol{\C4}}
					\pppeqbreak{-}
					\mvphantom
					\left(s_{\A\B}+s_{\B\C} + s_\ol{\C3}+s_\ol{14} + +s_\ol{49}+s_\ol{5\A}\right)
				\right)
				\ppeqbreak{+}\frac{s_\ol{69}+s_\ol{7\A}}{s_\ol{6\A}}\left(
					\frac{(s_\ol{\C3}+s_\ol{14})(s_\ol{5\A}+s_\ol{6\B})}{2s_\ol{\C4}}
					- \left(s_{45}+s_{\B\C} + s_\ol{\C3}+s_\ol{14} + s_\ol{5\A}+s_\ol{6\B}\right)
				\right)
			\right]
			\peqbreak{+} \frac{s_\ol{14}+s_{45}}{s_\ol{15}}\left[
				\mvphantom
				2\left(s_{67}+\ldots+s_{\B\C} + s_\ol{16}+s_\ol{6\B}\right)
				+ s_\ol{69}+\ldots+s_\ol{9\C}
				\ppeqbreak{-} 
				\frac{(s_\ol{6\B}+s_{\B\C})(s_{67}+\ldots+s_{9\A} + s_\ol{69}+s_\ol{7\A})}{s_\ol{6\A}}
				\ppeqbreak{-} 
				\frac{(s_\ol{16}+s_\ol{6\B})(s_{78}+\ldots+s_{\A\B} + s_\ol{7\A}+s_\ol{8\B})}{s_\ol{7\B}}
				\ppeqbreak{-} 
				\frac{(s_\ol{16}+s_{67})(s_{89}+\ldots+s_{\B\C} + s_\ol{8\B}+s_\ol{9\C})}{s_\ol{8\C}}
			\right]
			\peqbreak{+} \frac{s_{45}+\ldots+s_{78} + s_\ol{47}+s_\ol{58}}{s_\ol{48}}\left[
				\mvphantom
				\left(s_{9\A}+s_{\A\B}+s_{\B\C} + s_\ol{9\C}+s_\ol{\C3} + s_\ol{49}\right)
				\ppeqbreak{-} \frac{(s_\ol{49}+s_{9\A})(s_{\B\C}+s_\ol{\C3})}{s_\ol{\B3}}
			\right]
			\peqbreak{+} \frac{s_{56}+\ldots+s_{89} + s_\ol{58}+s_\ol{69}}{s_\ol{59}}\left[
				\mvphantom
				\left(s_{\A\B}+s_{\B\C} + s_\ol{\C3}+s_\ol{14} + s_\ol{49}+s_\ol{5\A}\right)
				\ppeqbreak{-} \frac{(s_\ol{49}+s_\ol{5\A})(s_{\B\C}+s_\ol{\C3}}{s_\ol{\B3}}
				- \frac{(s_\ol{5\A}+s_\ol{\A\B})(s_\ol{\C3}+s_\ol{14})}{s_\ol{\C4}}
			\right]
			\peqbreak{+} \frac{s_{67}+\ldots+s_{9\A} + s_\ol{69}+s_\ol{7\A}}{s_\ol{6\A}}\left[
				\mvphantom
				\left(s_{45}+s_{\B\C} + s_\ol{\C3}+s_\ol{14} + s_\ol{5\A}+s_\ol{6\B}\right)
				\ppeqbreak{-} \frac{(s_\ol{45}+s_\ol{5\A})(s_{\B\C}+s_\ol{\C3})}{s_\ol{\B3}}
				- \frac{(s_\ol{\C3}+s_\ol{14})(s_\ol{5\A}+s_\ol{6\B})}{s_\ol{\C4}}
			\right]
			\peqbreak{+} \frac{s_{78}+\ldots+s_{\A\B} + s_\ol{7\A}+s_\ol{8\B}}{s_\ol{7\B}}\left[
				\left(s_{45}+s_{56} + s_\ol{\C3}+s_\ol{14} + s_\ol{16}+s_\ol{6\B}\right)
				-\frac{(s_\ol{\C3}+s_\ol{14})(s_{56}+s_\ol{6\B})}{s_\ol{\C4}}
			\right]
			\peqbreak{+} \frac{(s_{89}+\ldots+s_{\B\C} + s_\ol{8\B}+s_\ol{9\C})
							(s_{56}+s_{67} + s_\ol{14}+s_\ol{47} + s_\ol{16})
				}{s_\ol{8\C}}
			\peqbreak{+} \frac{(s_{\B\C}+s_\ol{\C3})
				\left[2(s_{45}+\ldots+s_{9\A} + s_\ol{49}+s_\ol{9\A}) + s_\ol{47}+\ldots+s_\ol{7\A}\right]
				}{s_\ol{\B3}}
			\peqbreak{+} \frac{(s_\ol{\C3}+s_\ol{14})
				\left[2(s_{56}+\ldots+s_{\A\B} + s_\ol{5\A}+s_\ol{6\B}) + s_\ol{58}+\ldots+s_\ol{8\B}\right]
				}{s_\ol{\C4}}
		\right\} + [\Z_{12}].
	\label{eq:M12p2}
\end{align}}
We use the same abbreviations as above, with $\A,\B,\C=10,11,12$. Furthermore, we contract sums like $s_{12}+s_{23}+s_{34}+s_{45}$ to $s_{12}+\ldots+s_{45}$.

\bibliographystyle{myJHEP}
\bibliography{references}

\providecommand{\href}[2]{#2}\begingroup\raggedright\begin{thebibliography}{10}

\bibitem{gell-mann-levy}
M.~Gell-Mann and M.~Levy, \emph{{The axial vector current in beta decay}},
  \href{https://doi.org/10.1007/BF02859738}{\emph{Nuovo Cim.} {\bfseries 16}
  (1960) 705}.

\bibitem{weinberg-chpt}
S.~Weinberg, \emph{{Phenomenological Lagrangians}},
  \href{https://doi.org/10.1016/0378-4371(79)90223-1}{\emph{Physica} {\bfseries
  A96} (1979) 327}.

\bibitem{gasser-leutwyler-1}
J.~Gasser and H.~Leutwyler, \emph{{Chiral Perturbation Theory to One Loop}},
  \href{https://doi.org/10.1016/0003-4916(84)90242-2}{\emph{Annals Phys.}
  {\bfseries 158} (1984) 142}.

\bibitem{gasser-leutwyler-2}
J.~Gasser and H.~Leutwyler, \emph{{Chiral Perturbation Theory: Expansions in
  the Mass of the Strange Quark}},
  \href{https://doi.org/10.1016/0550-3213(85)90492-4}{\emph{Nucl. Phys.}
  {\bfseries B250} (1985) 465}.

\bibitem{Pich:2018ltt}
A.~Pich, \emph{{Effective Field Theory with Nambu-Goldstone Modes}},  in
  \emph{{Les Houches summer school: EFT in Particle Physics and Cosmology Les
  Houches, Chamonix Valley, France, July 3-28, 2017}}, 2018,
  \href{https://arxiv.org/abs/1804.05664}{{\ttfamily 1804.05664}}.

\bibitem{chpthomepage}
\href{http://home.thep.lu.se/\textasciitilde
  bijnens/chpt/}{http://home.thep.lu.se/\textasciitilde bijnens/chpt/}.

\bibitem{meson-meson}
J.~Bijnens and J.~Lu, \emph{{Meson-meson Scattering in QCD-like Theories}},
  \href{https://doi.org/10.1007/JHEP03(2011)028}{\emph{JHEP} {\bfseries 03}
  (2011) 028} [\href{https://arxiv.org/abs/1102.0172}{{\ttfamily 1102.0172}}].

\bibitem{pi-mass-decay}
J.~Bijnens and N.~Hermansson~Truedsson, \emph{{The Pion Mass and Decay Constant
  at Three Loops in Two-Flavour Chiral Perturbation Theory}},
  \href{https://doi.org/10.1007/JHEP11(2017)181}{\emph{JHEP} {\bfseries 11}
  (2017) 181} [\href{https://arxiv.org/abs/1710.01901}{{\ttfamily
  1710.01901}}].

\bibitem{Elvang:2013cua}
H.~Elvang and Y.-t. Huang, \emph{{Scattering Amplitudes}},
  \href{https://arxiv.org/abs/1308.1697}{{\ttfamily 1308.1697}}.

\bibitem{Kampf:2012fn}
K.~Kampf, J.~Novotny and J.~Trnka, \emph{{Recursion relations for tree-level
  amplitudes in the $SU(N)$ nonlinear sigma model}},
  \href{https://doi.org/10.1103/PhysRevD.87.081701}{\emph{Phys. Rev.}
  {\bfseries D87} (2013) 081701}
  [\href{https://arxiv.org/abs/1212.5224}{{\ttfamily 1212.5224}}].

\bibitem{Cheung:2015ota}
C.~Cheung, K.~Kampf, J.~Novotny, C.-H. Shen and J.~Trnka, \emph{{On-Shell
  Recursion Relations for Effective Field Theories}},
  \href{https://doi.org/10.1103/PhysRevLett.116.041601}{\emph{Phys. Rev. Lett.}
  {\bfseries 116} (2016) 041601}
  [\href{https://arxiv.org/abs/1509.03309}{{\ttfamily 1509.03309}}].

\bibitem{adler}
S.~L. Adler, \emph{{Consistency conditions on the strong interactions implied
  by a partially conserved axial vector current}},
  \href{https://doi.org/10.1103/PhysRev.137.B1022}{\emph{Phys. Rev.} {\bfseries
  137} (1965) B1022}.

\bibitem{Cheung:2016drk}
C.~Cheung, K.~Kampf, J.~Novotny, C.-H. Shen and J.~Trnka, \emph{{A Periodic
  Table of Effective Field Theories}},
  \href{https://doi.org/10.1007/JHEP02(2017)020}{\emph{JHEP} {\bfseries 02}
  (2017) 020} [\href{https://arxiv.org/abs/1611.03137}{{\ttfamily
  1611.03137}}].

\bibitem{Elvang:2018dco}
H.~Elvang, M.~Hadjiantonis, C.~R.~T. Jones and S.~Paranjape, \emph{{Soft
  Bootstrap and Supersymmetry}},
  \href{https://doi.org/10.1007/JHEP01(2019)195}{\emph{JHEP} {\bfseries 01}
  (2019) 195} [\href{https://arxiv.org/abs/1806.06079}{{\ttfamily
  1806.06079}}].

\bibitem{Low:2019ynd}
I.~Low and Z.~Yin, \emph{{Soft Bootstrap and Effective Field Theories}},
  \href{https://arxiv.org/abs/1904.12859}{{\ttfamily 1904.12859}}.

\bibitem{Osborn:1969ku}
H.~Osborn, \emph{{Implications of adler zeros for multipion processes}},
  \href{https://doi.org/10.1007/BF02755724}{\emph{Lett. Nuovo Cim.} {\bfseries
  2S1} (1969) 717}.

\bibitem{Susskind:1970gf}
L.~Susskind and G.~Frye, \emph{{Algebraic aspects of pionic duality diagrams}},
  \href{https://doi.org/10.1103/PhysRevD.1.1682}{\emph{Phys. Rev.} {\bfseries
  D1} (1970) 1682}.

\bibitem{Ellis:1970nt}
J.~R. Ellis and B.~Renner, \emph{{On the relationship between chiral and dual
  models}}, \href{https://doi.org/10.1016/0550-3213(70)90515-8}{\emph{Nucl.
  Phys.} {\bfseries B21} (1970) 205}.

\bibitem{Cachazo:2014xea}
F.~Cachazo, S.~He and E.~Y. Yuan, \emph{{Scattering Equations and Matrices:
  From Einstein To Yang-Mills, DBI and NLSM}},
  \href{https://doi.org/10.1007/JHEP07(2015)149}{\emph{JHEP} {\bfseries 07}
  (2015) 149} [\href{https://arxiv.org/abs/1412.3479}{{\ttfamily 1412.3479}}].

\bibitem{Gomez:2019cik}
H.~Gomez and A.~Helset, \emph{{Scattering equations and a new factorization for
  amplitudes. Part II. Effective field theories}},
  \href{https://doi.org/10.1007/JHEP05(2019)129}{\emph{JHEP} {\bfseries 05}
  (2019) 129} [\href{https://arxiv.org/abs/1902.02633}{{\ttfamily
  1902.02633}}].

\bibitem{Chen:2013fya}
G.~Chen and Y.-J. Du, \emph{{Amplitude Relations in Non-linear Sigma Model}},
  \href{https://doi.org/10.1007/JHEP01(2014)061}{\emph{JHEP} {\bfseries 01}
  (2014) 061} [\href{https://arxiv.org/abs/1311.1133}{{\ttfamily 1311.1133}}].

\bibitem{Chen:2014dfa}
G.~Chen, Y.-J. Du, S.~Li and H.~Liu, \emph{{Note on off-shell relations in
  nonlinear sigma model}},
  \href{https://doi.org/10.1007/JHEP03(2015)156}{\emph{JHEP} {\bfseries 03}
  (2015) 156} [\href{https://arxiv.org/abs/1412.3722}{{\ttfamily 1412.3722}}].

\bibitem{Du:2015esa}
Y.-J. Du and H.~Luo, \emph{{On single and double soft behaviors in NLSM}},
  \href{https://doi.org/10.1007/JHEP08(2015)058}{\emph{JHEP} {\bfseries 08}
  (2015) 058} [\href{https://arxiv.org/abs/1505.04411}{{\ttfamily
  1505.04411}}].

\bibitem{Low:2015ogb}
I.~Low, \emph{{Double Soft Theorems and Shift Symmetry in Nonlinear Sigma
  Models}}, \href{https://doi.org/10.1103/PhysRevD.93.045032}{\emph{Phys. Rev.}
  {\bfseries D93} (2016) 045032}
  [\href{https://arxiv.org/abs/1512.01232}{{\ttfamily 1512.01232}}].

\bibitem{Du:2016tbc}
Y.-J. Du and C.-H. Fu, \emph{{Explicit BCJ numerators of nonlinear simga
  model}}, \href{https://doi.org/10.1007/JHEP09(2016)174}{\emph{JHEP}
  {\bfseries 09} (2016) 174}
  [\href{https://arxiv.org/abs/1606.05846}{{\ttfamily 1606.05846}}].

\bibitem{Carrasco:2016ldy}
J.~J.~M. Carrasco, C.~R. Mafra and O.~Schlotterer, \emph{{Abelian Z-theory:
  NLSM amplitudes and $\alpha$'-corrections from the open string}},
  \href{https://doi.org/10.1007/JHEP06(2017)093}{\emph{JHEP} {\bfseries 06}
  (2017) 093} [\href{https://arxiv.org/abs/1608.02569}{{\ttfamily
  1608.02569}}].

\bibitem{Du:2016njc}
Y.-J. Du and H.~Luo, \emph{{Leading order multi-soft behaviors of tree
  amplitudes in NLSM}},
  \href{https://doi.org/10.1007/JHEP03(2017)062}{\emph{JHEP} {\bfseries 03}
  (2017) 062} [\href{https://arxiv.org/abs/1611.07479}{{\ttfamily
  1611.07479}}].

\bibitem{Cheung:2017yef}
C.~Cheung, G.~N. Remmen, C.-H. Shen and C.~Wen, \emph{{Pions as Gluons in
  Higher Dimensions}},
  \href{https://doi.org/10.1007/JHEP04(2018)129}{\emph{JHEP} {\bfseries 04}
  (2018) 129} [\href{https://arxiv.org/abs/1709.04932}{{\ttfamily
  1709.04932}}].

\bibitem{Low:2017mlh}
I.~Low and Z.~Yin, \emph{{Ward Identity and Scattering Amplitudes for Nonlinear
  Sigma Models}},
  \href{https://doi.org/10.1103/PhysRevLett.120.061601}{\emph{Phys. Rev. Lett.}
  {\bfseries 120} (2018) 061601}
  [\href{https://arxiv.org/abs/1709.08639}{{\ttfamily 1709.08639}}].

\bibitem{Low:2018acv}
I.~Low and Z.~Yin, \emph{{The Infrared Structure of Nambu-Goldstone Bosons}},
  \href{https://doi.org/10.1007/JHEP10(2018)078}{\emph{JHEP} {\bfseries 10}
  (2018) 078} [\href{https://arxiv.org/abs/1804.08629}{{\ttfamily
  1804.08629}}].

\bibitem{Rodina:2018pcb}
L.~Rodina, \emph{{Scattering Amplitudes from Soft Theorems and Infrared
  Behavior}}, \href{https://doi.org/10.1103/PhysRevLett.122.071601}{\emph{Phys.
  Rev. Lett.} {\bfseries 122} (2019) 071601}
  [\href{https://arxiv.org/abs/1807.09738}{{\ttfamily 1807.09738}}].

\bibitem{Mizera:2018jbh}
S.~Mizera and B.~Skrzypek, \emph{{Perturbiner Methods for Effective Field
  Theories and the Double Copy}},
  \href{https://doi.org/10.1007/JHEP10(2018)018}{\emph{JHEP} {\bfseries 10}
  (2018) 018} [\href{https://arxiv.org/abs/1809.02096}{{\ttfamily
  1809.02096}}].

\bibitem{Bjerrum-Bohr:2018jqe}
N.~E.~J. Bjerrum-Bohr, H.~Gomez and A.~Helset, \emph{{New factorization
  relations for nonlinear sigma model amplitudes}},
  \href{https://doi.org/10.1103/PhysRevD.99.045009}{\emph{Phys. Rev.}
  {\bfseries D99} (2019) 045009}
  [\href{https://arxiv.org/abs/1811.06024}{{\ttfamily 1811.06024}}].

\bibitem{Cornwell:1971sp}
D.~T. Cornwell, \emph{{The six pion amplitude to fourth order in momenta}},
  \href{https://doi.org/10.1016/0550-3213(71)90115-5}{\emph{Nucl. Phys.}
  {\bfseries B34} (1971) 125}.

\bibitem{Carrillo-Gonzalez:2019aao}
M.~Carrillo-Gonzalez, R.~Penco and M.~Trodden, \emph{{Shift symmetries, soft
  limits, and the double copy beyond leading order}},
  \href{https://arxiv.org/abs/1908.07531}{{\ttfamily 1908.07531}}.

\bibitem{Kampf:2013vha}
K.~Kampf, J.~Novotny and J.~Trnka, \emph{{Tree-level Amplitudes in the
  Nonlinear Sigma Model}},
  \href{https://doi.org/10.1007/JHEP05(2013)032}{\emph{JHEP} {\bfseries 05}
  (2013) 032} [\href{https://arxiv.org/abs/1304.3048}{{\ttfamily 1304.3048}}].

\bibitem{masterthesis}
M.~Sjö, \emph{{Flavour-ordering in the nonlinear sigma model with more
  derivatives and legs}},  Master thesis LU TP 19-21, Lund University, June
  2019.

\bibitem{supplementary}
 {File \texttt{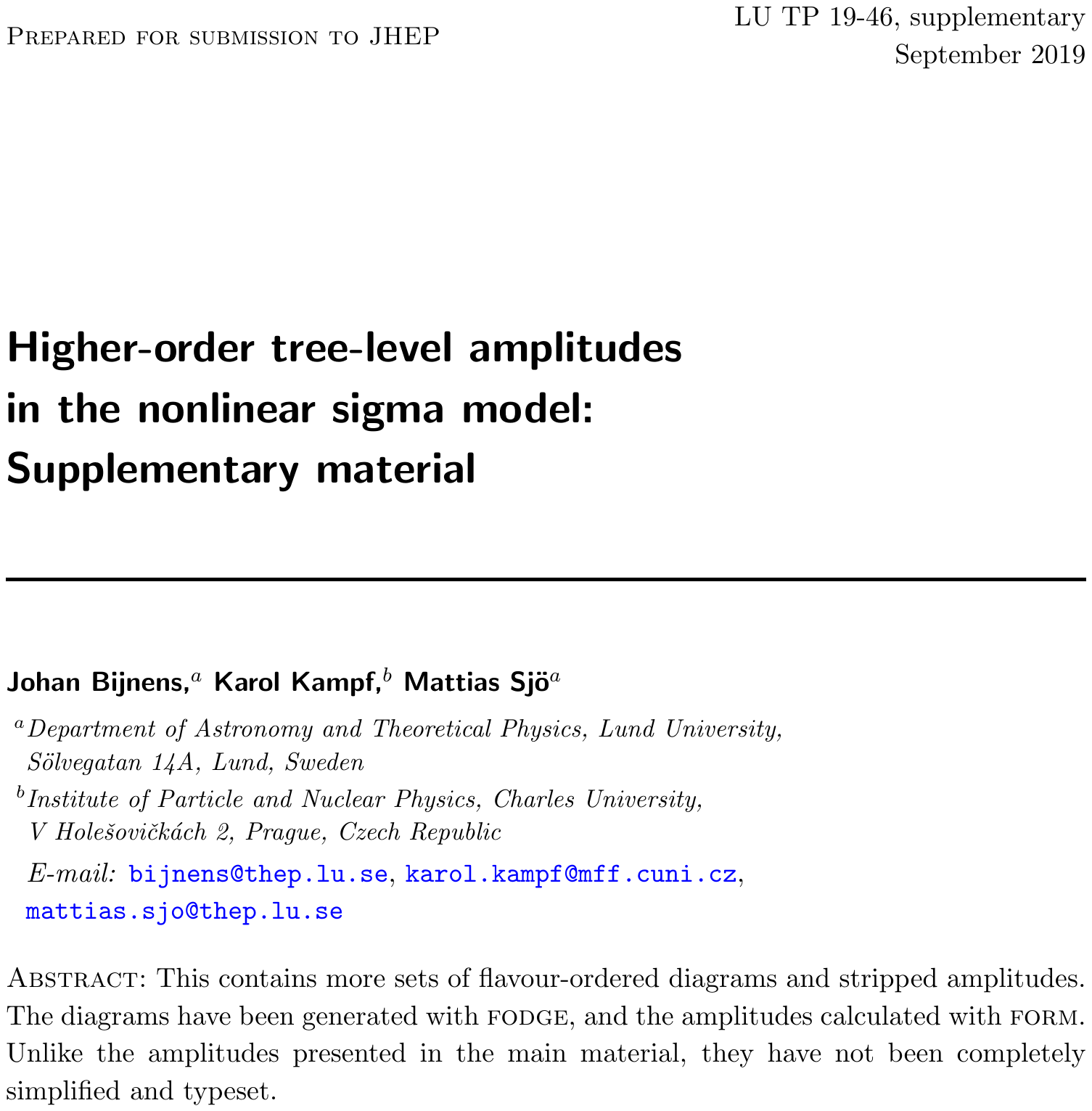} included with this submission}.

\bibitem{better-basis}
G.~Ecker, J.~Gasser, A.~Pich and E.~de~Rafael, \emph{{The Role of Resonances in
  Chiral Perturbation Theory}},
  \href{https://doi.org/10.1016/0550-3213(89)90346-5}{\emph{Nucl. Phys.}
  {\bfseries B321} (1989) 311}.

\bibitem{p6lagr}
J.~Bijnens, G.~Colangelo and G.~Ecker, \emph{{The Mesonic chiral Lagrangian of
  order p$^{6}$}},
  \href{https://doi.org/10.1088/1126-6708/1999/02/020}{\emph{JHEP} {\bfseries
  02} (1999) 020} [\href{https://arxiv.org/abs/hep-ph/9902437}{{\ttfamily
  hep-ph/9902437}}].

\bibitem{p8lagr}
J.~Bijnens, N.~Hermansson-Truedsson and S.~Wang, \emph{{The order p$^{8}$
  mesonic chiral Lagrangian}},
  \href{https://doi.org/10.1007/JHEP01(2019)102}{\emph{JHEP} {\bfseries 01}
  (2019) 102} [\href{https://arxiv.org/abs/1810.06834}{{\ttfamily
  1810.06834}}].

\bibitem{reuschle}
C.~Reuschle and S.~Weinzierl, \emph{{Decomposition of one-loop QCD amplitudes
  into primitive amplitudes based on shuffle relations}},
  \href{https://doi.org/10.1103/PhysRevD.88.105020}{\emph{Phys. Rev.}
  {\bfseries D88} (2013) 105020}
  [\href{https://arxiv.org/abs/1310.0413}{{\ttfamily 1310.0413}}].

\bibitem{colour-order}
T.~Schuster, \emph{{Color ordering in QCD}},
  \href{https://doi.org/10.1103/PhysRevD.89.105022}{\emph{Phys. Rev.}
  {\bfseries D89} (2014) 105022}
  [\href{https://arxiv.org/abs/1311.6296}{{\ttfamily 1311.6296}}].

\bibitem{mangano-parke}
M.~L. Mangano and S.~J. Parke, \emph{{Multiparton amplitudes in gauge
  theories}}, \href{https://doi.org/10.1016/0370-1573(91)90091-Y}{\emph{Phys.
  Rept.} {\bfseries 200} (1991) 301}
  [\href{https://arxiv.org/abs/hep-th/0509223}{{\ttfamily hep-th/0509223}}].

\bibitem{multiplet-1}
M.~Sjödahl and J.~Thorén, \emph{{Decomposing color structure into multiplet
  bases}}, \href{https://doi.org/10.1007/JHEP09(2015)055}{\emph{JHEP}
  {\bfseries 09} (2015) 055}
  [\href{https://arxiv.org/abs/1507.03814}{{\ttfamily 1507.03814}}].

\bibitem{adler-proof}
S.~Weinberg, \emph{The quantum theory of fields}, vol.~2. Cambridge university
  press, 1996.

\bibitem{eft-soft}
C.~Cheung, K.~Kampf, J.~Novotný and J.~Trnka, \emph{{Effective Field Theories
  from Soft Limits of Scattering Amplitudes}},
  \href{https://doi.org/10.1103/PhysRevLett.114.221602}{\emph{Phys. Rev. Lett.}
  {\bfseries 114} (2015) 221602}
  [\href{https://arxiv.org/abs/1412.4095}{{\ttfamily 1412.4095}}].

\bibitem{simplest-qft}
N.~Arkani-Hamed, F.~Cachazo and J.~Kaplan, \emph{{What is the Simplest Quantum
  Field Theory?}}, \href{https://doi.org/10.1007/JHEP09(2010)016}{\emph{JHEP}
  {\bfseries 09} (2010) 016} [\href{https://arxiv.org/abs/0808.1446}{{\ttfamily
  0808.1446}}].

\bibitem{Vermaseren:2000nd}
J.~A.~M. Vermaseren, \emph{{New features of FORM}},
  \href{https://arxiv.org/abs/math-ph/0010025}{{\ttfamily math-ph/0010025}}.

\bibitem{Kuipers:2012rf}
J.~Kuipers, T.~Ueda, J.~A.~M. Vermaseren and J.~Vollinga, \emph{{FORM version
  4.0}}, \href{https://doi.org/10.1016/j.cpc.2012.12.028}{\emph{Comput. Phys.
  Commun.} {\bfseries 184} (2013) 1453}
  [\href{https://arxiv.org/abs/1203.6543}{{\ttfamily 1203.6543}}].

\bibitem{Bijnens:2010xg}
J.~Bijnens and L.~Carloni, \emph{{The Massive O(N) Non-linear Sigma Model at
  High Orders}},
  \href{https://doi.org/10.1016/j.nuclphysb.2010.09.019}{\emph{Nucl. Phys.}
  {\bfseries B843} (2011) 55}
  [\href{https://arxiv.org/abs/1008.3499}{{\ttfamily 1008.3499}}].

\bibitem{Bijnens:2013yca}
J.~Bijnens, K.~Kampf and S.~Lanz, \emph{{Leading logarithms in N-flavour
  mesonic Chiral Perturbation Theory}},
  \href{https://doi.org/10.1016/j.nuclphysb.2013.04.012}{\emph{Nucl. Phys.}
  {\bfseries B873} (2013) 137}
  [\href{https://arxiv.org/abs/1303.3125}{{\ttfamily 1303.3125}}].

\bibitem{p6lagr-renorm}
J.~Bijnens, G.~Colangelo and G.~Ecker, \emph{{Renormalization of chiral
  perturbation theory to order $p^6$}},
  \href{https://doi.org/10.1006/aphy.1999.5982}{\emph{Annals Phys.} {\bfseries
  280} (2000) 100} [\href{https://arxiv.org/abs/hep-ph/9907333}{{\ttfamily
  hep-ph/9907333}}].

\end{thebibliography}\endgroup

\end{document}